\documentclass{aa}
\usepackage{txfonts}
\usepackage{graphicx, color}
\usepackage[colorlinks]{hyperref}
\usepackage{amsmath}
\usepackage{pdflscape}
\usepackage{multirow,bigdelim}
\usepackage{calc}
\usepackage{threeparttable}

\usepackage[normalem]{ulem}

\usepackage{caption}
\usepackage[labelformat = empty,position=top]{subcaption}
\usepackage{mathptmx}
\usepackage{multirow}

\definecolor{dgreen}{rgb}{0., 0.7, 0.}

\newcommand{\Msun}{\mbox{$\mathrm{M}_{\odot}$}}
\newcommand{\Lsun}{\mbox{$\mathrm{L}_{\odot}$}}

\newcommand{\Mi}{\mbox{$M_\mathrm{i}$}}

\newcommand{\Teff}{\mbox{$\mathrm{T}_{\mathrm{eff}}$}}
\newcommand{\Msunyr}{\mbox{$\mathrm{M}_{\odot}/\mathrm{yr}$}}

\usepackage[T1]{fontenc}
\usepackage{ae,aecompl}
{}

\usepackage{fixltx2e}
{\left\lbrace\begin{array}{@{}l@{}}}%
{\end{array}\right.}
\begin{document}

\title{PARSEC V2.0: Stellar tracks and isochrones of low and intermediate mass stars with rotation}
\titlerunning{\textsc{parsec V2.0} tracks and isochrones}
 % Society: \\ \LaTeXe\ style guide for authors}

\subtitle{}

%\pagerange{\pageref{firstpage}--\pageref{lastpage}} \pubyear{2002}

   \author{C. T. Nguyen
          \inst{1,3}
          \and
          G. Costa %         \inst{2}\fnmsep\thanks{Just to show the usage  of the elements in the author field}
         \inst{2,3,4}
         \and
          L. Girardi %         \inst{2}\fnmsep\thanks{Just to show the usage  of the elements in the author field}
         \inst{3}
         \and
          G. Volpato %         \inst{2}\fnmsep\thanks{Just to show the usage  of the elements in the author field}
         \inst{2}
         \and
          A. Bressan
          \inst{1}
          \and
          Y. Chen
         \inst{5,6}
         \and
          P. Marigo %         \inst{2}\fnmsep\thanks{Just to show the usage  of the elements in the author field}
         \inst{2}
          \and
          X. Fu
         \inst{1,7,8}
         \and
          P.~Goudfrooij \inst{9}
          }

        \institute{
        SISSA, Via Bonomea 265, I-34136 Trieste, Italy,\\
        \email{cnguyen@sissa.it}  %    \thanks{The university of heaven temporarily does not accept e-mails}
        \and
        Dipartimento di Fisica e Astronomia, Universit\`a degli studi di Padova,
        Vicolo Osservatorio 3, Padova, Italy,\\ \email{guglielmo.costa@unipd.it}
        %\email{paola.marigo@unipd.it}
        \and
        INAF Osservatorio Astronomico di Padova, Vicolo dell'Osservatorio n. 5, Padova, Italy
        %\email{leo.girardi@inaf.it}
        \and
        INFN - Padova, Via Marzolo 8, I--35131, Padova, Italy
        \and
        Anhui University, Hefei 230601, China
        \and
        National Astronomical Observatories, Chinese Academy of Sciences, Beijing 100101, China
%        \email{cy@ahu.edu.cn}
        \and
        Purple Mountain Observatory, Chinese Academy of Sciences, Nanjing 210023, China
        \and
        INAF Osservatorio di Astrofisica e Fisica dello Spazio, via Gobetti 93/3, 40129 Bologna, Italy
        \and
        Space Telescope Science Institute, 3700 San Martin Drive, Baltimore, MD 21218, USA
      }

\authorrunning{Nguyen et al.}

   \date{}

\hypersetup{
    linkcolor=blue,
    citecolor=blue,
    filecolor=magenta,      
    urlcolor=blue
}
% \abstract{}{}{}{}{}
% 5 {} token are mandatory

\abstract{We present a new comprehensive collection of stellar evolutionary tracks and isochrones for rotating low- and intermediate-mass stars assembled with the updated version of \textsc{parsec V2.0}. This version includes our recent calibration of the extra mixing from overshooting and rotation, as well as several improvements in nuclear reaction network, treatment of convective zones, mass loss and other physical input parameters. The initial mass of the stellar models covers the range from 0.09~\Msun\ to 14~\Msun, for  six sets of initial metallicity,  from Z=0.004 to Z=0.017.  Rotation is considered for stars above $\sim 1~\Msun$ with a smooth transition between non-rotating and extremely fast-rotating models, based on the initial mass.
For stars more massive than $\sim 1.3~\Msun$ the full rotation range, from low to the critical one, is considered. We adopt the solar-scaled chemical mixtures with Z$_\odot$ = 0.01524. All the evolutionary phases from the pre-main-sequence to the first few thermal-pulses on the asymptotic-giant-branch or central C exhaustion, are considered. 
The corresponding theoretical isochrones are further derived with \textsc{trilegal} code and are converted in several photometric systems, taking into account different inclination angles. Besides magnitudes, they also offer many other stellar observables in line with the data that are being provided by current large surveys. 
The new collection is fully integrated in a user friendly WEB interface for the benefit of easily performing stellar population studies.}%context

   \keywords{Stars: evolution - Stars: rotation - Stars: Hertzsprung-Russell and C-M diagrams - Stars: low-mass.}
   
   \maketitle
   
\section{Introduction}

The PAdova and tRieste Stellar Evolutionary Code, \textsc{parsec}, was first implemented in \citet{2012MNRAS.427..127B}, and then used in several works aimed to produce large grids of stellar evolutionary tracks and isochrones. For instance,
\citet{2014MNRAS.444.2525C} extended the calculation to very low-mass star models, \citet{2014MNRAS.445.4287T} and \citet{2015MNRAS.452.1068C} pursued massive stars up to 350~$\Msun$, and \citet{2018MNRAS.476..496F} studied the evolution with $\alpha$-enhanced compositions. Extended sets of isochrones using \textsc{parsec} tracks were described in \citet{2012MNRAS.427..127B} and \citet{2017ApJ...835...77M}. More recently, a significant development was presented in \citet{2019MNRAS.485.4641C}, who included the effects of rotation in the new version of the code 
\textsc{parsec V2.0}\footnote{Stellar tracks and isochrones computed in this work are available at the following links: \url{http://stev.oapd.inaf.it/PARSEC}, and \url{http://stev.oapd.inaf.it/cmd}, respectively.}. 

As described in \citet{1924MNRAS..84..665V, 1924MNRAS..84..684V, 1970A&A.....5..155K, 1992A&A...265..115Z, 1997A&A...321..465M, Chieffi2013, Chieffi2017} rotation might have a significant impact on the stellar structure induced by both geometrical distortion, extra-mixing, and enhanced mass loss rates. Observational evidence on the large fractions of rapidly rotating stars among the Milky Way field stars \citep[e.g.][]{2007A&A...463..671R} and in star clusters in Magellanic Clouds \citep[e.g.][]{2017ApJ...846L...1D, 2017NatAs...1E.186D},
suggests that rotation may indeed become an important driving agent for stellar evolution. Furthermore, it may be concurrent with other physical processes which drive extra-mixing, such as the convective overshooting \citep[see e.g.][]{Jermyn2018, 2019MNRAS.485.4641C}.
The effect of extra mixing caused by overshooting from the unstable core  has been introduced a few decades ago \citep[e.g.][]{1965ApJ...142.1468S,1975A&A....40..303M,Roxburgh1978,1981A&A...102...25B,Bressan86,Bertelli1984,Bertelli1990,Bressan1993,Meynet_etal94,  Fagotto1994a, Fagotto1994b,Girardi2000}, 
and is now incorporated in most libraries of stellar evolutionary tracks \citep[e.g.][]{2004ApJS..155..667D, 2004ApJ...612..168P,Weiss2008, Paxton2011ApJS, Paxton2018ApJS,2012A&A...541A..41M, 2012MNRAS.427..127B,  Bossini2015MNRAS, 2017ApJ...838..161S, 2018ApJ...856..125H}. 
Many authors also suggest a variation of the overshooting efficiency, usually parameterized by the efficiency parameter $\lambda_{\mathrm{ov}}$, with the initial mass \citep[see e.g. ][]{Pols1998}.  
Analysis of double-lined eclipsing binaries \citep[DLEBs;][]{2016A&A...592A..15C, 2017ApJ...849...18C, 2018ApJ...859..100C, 2019ApJ...876..134C} supports 
a growing efficiency in the mass range between $\sim 1-1.7~\Msun$ with a plateau 
in $\lambda_{\mathrm{ov}}$ above this mass range. 
However, the best fits of the DLEBs parameters require
a certain degree of stochasticity 
in some other important parameters, such as the mixing length scale, which in our opinion is difficult to accept, particularly in the case of binary components with the same mass. 
Indeed, \citet{2019MNRAS.485.4641C}  showed that the observations of DLEBs could be well explained with the interplay between a {\sl fixed} overshooting efficiency and a varying initial rotational velocity. In fact, the latter also depends on environmental conditions.
The results obtained by \citet{2019MNRAS.485.4641C}, can thus be considered as an important step in the calibration of the efficiency of the overshooting phenomenon, at least in the domain of low- and intermediate-mass stars.
This calibration has been subsequently supported by a combined analysis of Cepheids in the LMC star cluster NGC1866 and the color-magnitude diagrams (CMDs) of its multiple stellar populations \citep{2019A&A...631A.128C}. 

After these initial tests performed with the new code, we here present in this paper the new sets of evolutionary tracks and the corresponding isochrones for the \textsc{parsec} models with rotation. 
The initial mass range presented in this paper goes from $0.09~\Msun$ to $14~\Msun$. 
Models of more massive stars have already been computed 
for some particular purposes \citep{Spera2019MNRAS.485..889S,Costa2021MNRAS.501.4514C, Costa2022arXiv220403492C} but the full set including rotation is still in preparation and will be presented in a dedicated paper.

All the tracks start at the pre-main-sequence (PMS) phase and are terminated at a stage that depends on the initial mass: either at ages largely exceeding the Hubble time, or at the initial stages of thermally-pulsing asymptotic giant branch, or at carbon exhaustion for more massive stars. The tracks are computed with an initial metal content ranging from $Z=0.004$ to $0.017$, and with an initial He mass fraction following a linear enrichment law \citep{2012MNRAS.427..127B}. 
Tracks at lower metallicity are being computed with an enhanced partition, the details of these models with rotation will be presented in the coming project. 
For every metallicity, we consider the initial rotation rates from zero to the critical value. 
The theoretical isochrones are then derived and converted into several photometric systems. 

The structure of this paper is as follows. In Sect.~\ref{inputphysics} we review the main input physics used in the present calculations. We will pay particular attention to differences with respect to the previous non-rotating models, in the case of important changes of physical input. 
In Sect.~\ref{evoltrack} we describe the effects of rotational mixing on the evolution of our stellar models. 
In this section, we also compare our current models with previous non-rotating ones. We also perform some comparison with some existing models in the literature.
The corresponding isochrones and the quantities provided are described in detail in Sect.~\ref{iso}. 
Finally, in Sect.~\ref{Discussion}, we show a few preliminary fits to the observed CMDs of the two open clusters M67 and NGC 6633, and draw our main conclusions.

\section{Input physics}\label{inputphysics}

The input physics used in \textsc{parsec} is described in \citet{2012MNRAS.427..127B, 2014MNRAS.445.4287T, 2014MNRAS.444.2525C, 2018MNRAS.476..496F} and \citet{2019MNRAS.485.4641C, 2019A&A...631A.128C}. Here, we briefly summarise the main points that are more important for the low- and intermediate-mass stars dealt with in this paper.

\subsection{Solar metallicity, opacities, nuclear reactions, mixing length, equation of state}

The abundance of elements heavier than $^4$He in the Sun is still uncertain. The early compilation by \citet{1998SSRv...85..161G}, consisting of the abundances of 90 elements from lithium to uranium, yielded the solar metallicity $Z_\odot=0.017$. Later, \citet{2006NuPhA.777....1A, 2009ARA&A..47..481A} claimed lower values,  $Z_\odot=0.0122$ and $Z_\odot=0.0134$, respectively, or \citet{2009LanB...4B..712L} with $Z_\odot=0.0141$. Recent solar wind measurements give $Z_\odot=0.0196\pm 0.0014$ \citep{2016ApJ...816...13V}. 
In this paper, we adopt the solar-scaled mixtures by \citet{2011SoPh..268..255C} where the current solar metallicity is $Z_\odot=0.01524$, which is an intermediate value between those preferred by 
\citet{ 2009ARA&A..47..481A} and \citet{2016ApJ...816...13V}.
Further extensions to other metallicity ranges with more suitable input physics, e.g. $\alpha-$enhanced mixtures will be provided in the forthcoming works.

The Rosseland mean opacities, $\kappa_{\mathrm{rad}}$, are the same as those of \textsc{parsec} V1.2S. In the high temperature regime, $4.2\leq\log (T/\mathrm{K})\leq 8.7$, the opacity tables are provided by the Opacity Project At Livermore \citep[OPAL; see][]{1996ApJ...464..943I}, while in the low temperature regime $3.2\leq\log (T/K)\leq 4.1$ we generate the opacity tables with the \textsc{AESOPUS} tool \citep[see][for details]{2009A&A...508.1539M}. In the transition region $4.1\leq\log (T/K)\leq 4.2$, the opacities are linearly interpolated between the OPAL and \textsc{AESOPUS} values. The contribution from conduction is computed following \citet{2008ApJ...677..495I}.

The transport of convective energy is described by the mixing length theory of \citet{1958ZA.....46..245B}, adopting the value of mixing-length parameter $\alpha_{\mathrm{MLT}}=1.74$ calibrated on the solar model by \citet{2012MNRAS.427..127B} (see also \citealt{2019A&A...621A..84S} for more calibrations). It is interesting, however, to note that the variation of $\alpha_{\mathrm{MLT}}$ for different stars has been recently called out, e.g. \citet{2018ApJ...858...28V} suggests a dependency of $\alpha_{\mathrm{MLT}}/\alpha_{\mathrm{MLT}\odot}$ on gravity, effective temperature and metallicity, while the study of \citet{2020A&A...635A.176S} in FGK stars mainly focuses on the impact from metallicity. While these recent studies might be an inspiration for future works, at the moment we use the solar mixing-length parameter for all calculations, $\alpha_{\mathrm{MLT}}=\alpha_{\mathrm{MLT}\odot}=1.74$.

The equation of state in \textsc{parsec V2.0} is computed with the freely available \textsc{FREEEOS} code developed by A.W.~Irwin\footnote{http://freeeos.sourceforge.net/}. 

The nuclear reaction network, after the updates by \citet{2018MNRAS.476..496F} 
and by \citet{Costa2021MNRAS.501.4514C},
includes the p-p chains, the CNO tri-cycle, the Ne-Na and Mg-Al chains, 
$^{12}$C, $^{16}$O and $^{20}$Ne burning reactions,
and the $\alpha$-capture reactions up to $^{56}$Ni, for a total of 72 different reactions tracing 32 isotopes: 
$^1$H, D, $^3$He, $^4$He,   $^7$Li,   $^7$Be,  $^{12}$C, $^{13}$C, $^{14}$N, $^{15}$N, $^{16}$O, $^{17}$O, $^{18}$O, $^{19}$F, $^{20}$Ne, $^{21}$Ne, $^{22}$Ne, $^{23}$Na,  $^{24}$Mg, $^{25}$Mg, $^{26}$Mg, $^{26}$Al, $^{27}$Al, $^{28}$Si, 
$^{32}$S ,$^{36}$Ar,   $^{40}$Ca,    $^{44}$Ti,    
$^{48}$Cr,    $^{52}$Fe,    
$^{56}$Ni, $^{60}$Zn.  
We note that in the present calculations we do not go beyond the central carbon burning \citep[but see][for  massive stars]{Costa2021MNRAS.501.4514C}.

\subsection{Convective Overshooting}\label{overshootsection}

\textit{Core overshooting (COV)}: The convective unstable region is well defined within the framework of the Schwarzschild criterion \citep{Schwarzschild1958}. However, in reality, the convective elements can travel up to a certain point beyond the border of the unstable region until their velocity drops to zero. This phenomenon is called overshooting. The determination of the edge of the true convective core was described in \citet{1981A&A...102...25B}. In \textsc{parsec}, the overshooting parameter ($\lambda_{\mathrm{ov}}$) is taken across the unstable border; therefore, the core overshooting length is $l_{\mathrm{ov}} \sim \frac{1}{2}\lambda_{\mathrm{ov}} H_P$, where $H_P$ is the local pressure scale height.

\textit{Envelope overshooting (EOV)}: The overshooting downward from the base of the convective envelope has been invoked to explain the observations of the location of the Red-Giant-Branch bump (RGBB) or the extension of blue loops in the CMD \citep{1991A&A...244...95A, 2002ApJ...565.1231C, 2014MNRAS.445.4287T, 2018MNRAS.476..496F}. Solar calibration with helioseismic data have been done by \citet{2011MNRAS.414.1158C}.

In \textsc{parsec} models, the treatment of overshoot from both the convective core and envelope are related to the initial masses. For this reason, we will describe better the values of overshooting parameters that we use in our calculations in the next subsection.

\subsection{Mass Range for Core and Envelope Overshooting }\label{massrange}

\begin{figure}
	\includegraphics[width=\columnwidth]{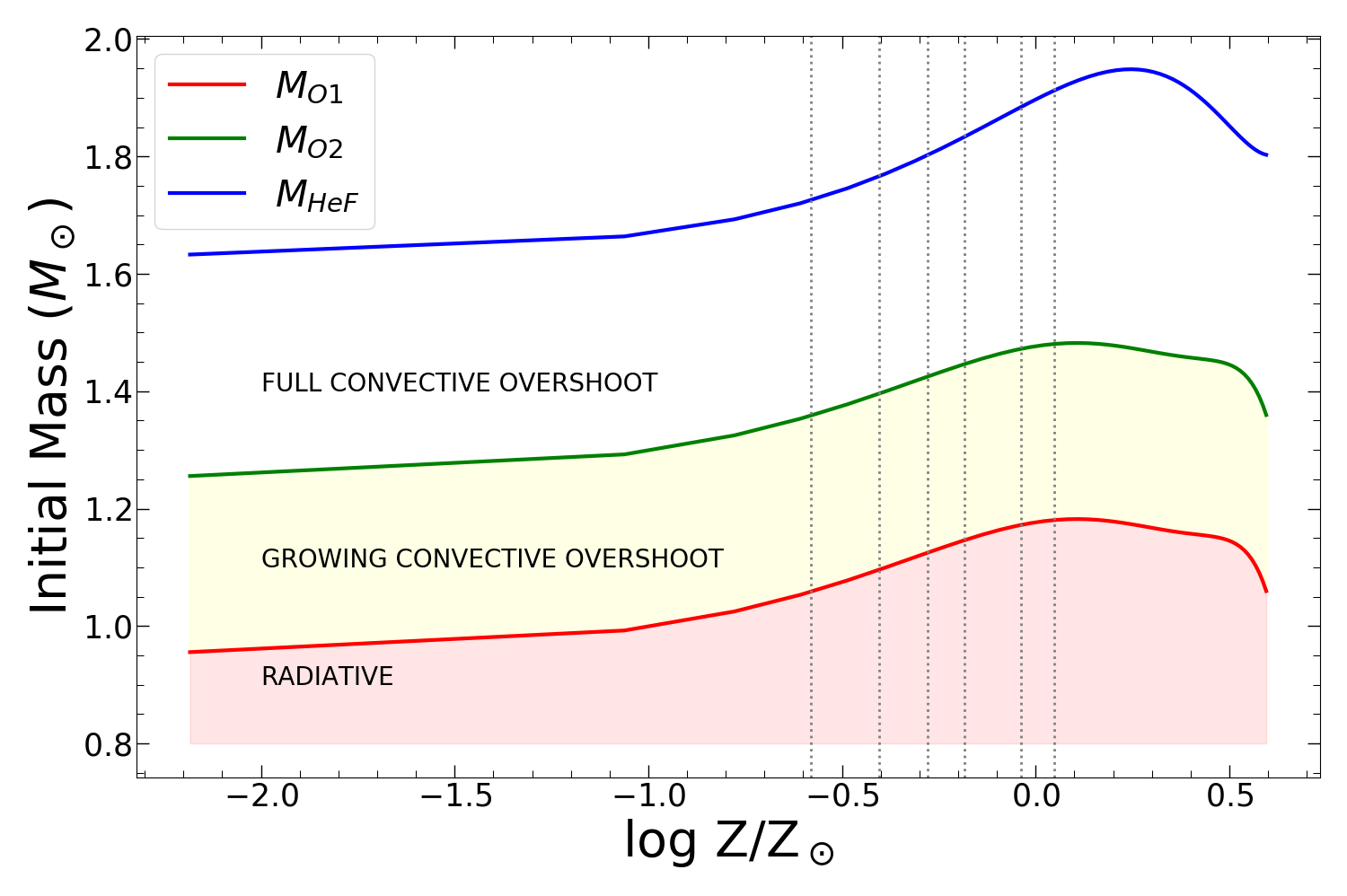}
    \caption{The transition masses ($M_{\mathrm{O1}}$, $M_{\mathrm{O2}}$, $M_{\mathrm{HeF}}$) as function of metallicity in the case of no rotation.
    The red and green lines indicate $M_{\mathrm{O1}}$ and $M_{\mathrm{O2}}$, respectively;
    The blue line indicates $M_{\mathrm{HeF}}$.
    The red area indicates models with a radiative core without convective overshooting. The yellow area delineates the region of growing overshooting efficiency, while full overshooting efficiency occurs in the region above the green line. The vertical dotted-grey lines places at the six computed initial metallicities. The solar metallicity is Z$_\odot=0.01524$ from \citet{2011SoPh..268..255C}.}
    \label{conv_slope}
\end{figure}

In a narrow interval of masses around $1~\Msun$, there is the transition between stars that burn central hydrogen in a radiative core and those that burn hydrogen in a convective core due to the predominance of the CNO-cycle over the pp-cycle. Assessing the efficiency of core overshooting in this mass range is a particularly delicate matter both theoretically \citep[e.g][]{1986A&AS...66..191B, 1990A&AS...85..845B, 1990A&A...240..262A, 1998A&A...334..953V, 2012A&A...541A..41M,2021A&A...646A.133H} and observationally \citep[e.g.][]{2014AJ....147...36T,2016A&A...592A..15C, 2018ApJ...859..100C,2021A&A...647A.187N}. This happens because the inclusion of overshooting modifies the structure and the following evolution of the stars in an irreversible way. In fact, the smallest stars in this mass range reach the zero-age main sequence with small convective cores that disappear as the central H-burning proceeds. If we apply an efficient overshooting to these models, the convective cores do not disappear, but instead tend to become larger and larger, producing a significantly different evolution. To avoid this behaviour that is not favoured by observations \citep{2019MNRAS.485.4641C,2000A&AS..141..371G}, we define the limiting mass $M_{\mathrm{O1}}$ as the largest initial mass of a star showing a vanishing convective core during the early hydrogen burning phase, calculated without overshooting. This mass depends on the initial chemical composition adopted. 
On the other hand, slightly above this mass limit, observations favour an already well-developed overshooting efficiency, with $\lambda_{\mathrm{ov}} = \lambda_{\mathrm{ov,max}}$. This second mass limit is defined as $M_{\mathrm{O2}} = M_{\mathrm{O1}} + 0.3\Msun$.

In \textsc{parsec V2.0}, we define an initial mass range where the transition from models with a radiative core to models with a fully-grown convective core takes place, $M_{\mathrm{O1}}\leq M_\mathrm{i} \leq M_{\mathrm{O2}}$. 
For initial masses $M_\mathrm{i}$ below $M_{\mathrm{O1}}$, the core is stable against convection and energy is transported by radiation.
For $M_\mathrm{i}$ between $M_{\mathrm{O1}}$ and $M_{\mathrm{O2}}$,
the overshooting parameter is let to increase linearly from $\lambda_{\mathrm{ov}}=0$ up to a maximum value $\lambda_{\mathrm{ov,max}}$, in order to have a smooth transition in the properties of the stars. For $M_\mathrm{i} \geq M_{\mathrm{O2}}$, the overshoot is applied with its maximum efficiency, $\lambda_{\mathrm{ov}}=\lambda_{\mathrm{ov,max}}=0.4$ following \citet{2019MNRAS.485.4641C}. This value corresponds to an overshooting length, $l_{ov}$, which extends about $0.2~H_P$ above the Schwarzschild border.

In this paper, we decided to switch on the EOV for low- and intermediate-mass stars only. 
For stars with mass $M_\mathrm{i}<M_{\mathrm{O1}}$ we adopted $\Lambda_e=0.5~H_P$ as inspired by \citet{2018MNRAS.476..496F}; for stars with mass $M_\mathrm{i}>M_{\mathrm{O2}}$ we applied the maximum efficiency as we did for COV, therefore $\Lambda_e=0.7~H_P$ as pointed out in \citet{1991A&A...244...95A} and \citet{2012MNRAS.427..127B}. In the transition region, $\Lambda_e$ of a star is linearly interpolated between $0.5~H_P$ and $0.7~H_P$. Table~\ref{overshoottable} shows the values of $\lambda_{\mathrm{ov}}$ and $\Lambda_e$ adopted for each initial mass.

\subsection{Rotation}\label{rotationsection}
We use \textsc{parsec V2.0} to compute evolutionary tracks of rotating low- and intermediate-mass stars. The code uses the methodology developed by \citet{1970stro.coll...20K} and \citet{ 1997A&A...321..465M}, implemented and described in \citet{2019MNRAS.485.4641C, 2019A&A...631A.128C}. The basic quantity describing the effect of rotation in the stellar structure is the angular rotation rate, $\omega$, defined as 
\begin{equation}
    \omega=\frac{\Omega}{\Omega_{\mathrm{c}}},\quad \Omega_{\mathrm{c}}=\left(\frac{2}{3}\right)^{3/2}\sqrt{\frac{GM}{R^3_{\mathrm{pol}}}}.
    \label{eq:omega}
\end{equation}
where $\Omega$ is the angular velocity, $\Omega_{\mathrm{c}}$ is the critical angular velocity (or break-up velocity), that is, the angular velocity at which the centrifugal force is equal to the effective gravity at the equator. $G$ is the gravitational constant, $M$ is the mass enclosed by $R_{\mathrm{pol}}$ that is the polar radius.

We consider a wide range of initial rotation rates, from non-rotating models ($\omega_\mathrm{i}=0$), to models initially very near the critical break-up rotational velocity ($\omega_\mathrm{i}=0.99$). 

It is commonly accepted that low-mass stars do not reach high values of the rotational speed, compared to intermediate and high-mass stars. 
\citet{2014ApJS..211...24M} reports a sample of the rotation period of more than 34,000 main-sequence (MS) stars. In their Figure~1, there is a clear trend for larger periods in smaller masses. 
This trend inevitably implies that stars with lower mass have lower initial rotational speeds.

On the other hand, rotation may reach high initial values for masses where convection is well developed \citep{2019MNRAS.485.4641C}. For this reason, in analogy to what we did for the efficiency of convective core overshooting, rotation was not considered for \Mi $\leq$ $M_{\mathrm{O1}}$ while, for \Mi $\geq$  $M_{\mathrm{O2}}$, models were computed for the following initial rotation rates: $\omega_\mathrm{i}$ = 0.0, 0.30, 0.60, 0.80, 0.90, 0.95, 0.99. For stars with an initial mass in the range $M_{\mathrm{O1}}$ $\leq$ \Mi $<$ $M_{\mathrm{O2}}$ we computed models with an initial rotation rate up to a maximum value of
\begin{equation}\label{omemax}
\omega_{\mathrm{i,max}}(M)\equiv 0.99\left(\frac{M-M_{\mathrm{O1}}}{M_{\mathrm{O2}}-M_{\mathrm{O1}}}\right).
\end{equation}

It is also important to mention that, in this version, the rotation is switched on a few models before the zero-age-main-sequence (ZAMS) phase. 
At this stage, the code computes the angular velocity $\Omega$ that corresponds to the initial rotation rate $\omega_\mathrm{i}$ and assigns it to each shell of the star, forcing a solid body rotation. 
From the ZAMS on, the solid body rotation constraint is relaxed, and the stellar rotation evolves accordingly with the conservation and the transport of angular momentum.

\subsection{Transport of angular momentum and chemical mixing}
The transport of angular momentum is treated based on the pure diffusive approximation \citep[][]{2000ApJ...528..368H}, where the total diffusion coefficient is contributed from there components,
\begin{equation}
    D = D_\mathrm{mix} + D_\mathrm{s.i.} + D_\mathrm{m.c.}.
\end{equation}
where $D_\mathrm{mix}$ is the diffusion coefficient in the convective zones. The last two terms are related to the shear instability and the meridional circulation. To compute the diffusion coefficient of the shear instability, we use the formula by \citet[][]{1997A&A...317..749T}, which reads
\begin{equation}
    D_\mathrm{s.i.}=\frac{8}{5}\frac{\mathrm{Ri}_\mathrm{c}(r \mathrm{d}\Omega_r/\mathrm{d}r)^2}{N_T^2/(K+D_\mathrm{h})+N_\mu^2/D_\mathrm{h}},
\end{equation}
where the Brunt-V\"{a}is\"{a}l\"{a} frequency is split into $N_\mathrm{T}^2$ and $N_\mu^2$, $\mathrm{Ri}_\mathrm{c}=1/4$ is the critical Richardson number, $K$ is the thermal diffusivity, and $D_\mathrm{h}$ is the coefficient of horizontal turbulence. 
When angular momentum transport is treated with the diffusive approach, the diffusion coefficient for the meridional circulation remains to be defined.
Some authors define it as the product of circulation velocity and its typical length scale \citep[][]{2000ApJ...528..368H}, while others use the same coefficient provided for chemical transport \citep[][]{2013ApJ...764...21C}. For the sake of simplicity, we decided to follow the latter approach. Therefore, we adopt the coefficient by \citet[][]{1992A&A...253..173C}, which reads
\begin{equation}
    D_\mathrm{m.c.} \simeq \frac{|rU|^2}{30D_\mathrm{h}},
\end{equation}
where $U$ is the radial component of the meridional circulation velocity \citep[see also][]{2009pfer.book.....M,2012MNRAS.419..748P}. 
It should be noted that, as discussed by \citet{1992A&A...253..173C} and \citet{1992A&A...265..115Z}, this coefficient takes into account the net effect of the meridional current and horizontal diffusion for chemical species. A more detailed description of angular momentum transport should include meridional circulation as an advective process. However, due to the difficult numerical implementation of the advective-diffusive treatment, and the fact that the angular momentum redistribution goes in the direction of flattening $\omega$ profiles during the main sequence phase \citep{2013ApJ...764...21C}, we decided to use the simpler diffusive approach.
Future \textsc{parsec} versions will include the full advective-diffusive treatment for angular momentum transport.
We refer to \citet{2019MNRAS.485.4641C} for more details on the numerical implementation of rotation.

Another important difference of this new version, \textsc{parsec V2.0}, with respect to \textsc{parsec V1.2S} concerns chemical mixing. While in the latter version the gas was chemically homogenized within convective regions, in the present version we adopt a diffusive approach and solve a unique equation for chemical variation due to nuclear reactions, turbulent motions, molecular diffusion, and rotational mixing. 
The turbulent diffusion coefficient is calculated with the usual approximation $D_T=\frac{1}{3} v_c l_c$
where the velocity of the eddies, $v_c$, and their mean free path, $l_c$ are obtained from the mixing length theory \citep{Bohm-Vitense1958}.

We note that, while all the above processes can be treated at once, molecular diffusion can be switched off in models where the core overshooting reaches its maximum efficiency (above $M\geq M_{\mathrm{O2}}$), because, in such models, the effects of molecular diffusion become negligible with respect to turbulent diffusion and, eventually, rotational mixing. This allows a speed-up of the calculations without loss of generality.
We will discuss later the effects brought by adopting either of the mixing schemes described above. 

\subsection{Mass loss rates}\label{masslossrate}

The effects of mass loss in the evolution of low and intermediate mass stars have been extensively studied in many papers \citep[e.g.][]{1975MSRSL...8..369R, 1995A&A...297..727B, 2005ApJ...630L..73S, 2011ApJ...741...54C, 2014ApJ...790...22R,2008ApJ...676..594K, 2008MNRAS.387.1693C, 2009ApJ...692.1013S}. In \textsc{parsec} models, as well as in previous models of the same group \citep{2000A&AS..141..371G, 2008A&A...484..815B}, mass loss was not activated in the calculations of the low-mass tracks but only at the stage of isochrone calculations. This approximation has been tested in many different applications and has always been considered acceptable from our group. It derives from the fact that the RGB evolution of low-mass stars is just marginally affected by this process, which eventually becomes important only very near to the RGB-tip. Therefore, mass loss just causes a decrease in mass between RGB and the stage of ``zero-age horizontal branch''. This decrease was easily taken into account when interpolating the helium-burning tracks to prepare isochrones. This method allows a great flexibility (different mass-loss prescriptions can be tested without recomputing the evolutionary tracks) and speed-up at the level of isochrone calculation. In the more advanced phases of low and intermediate-mass stars, typically from the early up to the end of the thermally-pulsing asymptotic giant branch (TP-AGB) phase, mass-loss is one of the main processes driving the evolution and cannot be neglected. However, the evolution of stars in these phases is computed subsequently with the \textsc{colibri} code \citep{2013MNRAS.434..488M}, where the most updated mass-loss rates are implemented. 

With the inclusion of rotation, mass-loss cannot be decoupled from evolution anymore and must be included in all evolutionary phases. This is because rotation may enhance the mass loss, affecting the evolutionary path of the star. This may become dramatic when the star is in proximity to the critical break-up velocity.

In this paper, we apply the \citet{1975MSRSL...8..369R,1977A&A....61..217R} law for non-rotating low-mass stars, which is
\begin{equation}\label{reimersmdot}
\dot{M}(\omega=0)=\eta\times 1.343\times 10^{-5}\frac{L^{1.5}}{m T^2_{\mathrm{eff}}},
\end{equation}
where $\dot{M}$ is the mass loss rate in $\Msunyr$, $L$ and $m$ are the luminosity and mass in solar units, respectively, and $\Teff$ is the effective temperature in K. $\eta$ is an efficiency coefficient which is generally calibrated against colour-magnitude diagrams of globular clusters, as for instance in \citet{1988ARA&A..26..199R} where the derived $\eta$ is $0.35$, or \citet{1982ApJS...48..161A} who claimed that $\eta=0.5-0.7$ fits well their data of the red globular clusters in the Magellanic Clouds. In this paper, we adopt $\eta=0.2$, as more recently indicated from the asteroseismic analysis of the two old open clusters NGC~6791 and NGC~6819  by \citet{2012MNRAS.419.2077M}. As described in \citet{2015MNRAS.452.1068C}, for non rotating intermediate-mass and massive stars we adopt the mass loss rate from \citet{1988A&AS...72..259D} and \citet{2001A&A...369..574V}, respectively, both corrected by a factor that assumes the same dependence on the surface metallicity, i.e. $\dot{M} \propto (Z/Z_\odot)^{0.85} $ \Msun/yr.

In the case of rotating stars, the mass-loss rates are enhanced by a factor that depends on the surface tangential velocity, $v$, as expressed in \citet{2019MNRAS.485.4641C,2019A&A...631A.128C}. By numerically solving the fluid equations of a radiation-stellar wind model, \citet{1986ApJ...311..701F} yield a relation where the mass loss rate of a rotating star is modified by a factor with respect to the mass loss of a non-rotating model, which is
\begin{equation}\label{mlossomega}
\dot{M}(\omega) = \dot{M}(\omega =0)\left(1-\frac{v}{v_{\mathrm{crit}}}\right)^{-\xi},
\end{equation}
where $\xi=0.43$ is provided in \citet{1993ApJ...409..429B} by fitting the numerical result of \citet{1986ApJ...311..701F}.
$\dot{M}(\omega=0)$ is the mass loss rate in case of zero rotation and $v_{\mathrm{crit}}$ is the surface critical velocity which is usually defined \citep{2000ApJ...528..368H} as
\begin{equation}
v_{\mathrm{crit}}^2=\frac{Gm}{r}(1-\Gamma_\mathrm{e}),
\end{equation}
$G$, $m$, $r$ are the gravitational constant, mass, and radius in solar units, respectively, and $\Gamma_\mathrm{e}$ is the Eddington factor. In this paper, the dependence of $\Gamma_\mathrm{e}$ with the angular velocity is neglected, instead, it should be considered for angular velocities near the critical one \citep{2000A&A...361..159M}.

\section{Evolutionary tracks}\label{evoltrack}

Before going into more detail on the analysis of our stellar evolutionary tracks, we summarise some of the main points on the adopted input physics. Firstly, 
we computed models with six initial metallicities: $Z=0.004$, $0.006$, $0.008$, $0.01$, $0.014$, $0.017$ which are relevant for the study of young and intermediate-age star clusters in the Milky Way disk and in the Magellanic Clouds. The initial helium mass fraction follows the enrichment law: $Y=Y_p+\frac{\Delta Y}{\Delta Z}Z$, where $Y_p=0.2485$ is the primordial He abundance \citep{2011ApJS..192...18K}; the helium-to-metal enrichment ratio $\Delta Y/\Delta Z=1.78$ is based on the solar calibration in \citet{2012MNRAS.427..127B}. More specifically, the corresponding initial He mass fraction is $Y= 0.256$, $0.259$, $0.263$, $0.267$, $0.273$, $0.279$; and the initial hydrogen abundance $X=0.740$, $0.735$, $0.729$, $0.723$, $0.713$, $0.704$.

Second, the initial rotation rate is parameterised by $\omega_{\mathrm{i}}$: for each set of metallicity above, we compute models with rotation rates going from zero to very near the critical point, $\omega_{\mathrm{i}}=$ 0.00, 0.30, 0.60, 0.80, 0.90, 0.95, 0.99. The treatment of rotation rate for each single star in terms of mass is described in Sect.~\ref{rotationsection}.

Third, the convective overshoot: we apply the overshoot from both the convective core and envelope in the calculations as described in Sects.~\ref{overshootsection} - \ref{massrange}.

Lastly, the mass intervals: For convenience, we will describe the evolutionary tracks in three mass ranges: i) very low-mass stars (VLMS) have initial masses $\Mi \lesssim M_\mathrm{vlm}$, where $M_\mathrm{vlm}$ is the smallest initial mass of a star that is able to ignite helium within the Hubble timescale. Stars with mass smaller than this limit spend their lifetime mainly on the hydrogen-burning phase; ii) low-mass stars (LMS) have initial masses between $M_\mathrm{vlm}$ and $M_{\mathrm{HeF}}$, which includes $M_{\mathrm{O1}}$ and $M_{\mathrm{O2}}$ as mentioned above; and iii) intermediate-mass stars (IMS) with $\Mi>M_{\mathrm{HeF}}$.  $M_{\mathrm{HeF}}$ is defined as the transition mass between stars that develop an electron-degenerate core after the main sequence and hence develop an extended RGB with an He-flash at its tip, and those that do not, hence quietly initiating He-core burning in a non-degenerate core. 

Figure~\ref{conv_slope} shows the dependence of $M_{\mathrm{O1}}$, $M_{\mathrm{O2}}$ and $M_{\mathrm{HeF}}$ as a function of metallicity, for non-rotating models. We also draw the six metallicities computed in this project. Table~\ref{overshoottable} lists the values of $\lambda_{\mathrm{ov}}$ and $\Lambda_e$ adopted for each initial mass. 

Finally, the database of all stellar evolutionary tracks that we produced in this work is available at \url{http://stev.oapd.inaf.it/PARSEC}.

\begin{table*}[!htbp]
\caption{The values of core and envelope overshooting parameters, $\lambda_{\mathrm{ov}}$ and $\Lambda_e/\mathrm{H}_p$ respectively, which correspond to each initial mass $M_\mathrm{i}/\Msun$ in six metallicities (Zs). The value of transition masses, $M_\mathrm{vlm}$, $M_\mathrm{O1}$ and $M_\mathrm{O2}$, of each metallicities is also noted.}
\label{overshoottable}
\centering
\begin{tabular}{c c c c c c c c c}
\hline\hline
\multirow{2}{*}{$\lambda_{\mathrm{ov}}$} & \multirow{2}{*}{$\Lambda_e/\mathrm{H}_p$} & Z$=0.017$ & Z$=0.014$ & Z$=0.01$ & Z$=0.008$ & Z$=0.006$ & Z$=0.004$ & \multirow{2}{*}{Note}\\ 
& & $\mathrm{M}_i/\mathrm{M}_\odot$ & $\mathrm{M}_i/\mathrm{M}_\odot$ & $\mathrm{M}_i/\mathrm{M}_\odot$ & $\mathrm{M}_i/\mathrm{M}_\odot$ & $\mathrm{M}_i/\mathrm{M}_\odot$ & $\mathrm{M}_i/\mathrm{M}_\odot$ & \\
\hline
$0.000$ & $0.000$ & $<0.80$ & $<0.80$ & $<0.75$ & $<0.75$ & $<0.70$ & $<0.70$ & $M_\mathrm{vlm}$ \\

$0.000$ & $0.500$ & $<1.18$ & $<1.16$ & $<1.14$ & $<1.14$ & $<1.09$ & $<1.06$ & \\

$0.000$ & $0.500$ & $1.18$ & $1.16$ & $1.14$ & $1.14$ & $1.09$ & $1.06$ & $M_\mathrm{O1}$ \\

$0.027$ & $0.513$ & $1.20$ & $1.18$ & $1.16$ & $1.16$ & $1.11$ & $1.08$ & \\

$0.053$ & $0.527$ & $1.22$ & $1.20$ & $1.18$ & $1.18$ & $1.13$ & $1.10$ & \\

$0.080$ & $0.540$ & $1.24$ & $1.22$ & $1.20$ & $1.20$ & $1.15$ & $1.12$ & \\

$0.107$ & $0.553$ & $1.26$ & $1.24$ & $1.22$ & $1.22$ & $1.17$ & $1.14$ & \\

$0.133$ & $0.567$ & $1.28$ & $1.26$ & $1.24$ & $1.24$ & $1.19$ & $1.16$ & \\

$0.160$ & $0.580$ & $1.30$ & $1.28$ & $1.26$ & $1.26$ & $1.21$ & $1.18$ & \\

$0.187$ & $0.593$ & $1.32$ & $1.30$ & $1.28$ & $1.28$ & $1.23$ & $1.20$ & \\

$0.213$ & $0.607$ & $1.34$ & $1.32$ & $1.30$ & $1.30$ & $1.25$ & $1.22$ & \\

$0.240$ & $0.620$ & $1.36$ & $1.34$ & $1.32$ & $1.32$ & $1.27$ & $1.24$ & \\

$0.267$ & $0.633$ & $1.38$ & $1.36$ & $1.34$ & $1.34$ & $1.29$ & $1.26$ & \\

$0.293$ & $0.647$ & $1.40$ & $1.38$ & $1.36$ & $1.36$ & $1.31$ & $1.28$ & \\

$0.320$ & $0.660$ & $1.42$ & $1.40$ & $1.38$ & $1.38$ & $1.33$ & $1.30$ & \\

$0.347$ & $0.673$ & $1.44$ & $1.42$ & $1.40$ & $1.40$ & $1.35$ & $1.32$ & \\

$0.373$ & $0.687$ & $1.46$ & $1.44$ & $1.42$ & $1.42$ & $1.37$ & $1.34$ & \\

$0.400$ & $0.700$ & $1.48$ & $1.46$ & $1.44$ & $1.44$ & $1.39$ & $1.36$ & $M_\mathrm{O2}$ \\

$0.400$ & $0.700$ & $>1.48$ & $>1.46$ & $>1.44$ & $>1.44$ & $>1.39$ & $>1.36$ & \\
\hline
\end{tabular}
\end{table*}

\subsection{Very-low-mass stars}
The \textsc{parsec} models for VLMS ($0.09~\Msun\leqslant \Mi \lesssim M_\mathrm{vlm}$) were described in \citet{2014MNRAS.444.2525C} and successfully calibrated against the mass-radius relation of a sample of eclipsing binaries. For this purpose, the authors slightly modified the $T-\tau$ relations provided by PHOENIX (BT-Settl) atmosphere models \citep[see][]{2009ARA&A..47..481A, 2012RSPTA.370.2765A}. 
After this calibration, the corresponding isochrones were able to reproduce well the very low ZAMS of old %metal-poor 
globular clusters NGC 6397 and 47 Tuc, and of the open clusters M67 and Praesepe. These models were also adopted to fit \textit{Gaia} DR2 CMD diagrams \citep{2018A&A...616A..10G}. 
Here, we will continue to use these very low-mass evolutionary tracks, referring to \citet{2014MNRAS.444.2525C} for all details. 

\begin{figure}
	\includegraphics[width=\columnwidth]{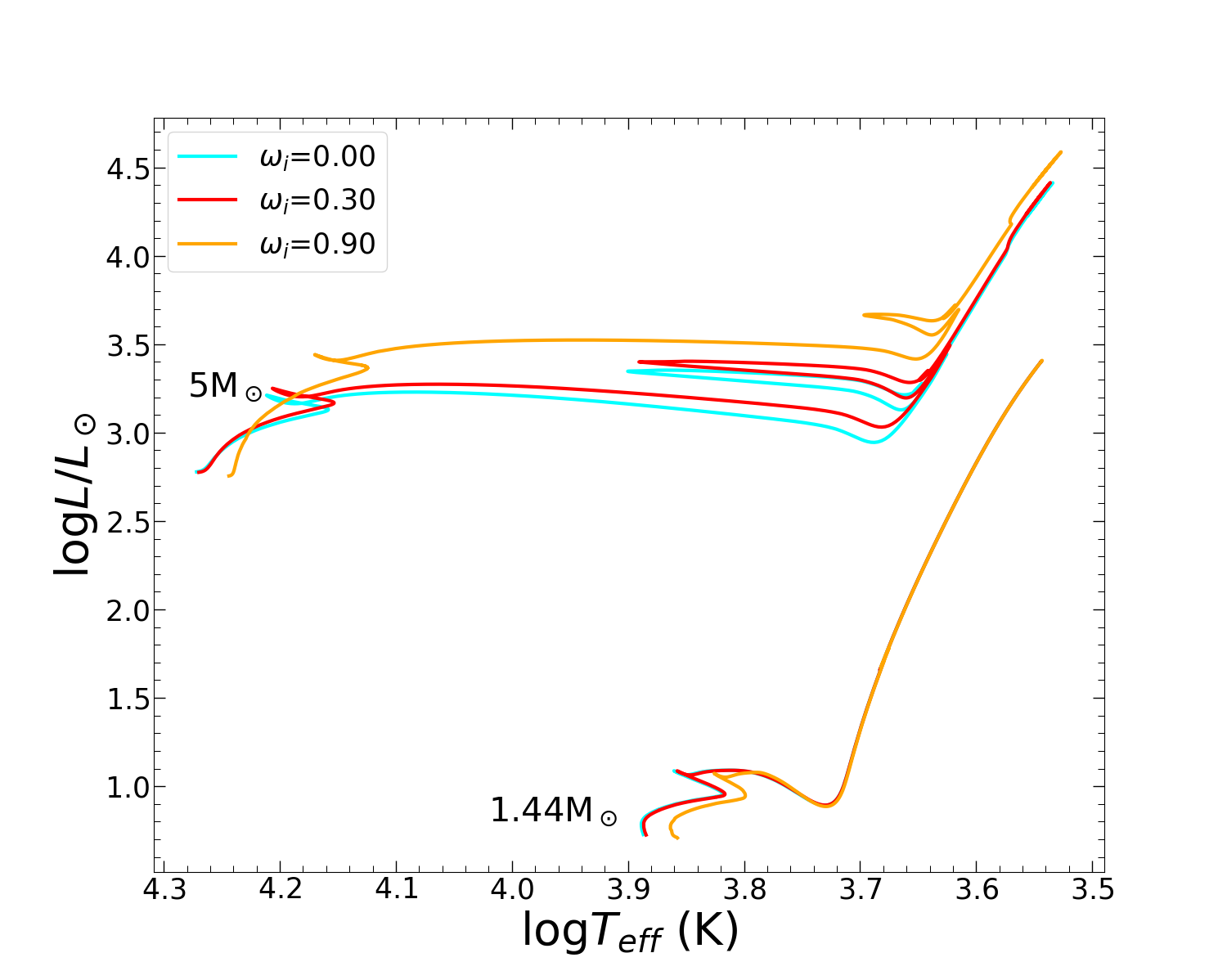}
    \caption{The Hertzsprung-Russell diagram (HRD) of stars with initial masses of $1.44~\Msun$ and $5~\Msun$ with the composition $Z=0.008$, $Y=0.263$, for initial rotation rates $\omega_\mathrm{i}=0.00$, $0.30$ and $0.90$ (cyan, red and orange lines, respectively). We do not plot the evolution on the PMS phase for sake of clarity. 
    }
    \label{HRD008}
\end{figure}

\subsection{Low-mass stars}

\subsubsection{From the pre-main-sequence to the tip of the RGB} We computed models of LMSs with initial masses in the interval from $M_\mathrm{vlm}$ to $M_\mathrm{HeF}$. The mass step is $0.05~\Msun$ in the mass range from $M_\mathrm{vlm}$ to $0.8~\Msun$, $0.02~\Msun$ for the range from $0.8~\Msun$ to $M_\mathrm{O2}$, and $0.1~\Msun$ for masses above $M_\mathrm{O2}$.
All the LMSs tracks begin from the PMS phase and end at the tip of the red-giant branch (TRGB), where the star ignites its central He under strongly degenerate conditions (the so-called He-flash).

Figure~\ref{HRD008} shows the HRD of $1.44~\Msun$ stars with different initial rotation rates. It should be noted that, for rotating stars, $\Teff$ is actually an average value over the isobaric surface; more precisely it is the value that a non-rotating star with the same ``volumetric radius'' would have to produce the same total luminosity. The volumetric radius is defined as the radius of a sphere with the same volume as that of the rotating star. 
The local effective temperature characterizing different points at the surface of the star, instead, is a quantity that varies along the co-latitude angle ($\theta=0^\circ$ aligns with the rotation axis), becoming cooler towards the equator. This can be explained by the proportionality between $T_\mathrm{eff}^4$ and effective gravity $g_\mathrm{eff}$, based on von Zeipel's theorem \citep{1924MNRAS..84..684V,2007A&A...470.1013E}. In turn, the local effective gravity is reduced by the centrifugal force, which is higher for a higher rotation rate. Therefore, as we see in Fig.~\ref{HRD008}, the higher the rotation rate, the cooler the star is (by means of the average value) during the MS. 
In the post-MS phases, the conservation of angular momentum forces the surface angular velocity to drop down when the star expands, hence causing the star to evolve along the same path of non-rotating stars.

\begin{figure}
	\includegraphics[width=\columnwidth]{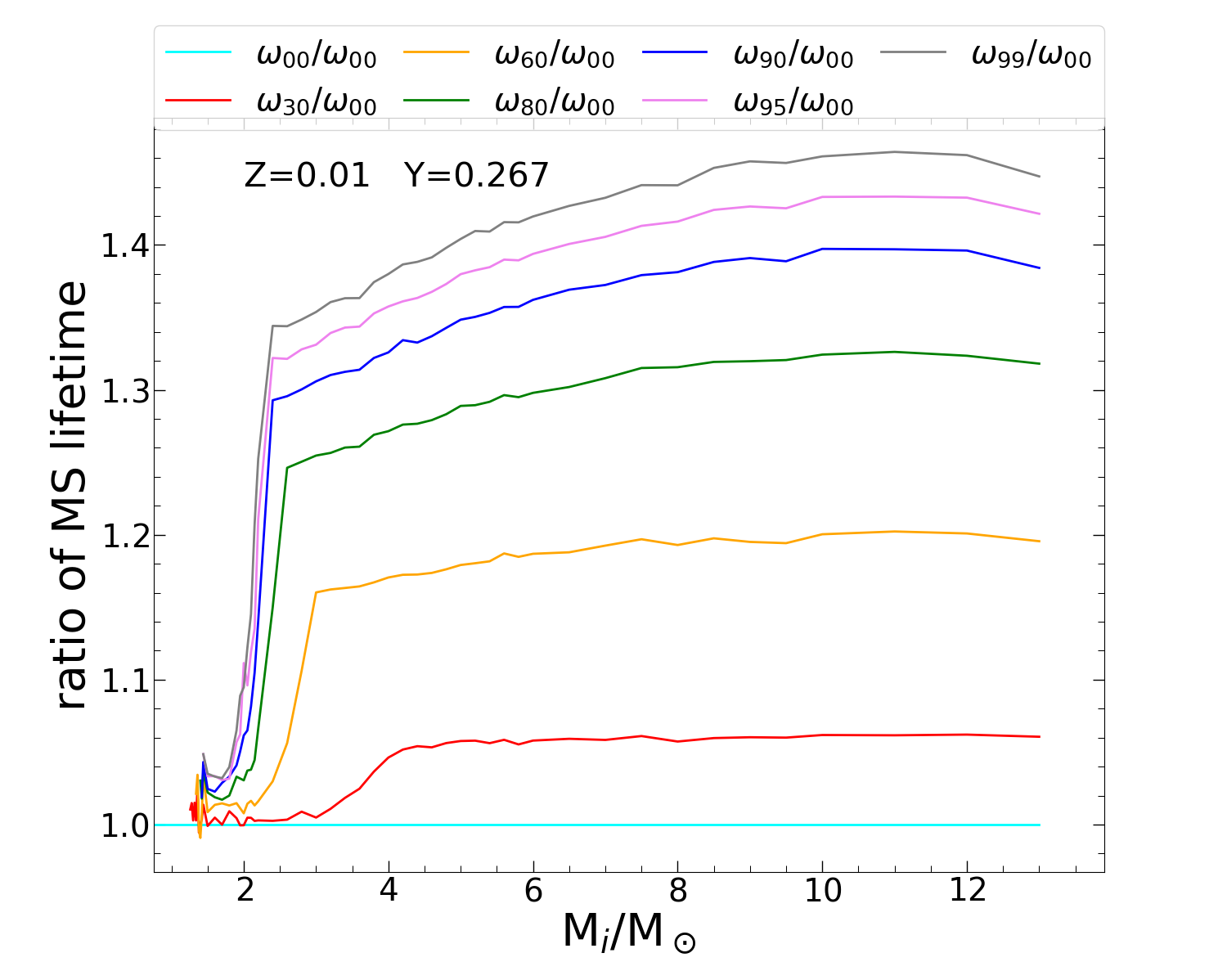}
    \caption{The ratio between the MS lifetimes between rotating and the non-rotating models of the same mass, as a function of initial mass in the set of $Z=0.01, Y=0.267$. This lifetime is measured from the ZAMS until the exhaustion of central-H ($X_\mathrm{c}<10^{-5}$). Different values of $\omega_\mathrm{i}$ are considered, going from  $0$ to $0.99$ as in the legend.
    }
    \label{MStau}
\end{figure}

\begin{figure}
	\includegraphics[width=\columnwidth]{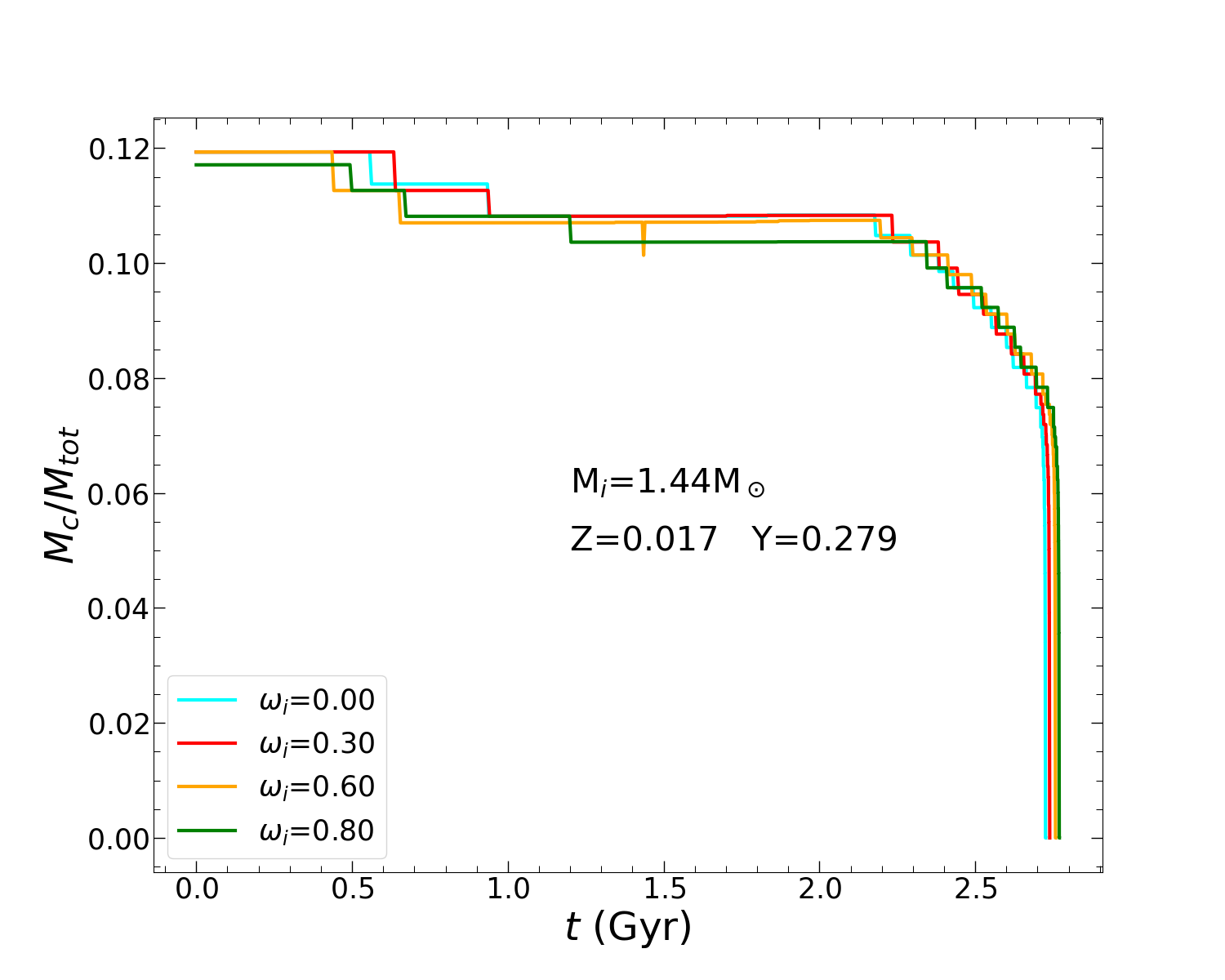}
    \caption{The time evolution of the mass of the convective core ($M_\mathrm{c}/M_\mathrm{tot}$) during the H-burning phase  for the model of  $\Mi=1.44~\Msun$ and $Z=0.017$. 
    }
    \label{conv_mass}
\end{figure}

Another effect of rotation is that the faster the stars rotate, the longer they stay in the MS phase \citep[][]{2010A&A...509A..72E, 2012A&A...537A.146E, 2019A&A...631A.128C}. Fig.~\ref{MStau} shows the ratio of the MS duration between models with different $\omega_\mathrm{i}$ and their standard non-rotating counterparts, $\omega_\mathrm{i}=0$, as a function of the initial mass $M_\mathrm{i}$ and for $Z=0.01$. 
We see that this ratio is higher than $1$ for all rotating models and becomes higher as $\omega_\mathrm{i}$ increases. 
In the low-mass range ($M\lesssim 1.8~\Msun$) the ratio remains modest, while it increases significantly in the domain of intermediate-mass and massive stars. 
This is understandable because of the lower efficiency of rotational mixing in LMSs with respect to intermediate and massive ones, as will be discussed later in Sect.~\ref{IMS}. 

We also find that in the low-mass range, the size of the convective core does not depend significantly on $\omega_\mathrm{i}$. This can be seen in Fig.~\ref{conv_mass} for the models of $M_\mathrm{i}=1.44~\Msun$ and $Z=0.017$.

\begin{figure}
	\includegraphics[width=\columnwidth]{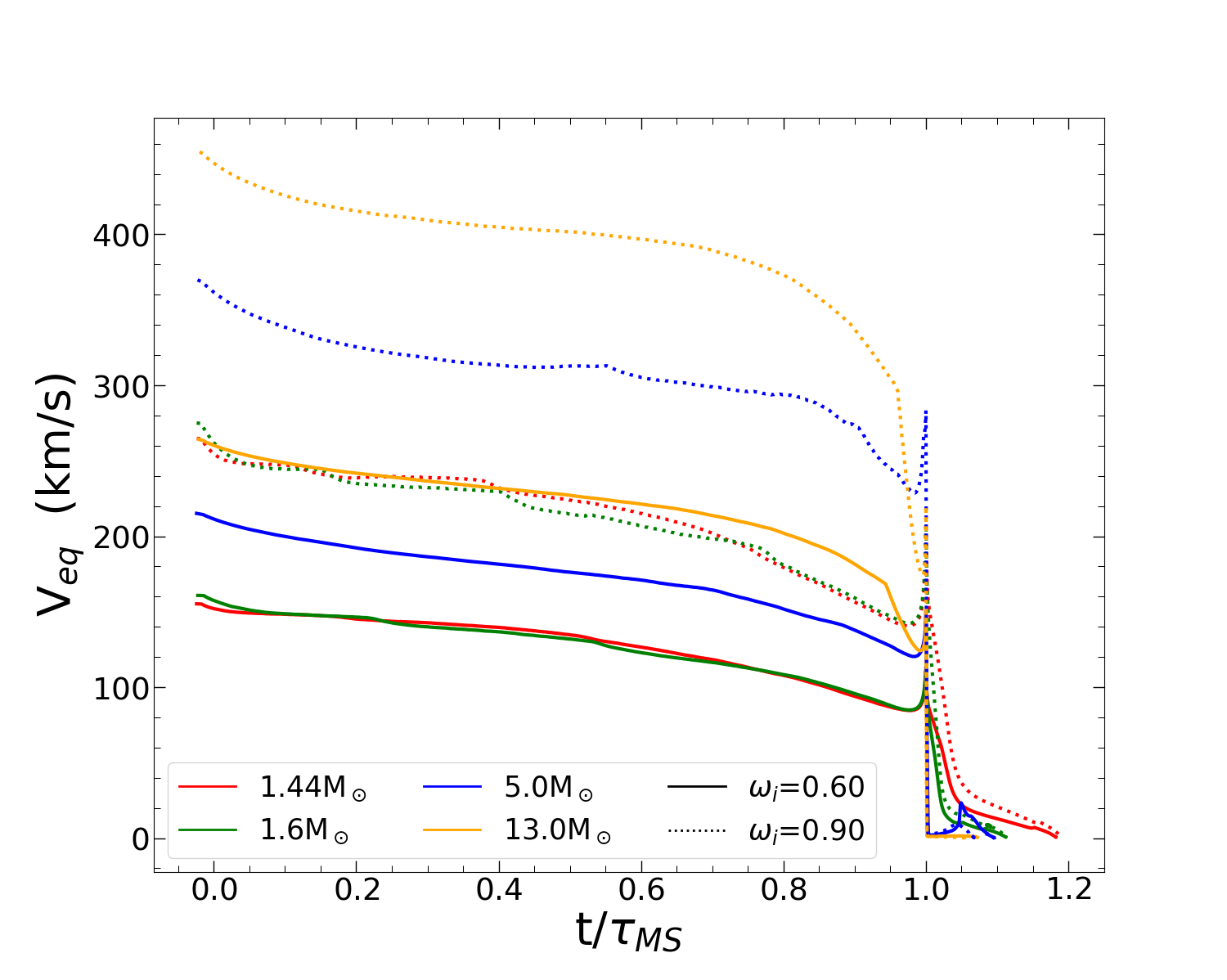}
    \caption{The evolution of surface tangential velocity at the equator, $V_\mathrm{eq}$, versus the time $t$ scaled to the MS duration $\tau_{MS}$. The four selected mass models: $1.44~\Msun$ (red), $1.6~\Msun$ (green), $5~\Msun$ (blue), $13~\Msun$ (orange) are shown, with two initial rotation rates $\omega_\mathrm{i}=0.60$ (solid-line) and $\omega_\mathrm{i}=0.90$ (dotted-line), from the set with $Z=0.014, Y=0.273$.}
    \label{Veq}
\end{figure}

After the formation of the H-exhausted core, the star enters into the sub-giant phase and then ascends the RGB. Expansion of the envelope leads to a decrease in surface rotation velocity. This impact is illustrated in Fig.~\ref{Veq}, where the equatorial tangential velocity drastically decreases after leaving the MS. The drop down on rotation rate results in an evolution as a non-rotating star, as already mentioned and as illustrated in the HRD of Fig.~\ref{HRD008}. We found that the luminosity at the tip of the RGB phase is almost the same for any applied rotational rates \citep[see also][]{2012A&A...537A.146E,2013A&A...558A.103G}

For instance, the TRGB luminosity of the Z=$0.004$, $M_\mathrm{i}=1.36~\Msun$ star is $\log L/\Lsun=3.38771$, $3.38934$ and $3.38919$ for the models with $\omega_\mathrm{i}=0.00$, $0.60$ and $0.90$, respectively. We see that the difference is less than $0.0016$ dex in any case, and this is due to the slightly heavier He-core mass discussed above. In general, we found that the TRGB luminosity of our models with $M\leq 1.5~\Msun$ is about $\log L/\Lsun\sim 3.385-3.420$, depending on the initial metallicity. This result is important in the context of the TRGB method of distance determinations, and the recent ``tension'' in the values of the Hubble constant $H_0$ \citep[see][and references therein for more details]{2019ApJ...882...34F, 2020ApJ...891...57F}.

\begin{figure}
	\includegraphics[width=\columnwidth]{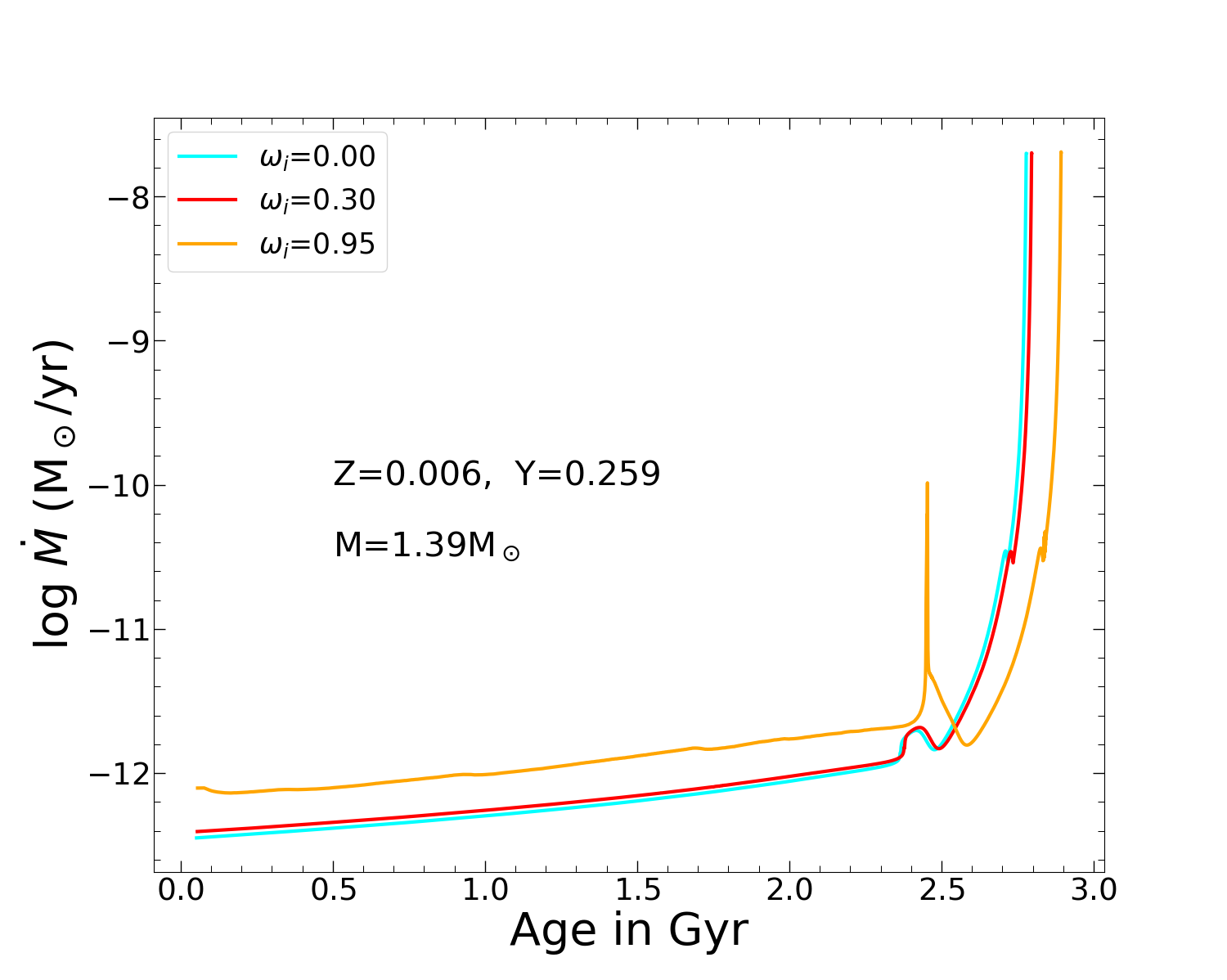}
    \caption{The mass loss rate as a function of time of star $M_\mathrm{i}=1.39~\Msun$, $Z=0.006$ for three rotation rates $\omega_\mathrm{i}=0.00$, $0.30$ and $0.95$ (cyan, red and orange lines, respectively).}
    \label{mlossrate}
\end{figure}

On the other hand, we also checked the effect of rotation on the mass loss rate. Fig.~\ref{mlossrate} shows the mass loss rate of the $1.39~\Msun$ model with metallicity $Z=0.006$, for three initial rotation rates, $\omega_\mathrm{i}=0.00$, $0.30$ and $0.95$. We plot $\log\dot{M}$ from the ZAMS to the end of the RGB. As expected from equation~(\ref{mlossomega}), during the MS phase the star with higher rotation rates has higher mass loss rates. However, in the RGB phase, because of the decline of the surface rotational velocity, it evolves as a non-rotating star but with a slightly older age.
In general, stars lose their mass at a rate of about $(0.6-6)\times 10^{-8}$ $\Msunyr$ at the TRGB stage. 
This result is based on the Reimers' law that we adopted in our models. Interesting alternative models for mass loss have been proposed, \citet{2007ApJS..171..520C} and \citet{2011ApJ...741...54C}, which will be the subject of other subsequent work.
We note that during the stellar contraction phase just after the end of the main sequence, the tangential velocity may reach its critical value, at least for models with the highest initial rotation rates. 
This is the case for the model with $\omega_\mathrm{i}=$0.95 shown in Fig.~\ref{mlossrate}. 
In this case, the mass loss as provided by equation~(\ref{mlossomega}), is enhanced by mechanical effects \citep{2013A&A...558A.103G,2019A&A...631A.128C} as shown by the relative peak of about two orders of magnitude with respect to the other tracks, before entering the RGB phase.

\begin{figure}
	\includegraphics[width=\columnwidth]{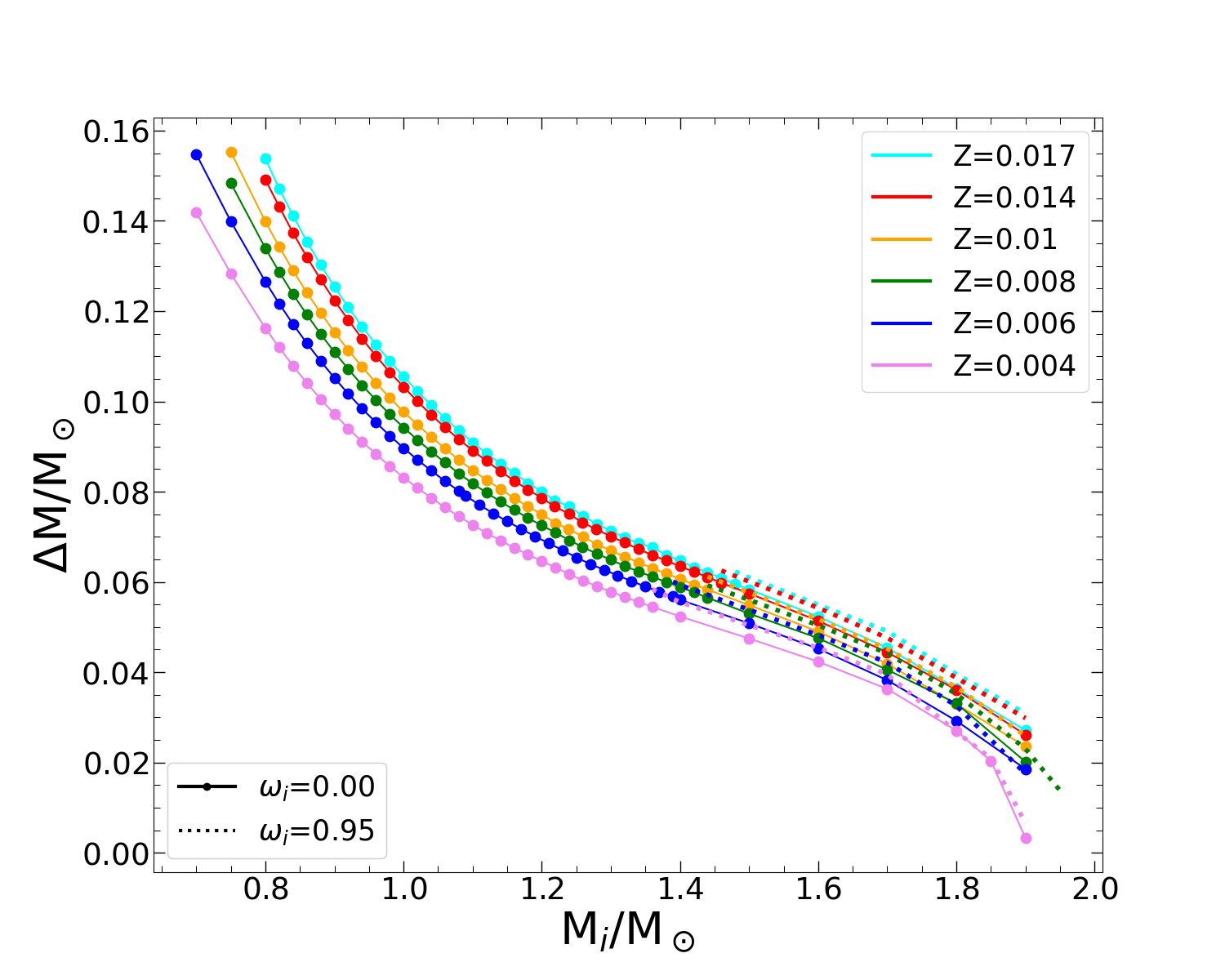}
    \caption{The total mass lost until the TRGB versus initial mass, for six different metallicity sets. Solid and dashed lines represents models with $\omega_\mathrm{i}=0.00$ and $\omega_\mathrm{i}=0.95$, respectively.
    }
    \label{Mdotcomp}
\end{figure}

Concerning the total mass lost on the RGB, we find that it is higher for the smaller initial masses. For non-rotating models of $Z=0.004$, the stars with initial masses $M_\mathrm{i}=0.9~\Msun$,  $1.16~\Msun$ and 
$1.36~\Msun$ lose about $11\%$,  $6\%$ and  $4\%$ of their initial mass, respectively.
The total mass lost by the stars at the RGB-tip is illustrated in Fig.~\ref{Mdotcomp} for all six sets of metallicity and for two initial rotation rates, $\omega_\mathrm{i}=0.00$ and $0.95$ (the solid- and dashed-lines, respectively). From this figure we also see that the key role in the total mass lost by the stars is taken, in decreasing order, by mass, metallicity, and rotation.  

\begin{figure}
	\includegraphics[width=\columnwidth]{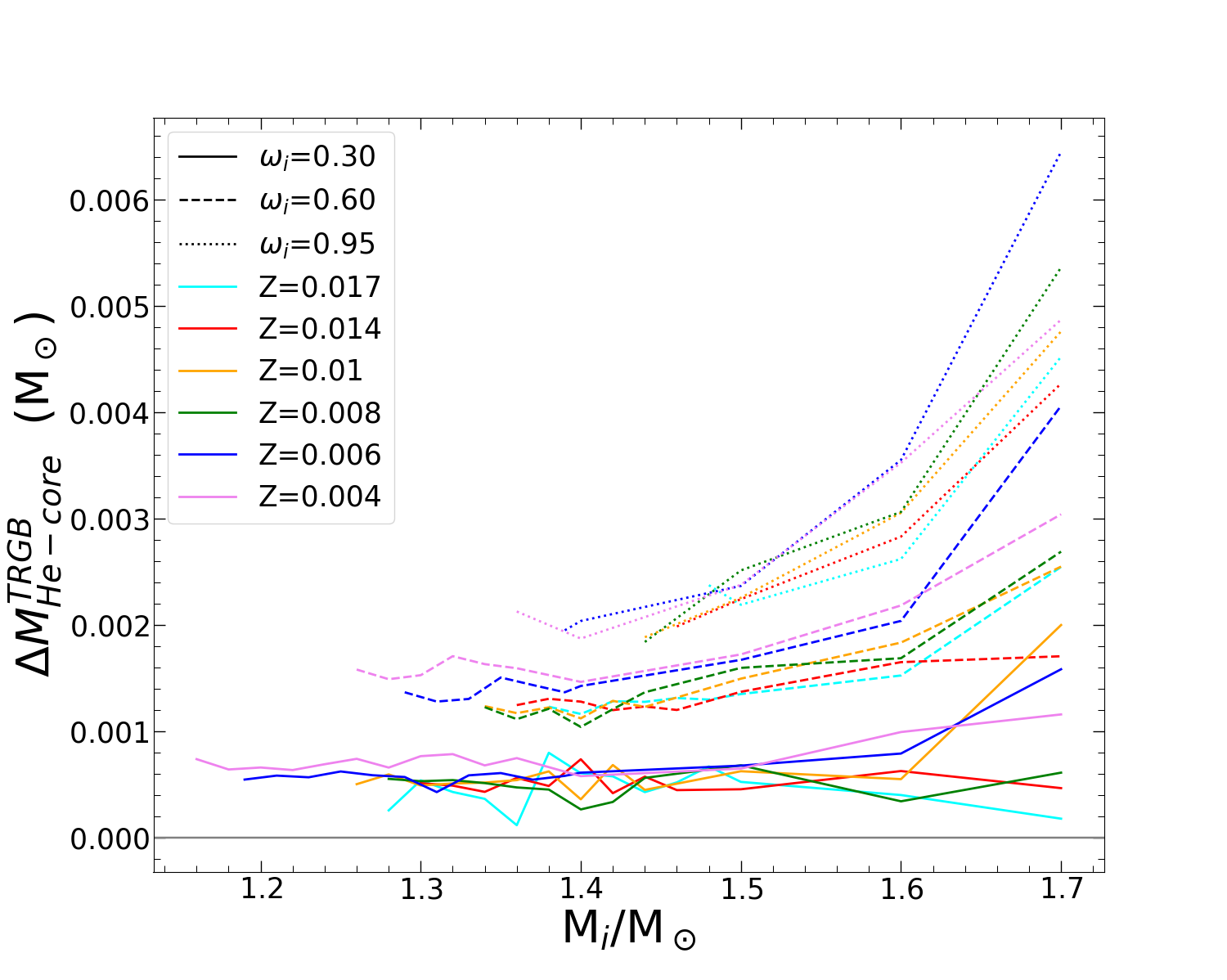}
    \caption{The difference in He-core mass at the tip of RGB phase between the rotating models and their non-rotating counterparts, $\Delta M_\mathrm{He-core}^\mathrm{TRGB}$. Three initial rotating rates are considered: $\omega_\mathrm{i}=0.30, 0.60, 0.95$ (solid, dashed and dotted line, respectively). The colours represent different initial metallicities. The grey-solid line marks the reference line for $\omega_\mathrm{i}=0$ models.}
    \label{DeltaMc_TRGB}
\end{figure}

\begin{figure}
	\includegraphics[width=\columnwidth]{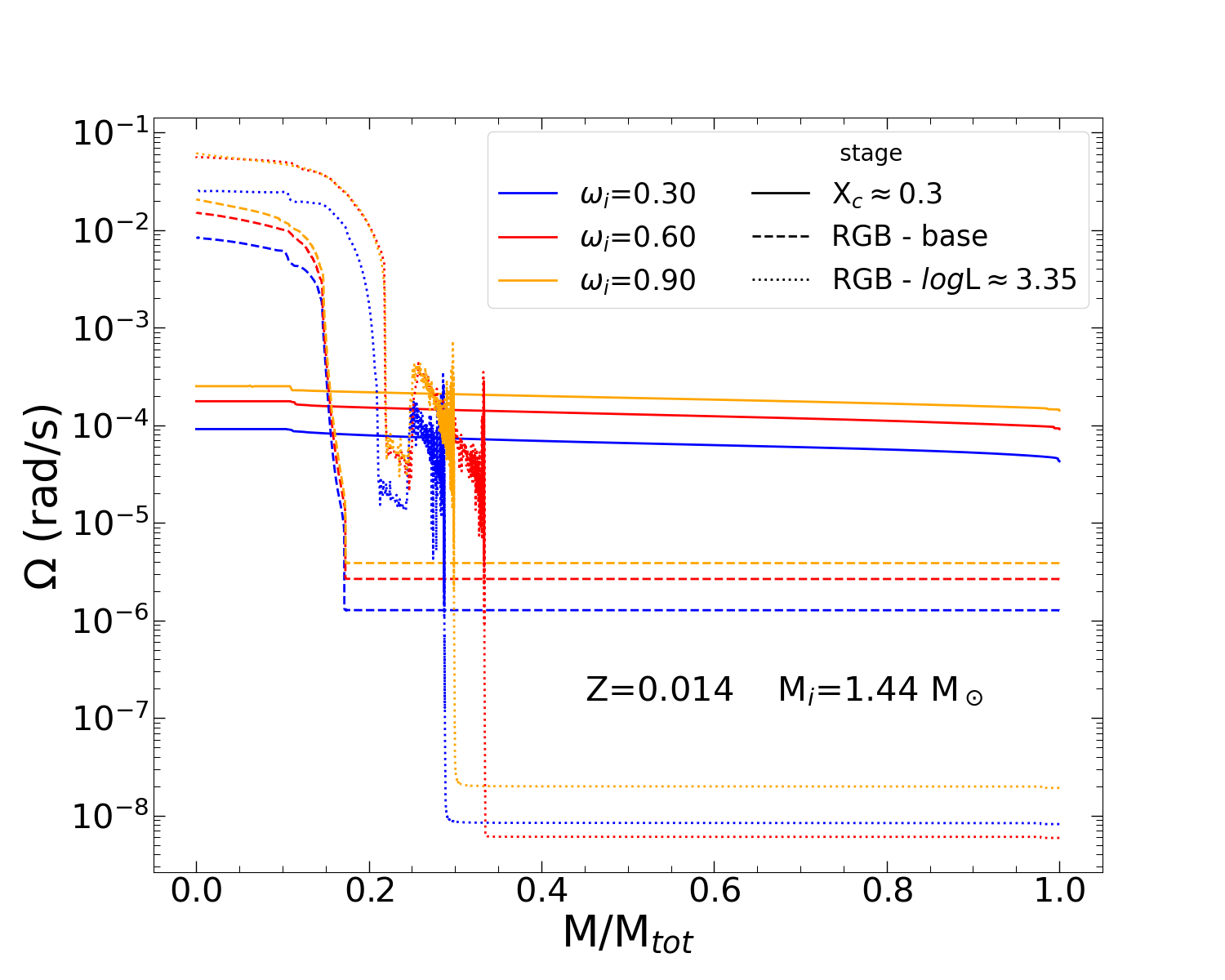}
    \caption{The internal distribution of angular velocity, $\Omega$. Three selected stages of models with $Z=0.014$, $M_\mathrm{i}=1.44~\Msun$ are presented: at which central hydrogen $X_\mathrm{c}\approx 0.3$ during the MS (solid lines), at the base of the RGB phase (dashed lines) and near the RGB's tip with $\log L\approx 3.35$ (dotted lines). The colours represent different selected initial rotation models.}
    \label{omega_distribution}
\end{figure}

Fig.~\ref{DeltaMc_TRGB} shows the difference in He-core mass at the TRGB between rotating models and their non-rotating counterparts, for three values of $\omega_\mathrm{i}=0.30, 0.60, 0.95$ and for six metallicities. The higher the initial rotation rate, the larger the He-core mass the star has at the tip, at any metallicity. While the surface rotation at this stage is small even for the largest $\omega_\mathrm{i}$, in the core it is still significant, as can be seen in  Fig.~\ref{omega_distribution}.
The larger the rotation, the less concentrated is the core, and a larger core mass is needed to reach the conditions for He ignition. At the larger initial masses, there is also a contribution of the more efficient rotational mixing during the main sequence phase.
In general, the difference is $\leq 0.006~\Msun$, depending on $\omega_\mathrm{i}$. We note that these differences might affect the location of red clump stars in the HR diagram. This issue will be further investigated in a subsequent work.

\subsubsection{From the ZAHB to the TP-AGB} 
\begin{figure}
	\includegraphics[width=\columnwidth]{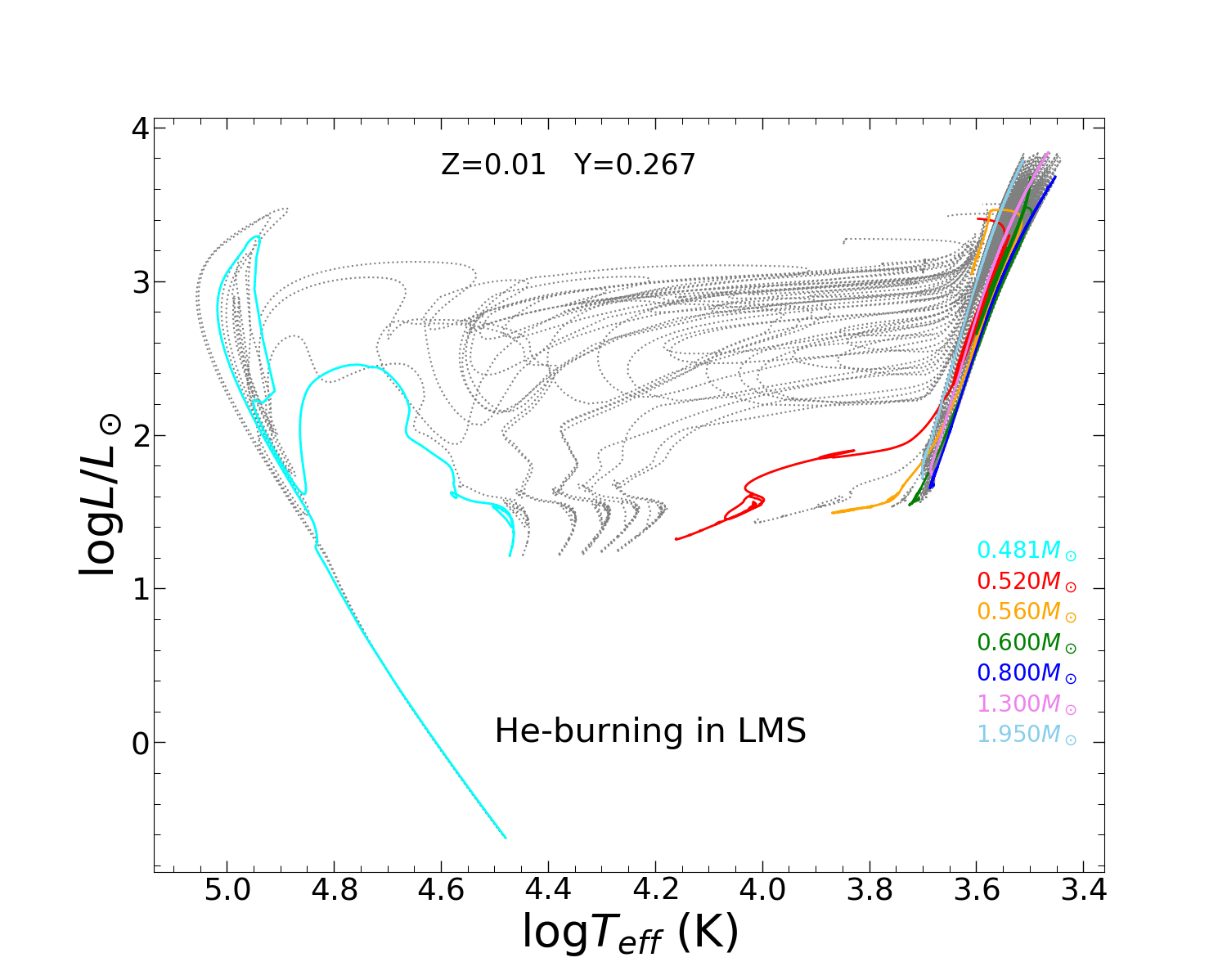}
    \caption{The He-burning phase of low-mass-stars in the set with $Z=0.01$, $Y=0.267$. Tracks with ZAHB masses between 0.481 and 1.95 \Msun\ were computed in this case. A few of these tracks are marked with a different colour (with their mass in $\Msun$ indicated in the legend), for reference.
    }
    \label{HB_HRD}
\end{figure}

Low-mass stars develop an electron-degenerate core and climb the RGB until they undergo the He-flash. The latter requires large amounts of CPU times to be computed in detail \citep[see][for more details]{2012sse..book.....K,2008A&A...490..265M}. Therefore, the computation of the evolutionary track is interrupted during the He-flash, and restarted from a zero-age horizontal branch (ZAHB) model with the same He-core mass and surface chemical composition as the last RGB model. The initial ZAHB model is built following the method described in \citet{2012MNRAS.427..127B}, taking into account the fraction of He that has been burned into carbon during the flash so that the degenerate core is lifted into a non-degenerate state. Then, the star is evolved along the Horizontal Branch (HB) and the evolution is terminated again after it experiences a few pulses of the TP-AGB phase. The evolutionary tracks in the HR diagram during the post-ZAHB phases of LMS are illustrated in Fig.~\ref{HB_HRD}, just for a single set of metallicity. Similar grids are available for all metallicities. 

It is important to point out a few details in these calculations: 
\begin{enumerate}
\item Rotation is turned off for the entire evolution beyond the ZAHB, simply because at those stages the rotational velocities have become small enough to not imply significant evolutionary effects. 
\item With respect to the previous version of \textsc{parsec}, the new tracks include mass loss starting from the ZAMS. Thus, for any given initial metallicity,
we have different relations $M_\mathrm{TRGB}(M_{i})$ for different $\omega_\mathrm{i}$. These relations are merged to obtain a complete unique sequence of $M_\mathrm{ZAHB}(M_{i})$, with $M_\mathrm{ZAHB}^i$ spanning the range from the largest $M_\mathrm{TRGB}$ to the lowest value compatible with the thinnest envelope mass along the ZAHB sequence. We also pay attention to carefully sample the mass interval close to $M_\mathrm{HeF}$.
We then interpolate on the sequence of non-rotating models to obtain a unique complete $M_\mathrm{core}(M_\mathrm{ZAHB}^i)$ relation, which is used to construct the ZAHB model sequence.
\end{enumerate}

\subsection{Intermediate-mass and massive stars}\label{IMS}

Intermediate-mass stars (IMSs) are defined as having masses larger than the $M_{\mathrm{HeF}}$ limit and smaller than the $M_{\mathrm{up}}$ threshold for C ignition in the core. Both limits depend on the initial metallicity and the rotation rate. Massive stars are computed up to 14 \Msun, leaving more massive stars to a dedicated paper, which is in preparation. Models with initial masses between $M_{\mathrm{HeF}}$ and 2.2 \Msun\ are computed with a mass step $\Delta\mathrm{M_i}$ = 0.05 \Msun ; from 2.2 \Msun\ up to 6 \Msun , $\Delta\mathrm{M_i}$ = 0.2 \Msun ; up to 10 \Msun, $\Delta\mathrm{M_i}$ = 0.5 \Msun\ and $\Delta\mathrm{M_i}$ = 1 \Msun\ above $\mathrm{M_i}$ = 10 \Msun . All these tracks start on the PMS phase and are interrupted either 
after the first few thermal pulses along the AGB, or after the ignition of carbon in the core.
All evolutionary tracks in this mass range have been computed with the maximum overshooting efficiency, i.e. with $\lambda_\mathrm{ov}=0.4$ and $\Lambda_\mathrm{e}=0.7~H_P$ and for all initial rotation rates from $\omega_\mathrm{i}=0.00$ to $0.99$ (Sect.~\ref{rotationsection}). The mass loss rates of rotating stars follow the description in Sect.~\ref{masslossrate}, while the formulation of \citet{1988A&AS...72..259D} was adopted for non-rotating models.

\begin{figure}
	\includegraphics[width=\columnwidth]{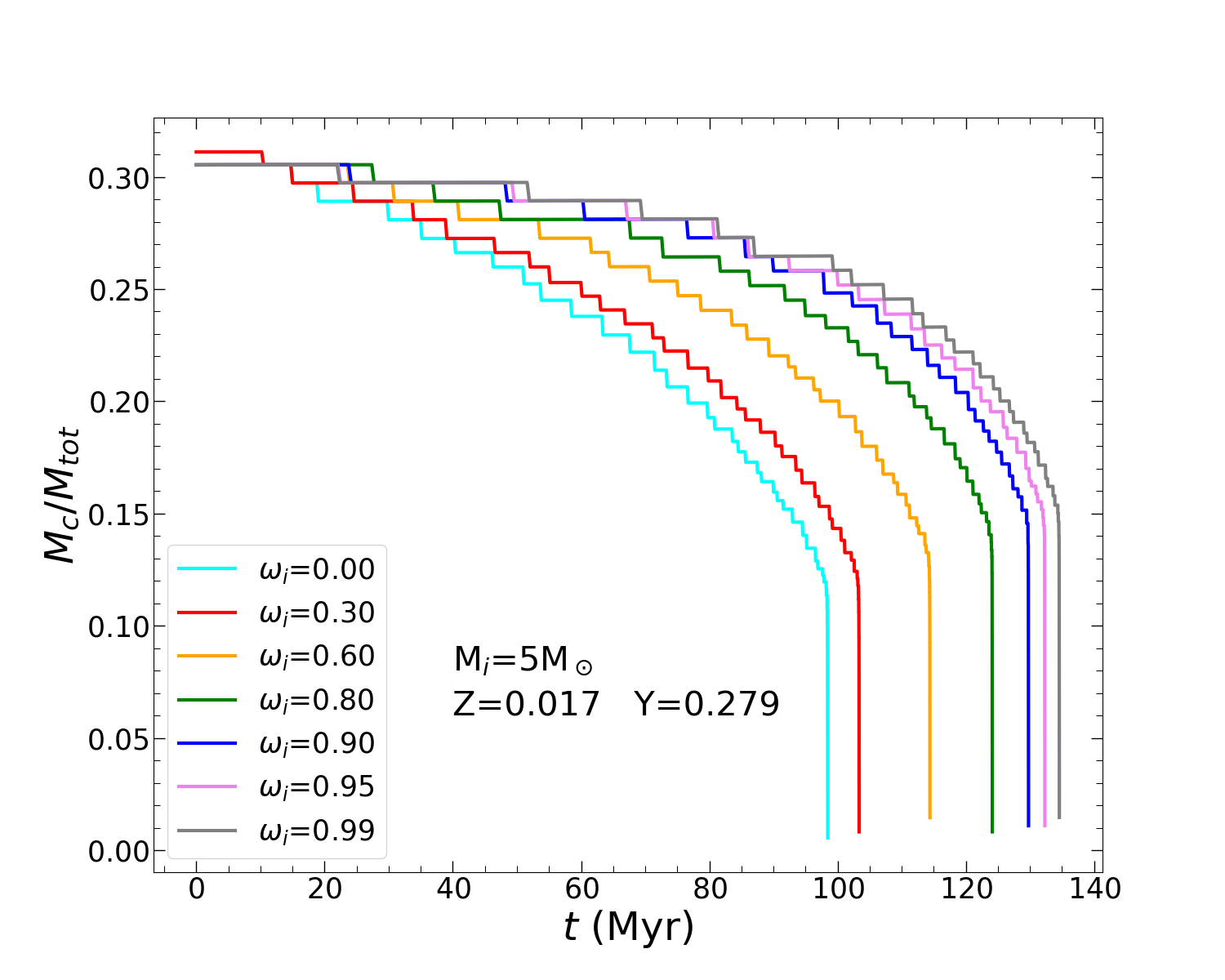}
    \caption{The evolution of the mass of the convective core (M$_c$/M$_{tot}$) during the H-burning phase for the model of  $M_i=5.0\Msun$ and $Z=0.017$. 
    }
    \label{conv_mass_5.0}
\end{figure}

In Figure~\ref{HRD008} we have already compared the evolution of a 5~\Msun\ model calculated with three different rotation rates, $\omega_\mathrm{i}$ = 0.0, 0.30, 0.90, with that of a 1.44~\Msun\ model with the same $\omega_\mathrm{i}$. Rotation impacts the evolution of IMSs in a way different from the LMSs. 
At the beginning of the evolution, only the geometrical effects of rotation are visible: in both cases, the models that rotate faster are less luminous and cooler.
As evolution proceeds,
IMSs develop a convective core surrounded by a radiative envelope where the meridional circulation works efficiently. As a result, rotational mixing provides more fresh fuel to the central core, and hence a more massive core is built up (see Fig.~\ref{conv_mass_5.0}). This causes the IMSs models that rotate faster to become more luminous and to increase their MS lifetimes significantly (as shown in Fig.~\ref{MStau}). Due to the larger core masses, the higher luminosity is maintained during all post-MS evolutionary phases. 
In contrast, in low-mass models even in the case with the largest rotation rate, the growth of the core is never so high to make it more luminous than the non-rotating one. 
At lower masses, rotation affects more the effective temperature than the luminosity.

\begin{figure}
	\includegraphics[width=\columnwidth]{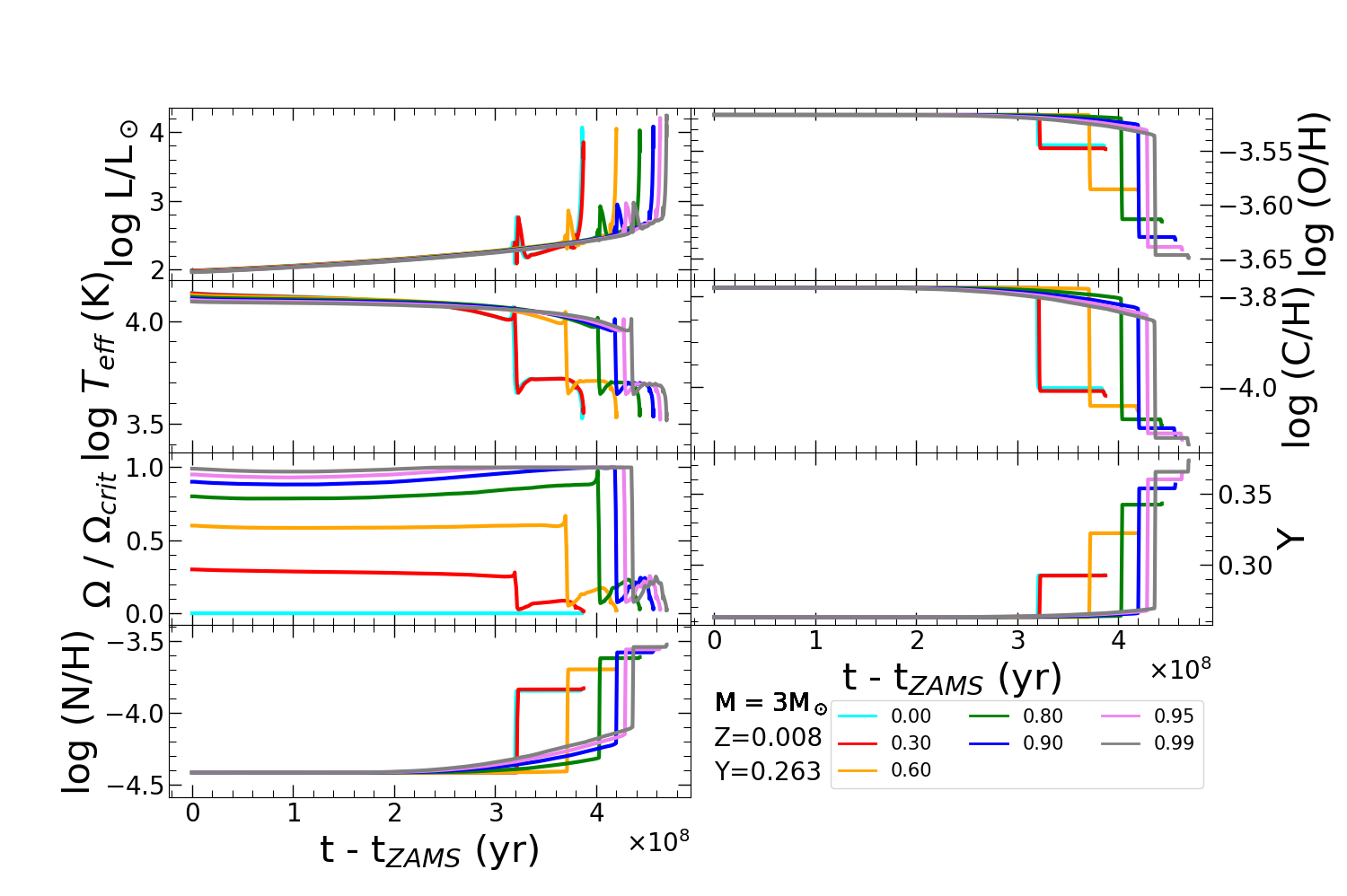}
	\includegraphics[width=\columnwidth]{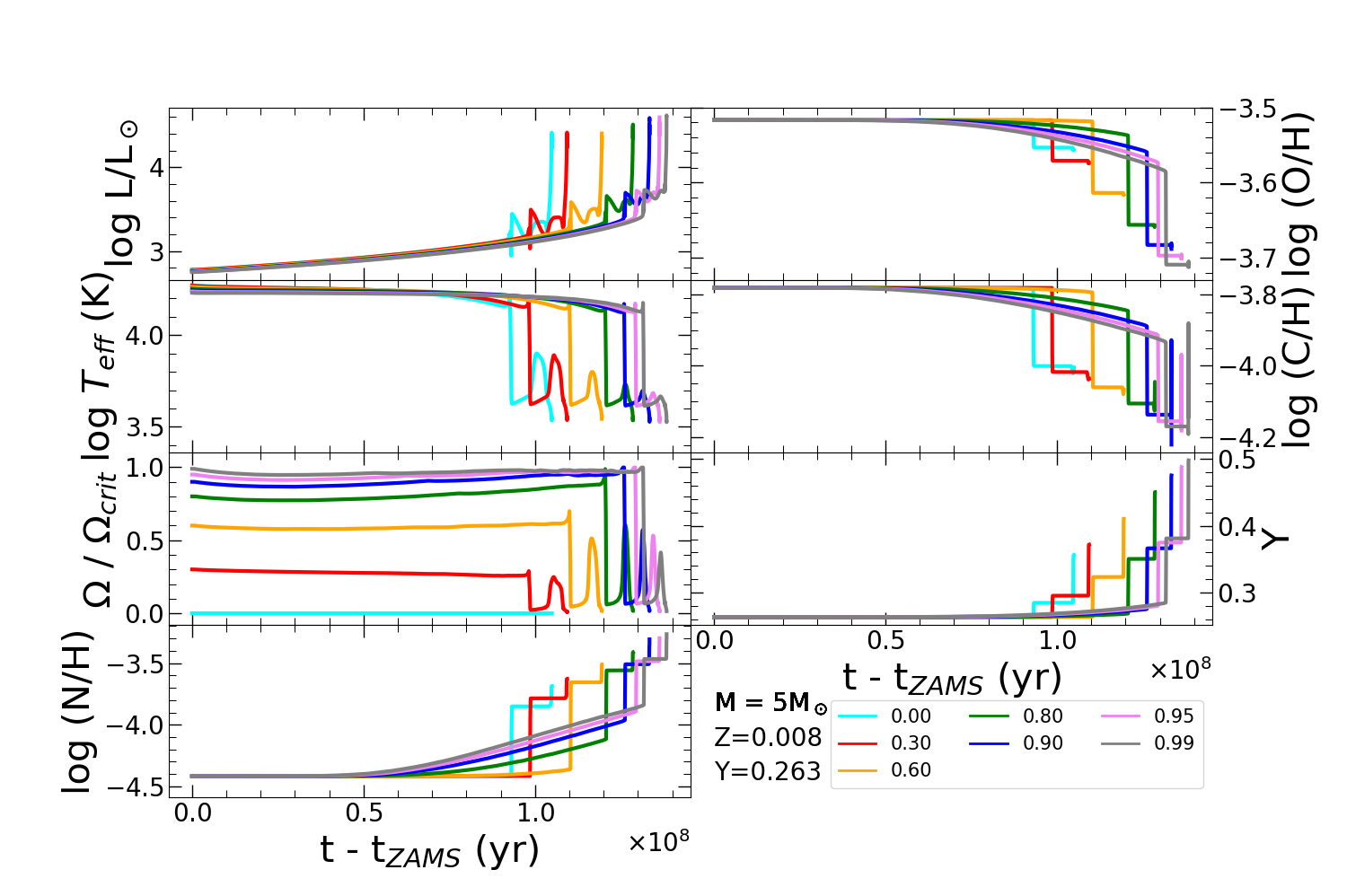}
	\includegraphics[width=\columnwidth]{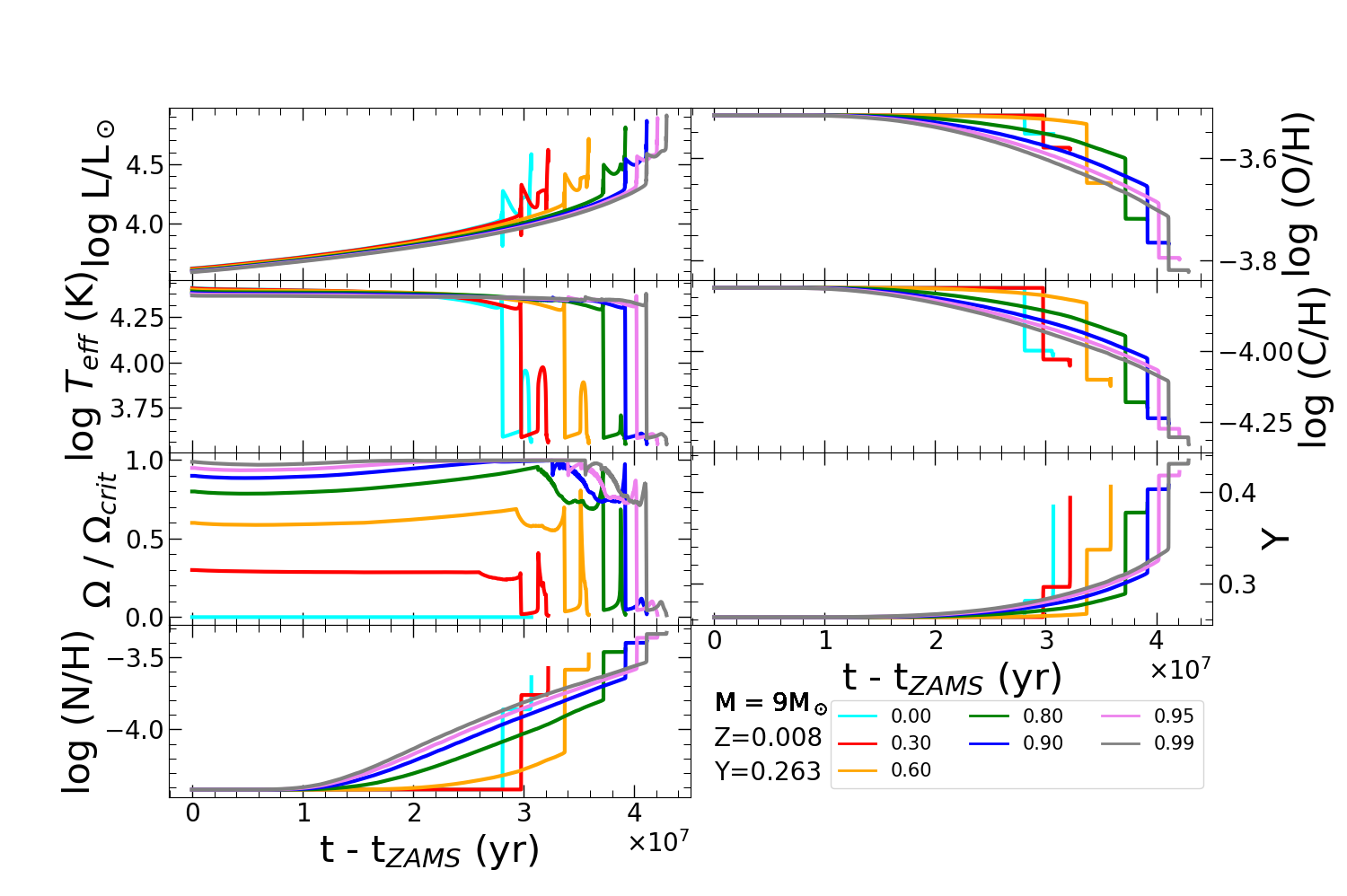}
    \caption{The evolution of $\log L$, $\log T_\mathrm{eff}$, rotation rate $\omega$ and surface abundances of N, C, O, and He at the surface for many initial rotation rates from $\omega_\mathrm{i}=0.00$ to $0.99$ (from cyan to grey colours respectively), for the cases of 3, 5 and $9~\Msun$ stars (in the three sets of panels from top to bottom, respectively) with $Z=0.008, Y=0.263$. The abundances of N, O, and C are by number, and relative to the hydrogen abundance. For He, instead, we present the surface mass fraction $Y$.
    }
    \label{LTOAs_evolution}
\end{figure}

Another consequence of rotational mixing during the MS phase is the transport of nuclear-burned products from the central region to the surface. This effect does not occur in non-rotating stars, until dredge-up events occur when the stars become red giants. In rotating stars, instead, significant mixing can occur at much earlier stages. The most evident effect of this mixing is an enhancement in the surface nitrogen and helium, followed by a depletion of both oxygen and carbon. Fig.~\ref{LTOAs_evolution} shows the evolution of He, C, N, O abundance, luminosity, effective temperature, and $\omega$, in three stars of mass $3~\Msun$, $5~\Msun$, and $9~\Msun$, for several initial rotation rates.
The faster the star rotates on the MS, the more N and He appear at the surface, and the more C and O are depleted. The increase/decrease in surface abundances develops gradually during the MS but suddenly jumps up/down during the first dredge-up event that occurs after the end of the MS, when the star becomes a red giant. Afterwards, rotational mixing is no longer efficient, and the surface abundances remain constant until, eventually, the advent of the second dredge-up, which affects the IMS of higher mass, after the core-helium burning phase. 

As can be seen in Figs.~\ref{HRD008} and \ref{LTOAs_evolution}, the increased rotation rates on the MS also reduce the extension (in \Teff) of the blue loop during the central He-burning phase. This is also an effect of the enhanced mixing caused by rotation \citep[see the discussion of][and references therein]{2019A&A...631A.128C}.

\begin{figure}
	\includegraphics[width=\columnwidth]{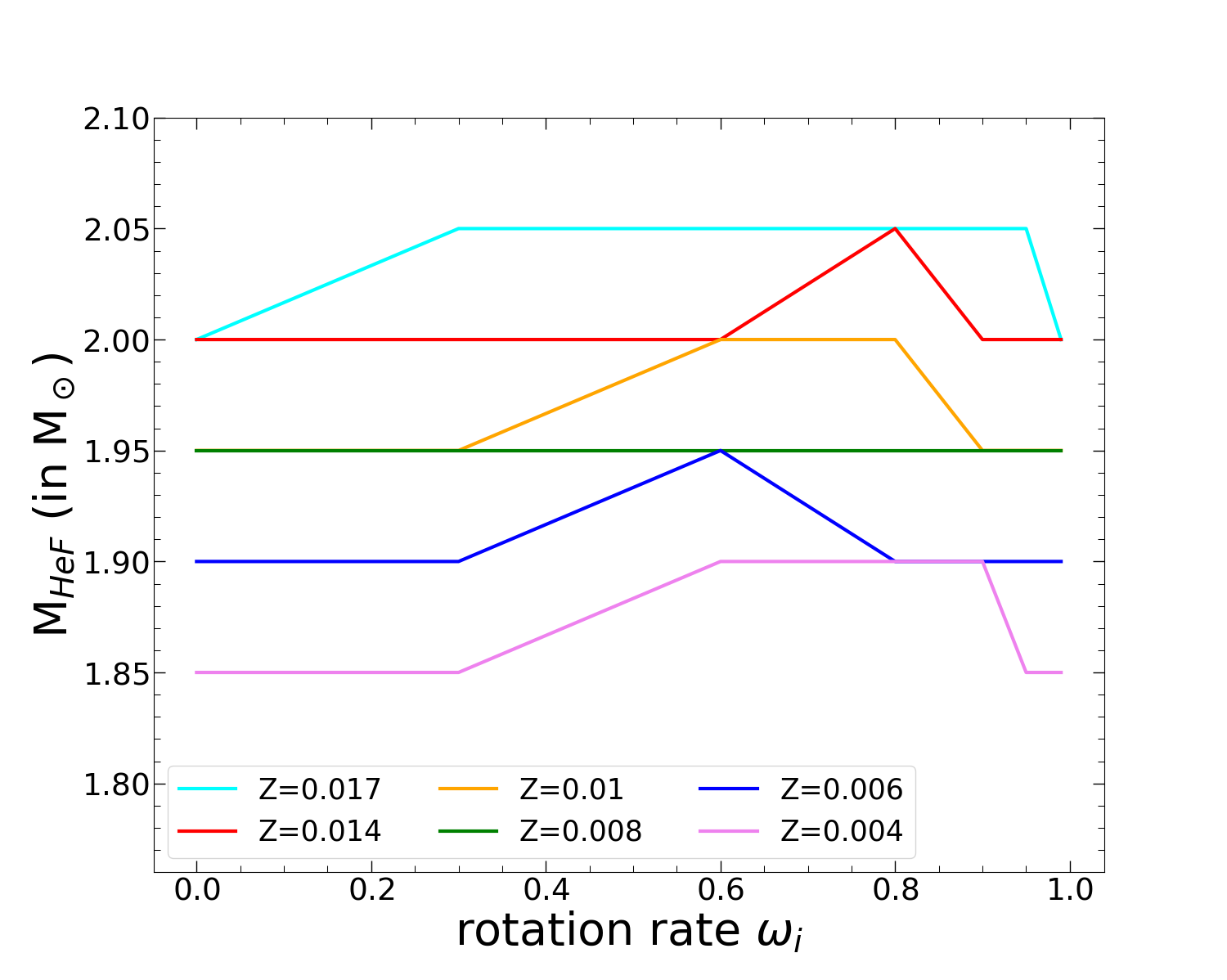}
    \caption{The maximum mass of which the star burns He in the central region under a strongly degenerative condition, $M_\mathrm{HeF}$, versus the initial rotation rates. The colour refers to six computed metallicities in this paper as written in label.}
    \label{Mhefvsomega}
\end{figure}

The transition between LMS and IMS is set at a mass $M_{\mathrm{HeF}}$, above which He ignition takes place quietly in a non-degenerate core. Fig.~\ref{Mhefvsomega} shows the value of $M_{\mathrm{HeF}}$ as a function of initial rotation rates for the six metallicities computed in this paper. The plot has a resolution of 0.05~\Msun, which is the mass separation between successive tracks computed around this mass range. First, lines of different colours illustrate the well-known dependency of $M_\mathrm{HeF}$ on the initial metallicity. Second, the dependency of $M_\mathrm{HeF}$ on initial rotation rate. For instance, the $Z=0.004$ models have $M_\mathrm{HeF}$ values of either $1.85~\Msun$ or $1.90~\Msun$, while those with $Z=0.017$ the values of 2.00 or 2.05 \Msun. We find that at increasing rotation rates, the value of $M_\mathrm{HeF}$ also tends to increase with respect to the non-rotating models. However, when the initial rotation rate increases to values close to the critical break-up velocity, $M_\mathrm{HeF}$ declines again, returning to the value of non-rotating stars. This behaviour of $M_\mathrm{HeF}$ cannot be discussed in much detail because all changes occur within the mass separation step of $0.05~\Msun$. However, it is remarkable that rotation appears to have a limited impact on $M_\mathrm{HeF}$.

\subsection{Comparison with \textsc{parsec V1.2S} and other databases}
\begin{figure*}
	\includegraphics[width=\textwidth]{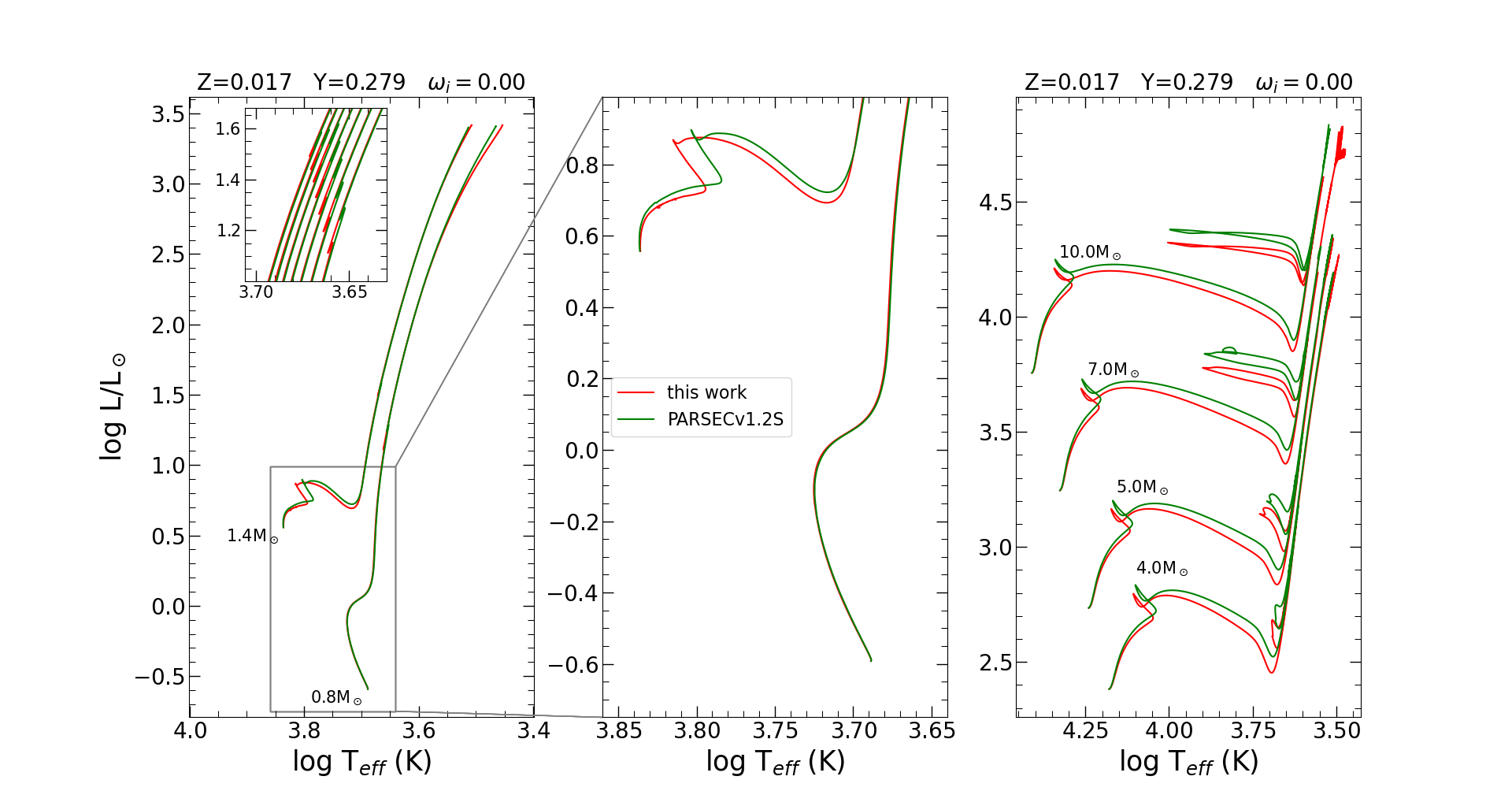}
    \caption{HRD comparing tracks between \textsc{parsec v2.0} (this work, red lines) and the previous version (\textsc{parsec v1.2S}, green lines) for non-rotating stars of $Z=0.017$ and $Y=0.279$. The left-hand panel shows the HRD of two low-mass stars with $0.8$ and $1.4~\Msun$. Their PMS phase is not shown, because it is essentially the same in the two versions. The inset details the region around the RGB bump for tracks in the mass range from 0.8 to 1.4~$\Msun$ with a step of 0.1~$\Msun$. The middle panel zooms in the MS regions of the $0.8$ and $1.4~\Msun$ tracks. The right panel instead compares intermediate-mass models for four different masses as indicated.}
    \label{comp_v1.2}
\end{figure*}

In Figure~\ref{comp_v1.2} we show a comparison between the selected tracks calculated with the new version of the code, \textsc{parsec V2.0}, hereafter PS2, and with the older version, \textsc{V1.2S}, hereafter PS1. In both versions, we use the same initial chemical composition. 
In the leftmost panel, we show the case of a low-mass star with \Mi = 0.8~\Msun. Since this star does not possess a convective core during the H-burning phase, the HR diagram is the same for the two versions. 

Instead, we recall that during the RGB evolution in the older version, PS1, overshooting at the bottom of the convective envelope has not been considered in the mass range ($\leq M_\mathrm{O1}$), producing RGB bumps (RGBBs) that were too luminous with respect to the observed ones \citep{2018MNRAS.476..496F}. 
To cope with this evident discrepancy, in the new version PS2, we include EOV in low-mass stars as described in Sects.~\ref{overshootsection} - \ref{massrange}. The effect of adding an extra mixing at the bottom of the convective envelope is highlighted in the inset of the left panel in Fig.~\ref{comp_v1.2}, where the RGBBs of stars with masses $0.8~\Msun$ $\leq$ \Mi\ $\leq$ $1.4~\Msun$, are shown.

However, in the mass range where stars develop a convective core in the main sequence (as in the case of \Mi\ = 1.4\Msun), COV and EOV are fully considered in both versions of the \textsc{parsec} code. In PS2, we adopt a smaller value of the maximum core overshooting parameter $\lambda_\mathrm{ov,max}=0.4$, instead of the $\lambda_\mathrm{ov,max}=0.5$ in PS1. Furthermore, in PS2 we adopt a diffusive treatment for convective mixing, where the diffusion equations are coupled with the nuclear reaction rates for all elements in the turbulent regions. In the PS1 version, the convective zones are ``instantaneously" homogenized at every time step. These differences already affect the MS phase of stars with convective cores (with $M_\mathrm{i}>M_\mathrm{O1}$), as shown in the middle panel in Fig.~\ref{comp_v1.2}. The PS2 track with 1.4~\Msun\ presents a hotter and slightly fainter MS phase, and a fainter sub-giant phase.

\begin{figure}
	\includegraphics[width=\columnwidth]{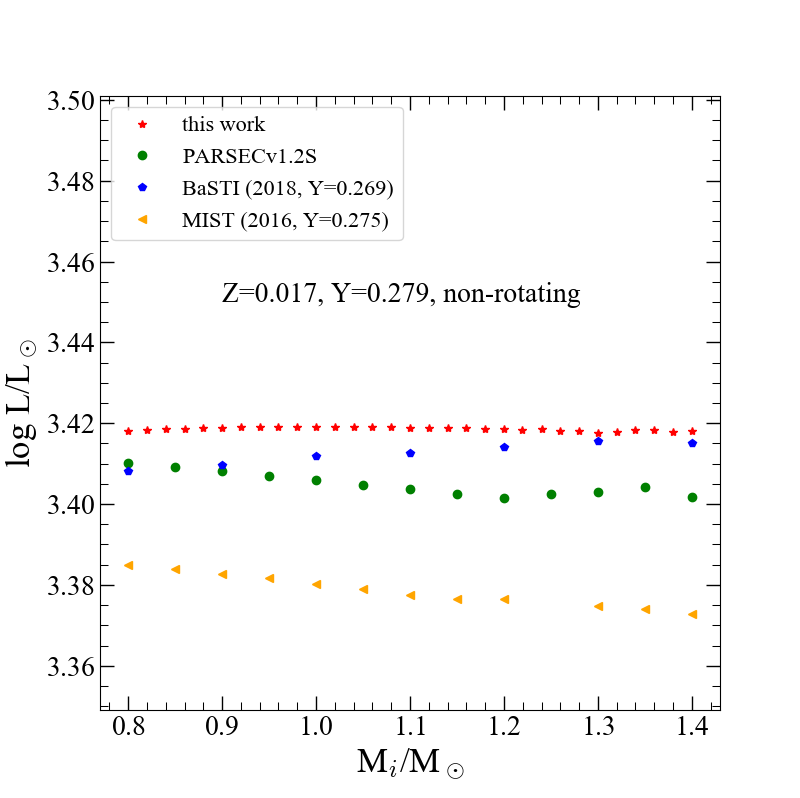}
    \caption{The luminosity at the tip of RGB as a function of initial mass for the tracks produced in this work (with $\omega_\mathrm{i}=0$; red stars), in parsec v1.2S (green circles), and BaSTI (blue pentagons). The BaSTI tracks for solar-scaled composition are taken from \citet{2018ApJ...856..125H} with $Z=0.01721$, $Y=0.2695$.}
   \label{comp_tip}
\end{figure}
On the other hand, the RGB phase has the same slope in both versions of \textsc{parsec} tracks. The new tracks show a brighter and cooler RGB-tip. These differences in the TRGB are caused by the more massive He-core and the more extended envelope at the tip. This is mainly due to the different overshoot parameters used in PS2 and the fact that in this new calculation mass loss was implemented along the evolution while, in PS1, models were evolved at constant mass, and mass loss was applied at the stage of isochrone calculation only. 

In Fig.~\ref{comp_tip}, we compare the luminosity at the RGB-tip of the PS2 models with that of PS1, BaSTI (Bag of Stellar Tracks and Isochrones, \citealt{2018ApJ...856..125H}) and MIST (MESA Isochrones and Stellar Tracks, \citealt{2016ApJ...823..102C}) evolutionary tracks.
The latter two databases also include convective overshooting and diffusion in their models. We can see that PS2 predicts a quite constant TRGB luminosity and, generally, above the luminosity of other models shown in the plot. The difference between the new and old versions of \textsc{parsec} is about $\sim 0.01-0.02$ dex. BaSTI gives an increased trend of TRGB luminosity with initial masses, which is in contrast with the trend from MIST. PS2's TRGBs are $\sim 0.04$~dex brighter than MIST. 

The right-hand panel of Figure~\ref{comp_v1.2} shows the comparison between non-rotating models of intermediate-mass stars of the two \textsc{parsec} code versions. In this case, the impact of the core overshooting parameter is clear. The difference between PS1 and PS2's tracks starts from the MS and continues up to the He-burning phase. In particular, the new tracks are less luminous than the previous ones, due to their smaller $\lambda_\mathrm{ov,max}$ value.

\begin{figure}
	\includegraphics[width=\columnwidth]{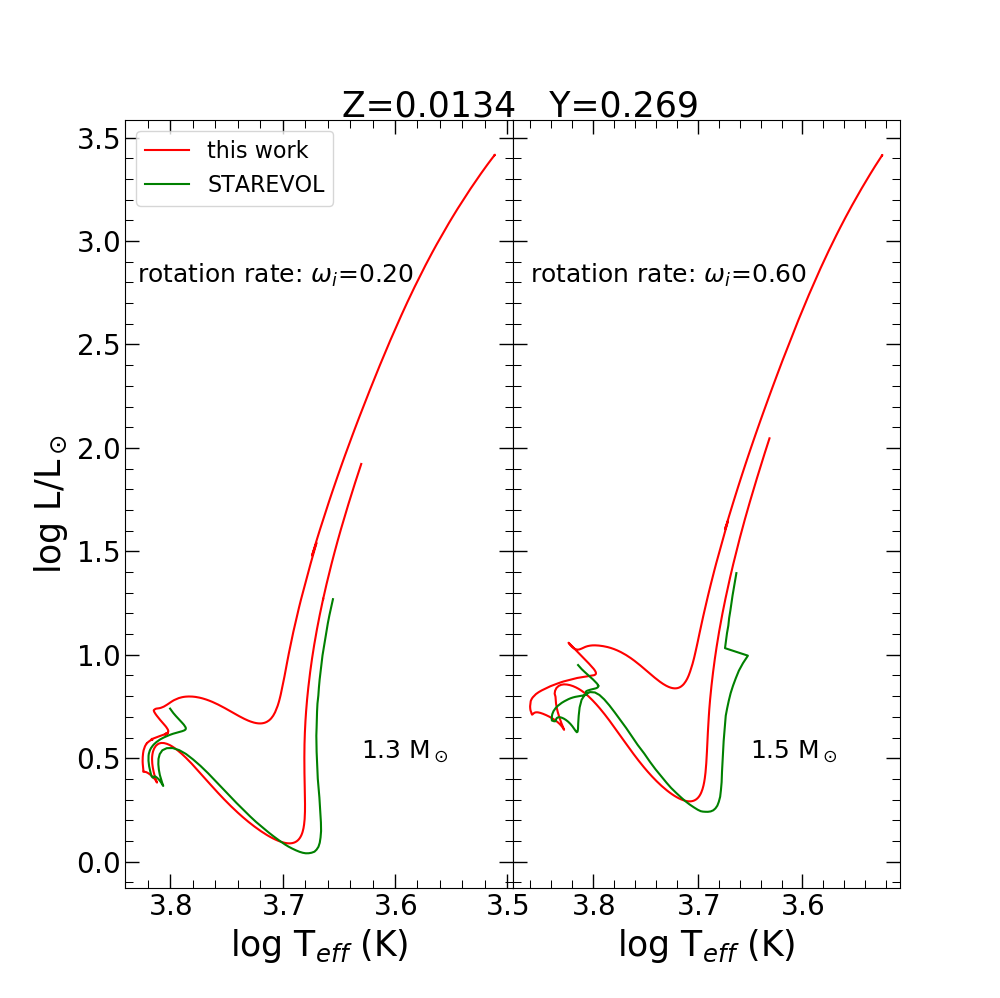}
    \caption{The comparison on HRD of rotating tracks between this work (red lines) and \textsc{starevol} (green lines) with the same $Z=0.0134$ and $Y=0.269$. Left-hand panel: slow-rotating tracks with $\omega_\mathrm{i}=0.20$ of $1.3~\Msun$ star. Right-hand panel: the same as the left panel but with $\omega_\mathrm{i}=0.60$ and $\Mi=1.5~\Msun$.}
    \label{comp_starevol}
\end{figure}

Recently, \citet{2019A&A...631A..77A} published grids of \textsc{starevol} models in which rotation is included for masses from $0.2~\Msun$ to $1.5~\Msun$. \textsc{starevol} tracks are provided for three values of initial rotation rates, $\omega_\mathrm{i}=$ 0.20, 0.40 and 0.60, while in this work we provide $\omega_\mathrm{i}=$ 0.30, 0.60, 0.80, 0.90, 0.95, 0.99. To facilitate the comparison, we perform a few \textsc{parsec V2.0} calculations with exactly the same initial composition ($Z=0.0134$ and $Y=0.269$) and the same initial rotation rate as \textsc{starevol}. Fig.~\ref{comp_starevol} compares the models of $1.3~\Msun$ and $1.5~\Msun$ produced by both \textsc{parsec v2.0} (red line) and \textsc{starevol} (green line).
The differences between our and \textsc{starevol} models are significant. First, the \textsc{starevol} tracks evolve until the end of the MS phase, while our tracks extend up to the He-flash. Second, for the same initial mass, rotation rate, and composition, our MS stars are hotter and brighter. This might be explained by the many differences in the input physical parameters between the two codes. For example, \citet{2019A&A...631A..77A} do not include overshooting in their calculations, while we consider it for both the convective core and the envelope. Third, 
there are differences in the implementation of rotation in each codes, namely, \textsc{starevol} implements rotation from the PMS while we assign the rotation (and let it evolve) just before the ZAMS. It is also worth mentioning that there are other differences between the two codes, e.g. they adopt the mixing-length parameter $\alpha_{\mathrm{MLT}}=1.973$ and the nuclear reaction rates from the NACRE II database \citep{2013A&A...549A.106X}.

However, despite the differences listed above, the two codes give similar ages at the terminal-age-MS (TAMS). For instance, for the $1.3~\Msun$ star with $\omega_\mathrm{i}=0.2$ PS2 gives 3.74 Gyr while \textsc{starevol} gives 3.94 Gyr. 

Another similarity is in the mass loss rates: at the TAMS, the $1.5~\Msun$ star with $\omega_\mathrm{i}=0.6$ loses its mass with a rate of $\log \dot{M}=-11.70$ ($M_\odot/yr$) in PS2 while \textsc{starevol} gives $\log \dot{M}=-11.62$ ($\Msun/yr$), even though the codes use different schemes for the mass loss rate. In particular, we use the enhanced formula from the Reimers law with $\eta=0.2$ for rotating stars while \textsc{starevol} uses the recipe of \citet{2011ApJ...741...54C}. However, we should note that at these early stages, the mass loss does not play a crucial role yet.

\begin{figure*}
\includegraphics[width=\textwidth]{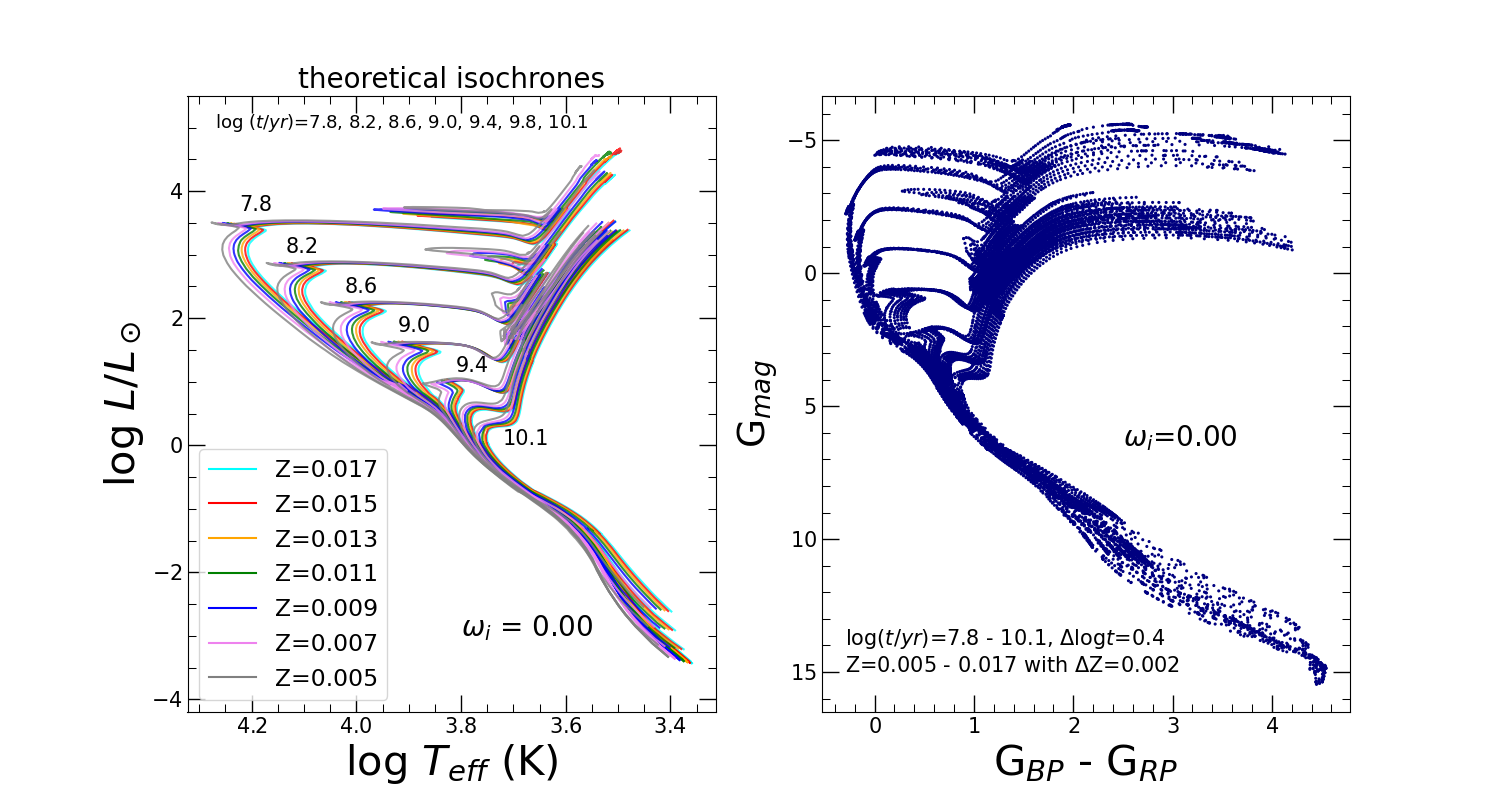}
\caption{Theoretical isochrones calculated with the \textsc{trilegal} code. Left-hand panel: isochrones of non-rotating models for the ages $\log(t/\mathrm{yr})=7.8, 8.2, 8.6, 9.0, 9.4, 9.8, 10.1$, and seven different metallicities from $0.005$ to $0.017$, are shown in different colours from gray to cyan, as in the legend. Right-hand panel: the corresponding colour-magnitude diagram in \textit{Gaia}-passbands of the theoretical isochrones shown in the left panel.
}
\label{iso_result_rot000}
\end{figure*}

\begin{figure*}
\includegraphics[width=\textwidth]{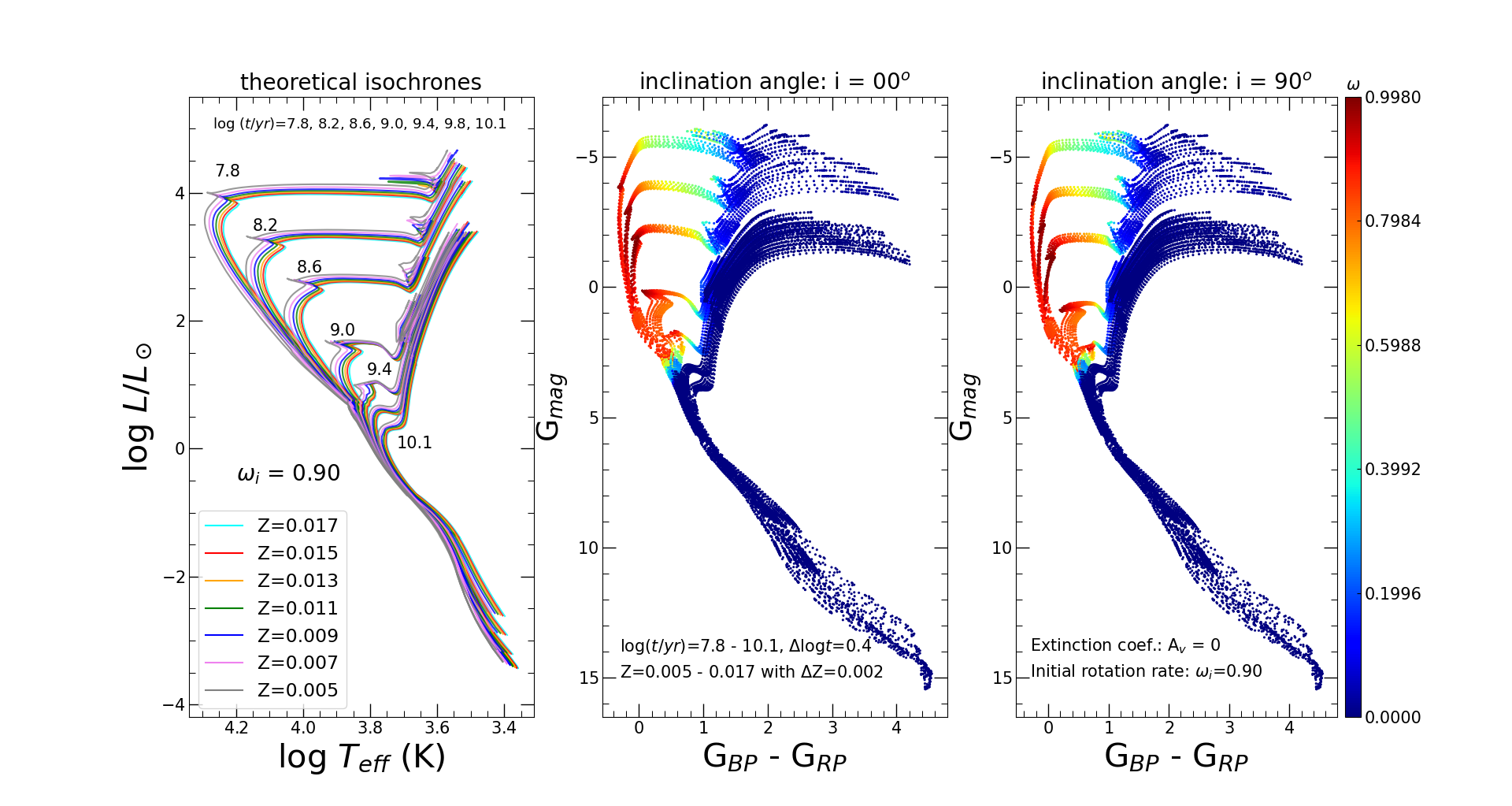}
\caption{Left-hand panel: theoretical isochrones with the same parameter of $\log t$s and $Z$s as in Fig.~\ref{iso_result_rot000} but for $\omega_\mathrm{i}=0.90$ set. The middle and right-hand panels: the corresponding colour-magnitude diagram in the \textit{Gaia}-passbands for two values of the inclination angle: $i=0^\circ$ (pole-on) and $i=90^\circ$ (equator-on). Changes in the rotation rate along the isochrones are indicated by the colour bar. 
}
\label{iso_result}
\end{figure*}

\section{Isochrones}\label{iso}

For all sets of evolutionary tracks described in the previous sections, we have constructed the corresponding isochrones. 
The initial phase begins from the PMS, and the final stage is either the beginning of the TP-AGB phase for low- and intermediate-mass models or the C-exhaustion for higher masses. 
As explained in Sect.~\ref{rotationsection}, at a given initial metallicity and rotation rate, a certain number of low-mass models were not computed, due to our choice of decreasing $\omega_\mathrm{max}$ at decreasing 
$M_\mathrm{i}$, in the transition towards low-mass stars (Eq.~\ref{omemax}). While computing the isochrones, the missing tracks of a given $\omega_\mathrm{i}$ are replaced by the track with the nearest initial mass in the set of tracks with the same metallicity and with $\omega_\mathrm{i}$ immediately smaller. This ensures that the isochrones gradually shift from the required $\omega_\mathrm{i}$ to the non-rotating case in the mass interval between $M_\mathrm{O2}$ and $M_\mathrm{O1}$.

After selecting all the stellar tracks in each set, based on the initial metallicity and rotation rate, the computation of isochrones proceeds in the following steps: first, the computed stellar evolutionary tracks in each set are homogeneously divided into phases separated by a few characteristic ``equivalent evolutionary points''. Then, for a given age, the isochrone is constructed by interpolating all stellar properties between points of different initial mass but equivalent evolutionary stage. More details of the interpolation scheme can be found in \citet{1990A&AS...85..845B} (see also \citealt{2008A&A...484..815B}). In this paper, the isochrones are produced by a recent version of the \textsc{trilegal} code \citep{2005A&A...436..895G,2017ApJ...835...77M}, which interpolates all the additional quantities needed to characterize rotating stars. 
Several isochrones have been produced with metallicity in the range from $0.004$ to $0.017$ in steps of $0.001$ and ages in the range from $10$ Myr to $\sim 13$~Gyr at intervals of $0.05$ in the scale $\log$ and for the seven sets of initial rotation rates from zero to $\omega_\mathrm{i}=0.99$. As an example, the left-hand panel of Fig.~\ref{iso_result_rot000} shows the theoretical isochrones of non-rotating stars for several ages and metallicities.

The theoretical isochrones provide the intrinsic properties of the stars, such as, for instance, the luminosity, mean effective temperature, angular velocity, radius at pole and equator, etc. Then they are complemented with photometric magnitudes in several filters for comparison with observed colour-magnitude diagrams. For non-rotating stars, this is usually done by using tables of bolometric corrections (BCs) as a function of effective temperature, surface gravity and metallicity \citep[see][]{2002A&A...391..195G}; eventually these tables also consider the interstellar extinction in a star-to-star basis, as in \citet{2008PASP..120..583G}. The right-hand panel of Fig.~\ref{iso_result_rot000} shows non-rotating isochrones in the \textit{Gaia} passbands, corresponding to those shown in the left-hand panel, where \textit{Gaia} EDR3 photometry is adopted \citep[see][]{2021yCat..36490003R}.

BC tables for rotating stars have at least two more parameters than those for non-rotating stars: the rotation rate $\omega$ and the inclination angle, $i$, of the line of sight with respect to the stellar rotation axes. Such BC tables are described in \citet{2019MNRAS.488..696G}. They are already implemented in the \textsc{ybc} database\footnote{\url{https://sec.center/YBC}} of BCs by \citet{2019A&A...632A.105C}, and in the \textsc{trilegal} code we use to produce the present isochrones. The left-hand panel of Fig.~\ref{iso_result} shows some selected rotating isochrones. The two panels on the middle and right-hand side of Fig.~\ref{iso_result} illustrate the result of applying the BCs to isochrones with rotation $\omega_\mathrm{i}$, and how the photometry changes when observing rotating stars from $i=0^\circ$ (pole-on) and $i=90^\circ$ (equator-on). The changes in the photometry are the most remarkable for the stars close to the upper main sequence, since those are the stars which still retain a large fraction of their initial rotational velocity.

For cool red giants, the BC tables for rotating stars do not cover the complete range of low effective temperatures that might be necessary to build the isochrones containing fast-rotating stars. For instance, for $\omega=0.9$ and $\log g=2$ the BC tables defined in \citet{2019MNRAS.488..696G} are limited to effective temperatures above $\sim4000$~K. 
Fortunately, inspection of our final isochrones reveals that this limitation is not a practical problem: it turns out that all giants with $\Teff$ smaller than $\sim5000$~K are slow rotators, with $\omega\lesssim 0.2$.
Since these slow rotators have nearly spherical configurations, we decide to apply the BC tables for non-rotating stars from \citet{2019A&A...632A.105C} (YBC) to all stars with $\Teff<5250$~K, for all values of $\omega_\mathrm{i}$. This choice ensures a smooth behaviour of the colours, as can be appreciated in the middle and right panels of Fig.~\ref{iso_result}.

The database of isochrones in several photometry systems is available in \url{http://stev.oapd.inaf.it/cmd}.

\section{Discussion and Conclusions}\label{Discussion}

We have presented a new library of evolutionary tracks with rotation for low- and intermediate-mass stars produced with \textsc{parsec V2.0}. Masses from 0.09 \Msun\ to 14 \Msun\ and metallicity Z between Z=0.004 and Z=0.017 are considered, for seven values of the initial rotation rate in the range $\omega_\mathrm{i}=0.00-0.99$.
The major differences between the last version of \textsc{parsec, V2.0} and the previous one are, i) the inclusion of rotation; ii) The inclusion of mass loss along the evolution of all the stars because, for rotating models, it constitutes an important sink of angular momentum \citep{1986ApJ...311..701F};  iii) the treatment of turbulent mixing as a diffusive process together with rotational mixing, nuclear processing, and, molecular diffusion (for low-mass stars). 
In particular, concerning the last point, we recall that,
to estimate the efficiency of overshooting from the convective core, we were guided by the work of \citet{2019MNRAS.485.4641C}, where the maximum core overshooting parameter
has been calibrated in a well-studied sample of eclipsing binary systems \citep{2018ApJ...859..100C}, obtaining $\lambda_\mathrm{ov}=0.4$.

We also calculated the isochrones up to the beginning of the TP-AGB phase or up to the end of the central C-burning phase. Using the \textsc{trilegal} code, they can be interpolated in metallicity, between $Z=0.004$ and $0.017$ , and in the age range $7.0\leq \log (t/yr)\leq 10.1$.

\begin{figure}
    \includegraphics[width=\columnwidth]{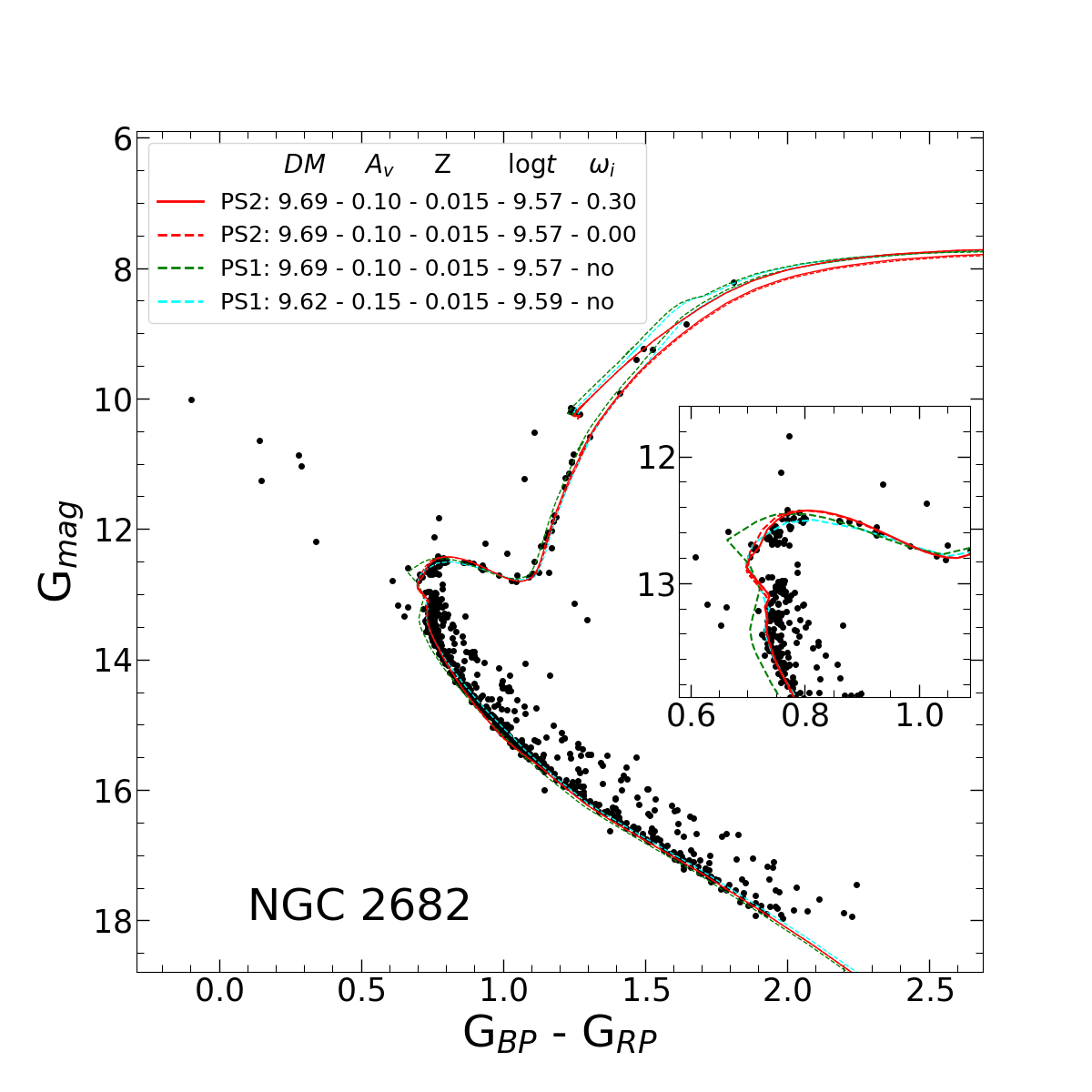}
    \caption{The CMD of open cluster M67 (NGC 2682) from GDR2 data, overplotted by the isochrones that are produced in this work (solid- and dashed-red lines, labeled by PS2) and those from previous version \textsc{parsec V1.2S} (dashed-green and cyan lines, labeled by PS1). The parameters of the isochrones, DM=$(m-M)_0$, $A_\mathrm{V}$, Z, $\log t/yr$ and $\omega_\mathrm{i}$, are displayed in the legend. The inset figure zooms into the turn-off region of this cluster.
    }
    \label{fit_M67}
\end{figure}

\begin{figure}
	\includegraphics[width=\columnwidth]{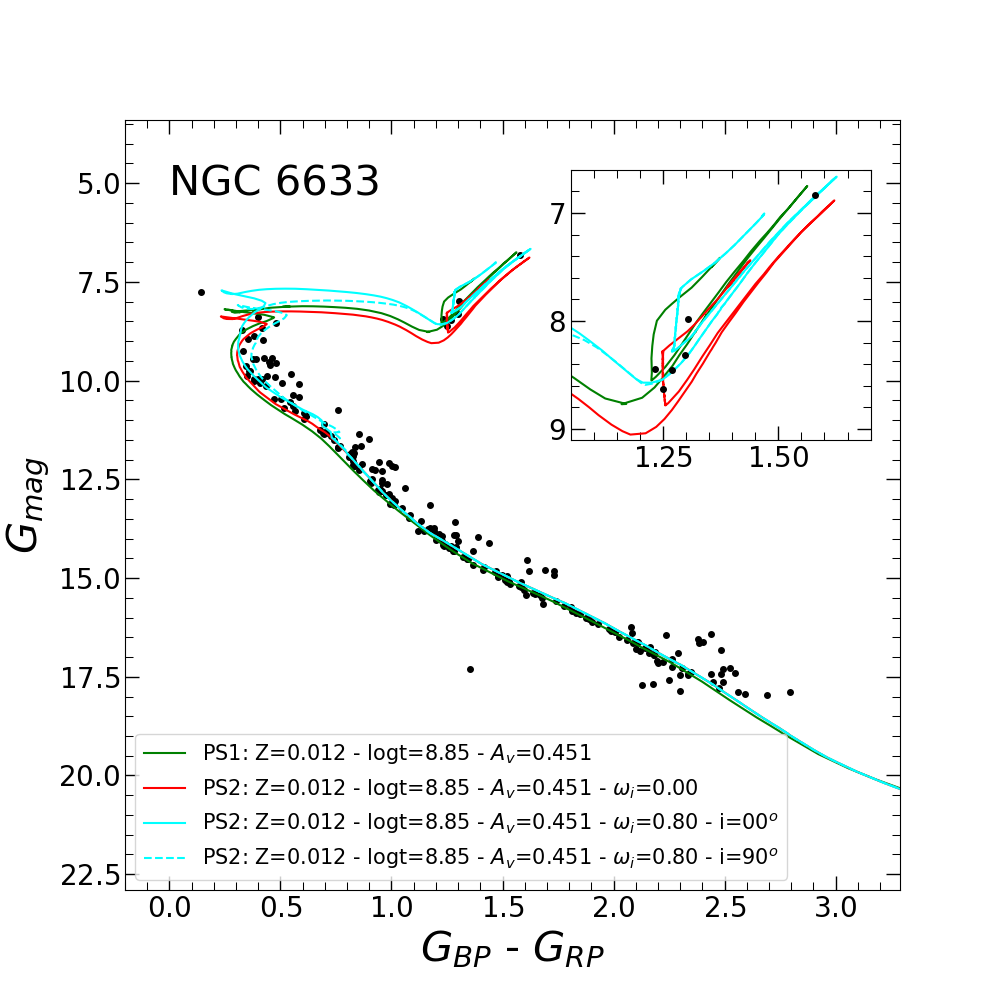}
    \caption{The CMD of the open cluster NGC 6633 from GDR2. The displayed isochrones are for metallicity $Z=0.012$, $\log (t/\mathrm{yr})=8.85$, $(m-M)_0=7.841$ and $A_\mathrm{V}=0.451$ mag. The red-line represents the non-rotating case while the two cyan-lines are for rotating isochrones with the same $\omega_\mathrm{i}=0.80$, 
    and inclination angles $i=0^\circ$ (solid-line) and $i=90^\circ$ (dashed-line), respectively. The green line is the isochrone obtained with \textsc{parsec V1.2S} assuming the same parameters of the previous non-rotating case.}
    \label{fit_NGC6633}
\end{figure}

To illustrate some important consequences of the above differences, we show in 
Fig.~\ref{fit_M67} and Fig.~\ref{fit_NGC6633} two preliminary fits to the observed CMD of the open clusters M67 and NGC 6633, respectively.
M67 (NGC 2682) is a well known test-bench for study the internal physics of stellar models of low-mass stars with typical turn-off masses around 1.2 \Msun. In particular, its CMD was used to calibrate the efficiency of convective overshooting, due to the well-developed convective core in stars around its turn-off region. Furthermore, the cluster, together with other known open clusters, was also used to obtain the age-metallicity relation for the Milky Way disc stars \citep[e.g.][]{2022A&A...660A.135V}. Its age has been repeatedly estimated over the years: \citet{2009ApJ...698.1872S} reported an age between 3.5 and 4.0 Gyr; from the asteroseismic properties of the giant and red clump stars, \citet{2016ApJ...832..133S} derived an age of the cluster of $3.46\pm 0.13$ Gyr; more recently, using the data from the \textit{Gaia} Second Data Release (hereafter GDR2), \citet{2019A&A...623A.108B} derived 
a distance modulus $(m-M)_0=9.726$ mag,  an interstellar extinction coefficient $A_\mathrm{V}=0.115$ mag and an age of 
3.639~$\pm$~0.017 Gyr
\citep[see][for more details about GDR2]{2018A&A...616A..17A,2018A&A...616A..10G,2018A&A...616A...1G,2018A&A...616A...2L}. 

The M67 CMD, shown in Fig.~\ref{fit_M67}, has been obtained from the data provided by \citet{2018A&A...618A..93C} who determined 
photometry, memberships, mean distances, and proper motions of stars in 1229 open clusters. It should be noted that the stars are limited to apparent $G\lesssim 18$ mag to keep the photometric precision in \textit{Gaia}'s passbands at the level of a few millimag \citep[see][for more details]{Godoy-Rivera2021ApJS,2019A&A...623A.108B,2018A&A...616A...4E}.
Also plotted in Fig.~\ref{fit_M67} are a number of our isochrones selected with the following criteria. Lines labeled PS2 indicate our best fit \textsc{parsec V2.0} isochrones. The fit has been obtained by adopting the distance modulus obtained by \citet{2019A&A...623A.108B},  $(m-M)_0=9.726$ mag, but correcting it for a zero-point offset of  -30$\mu$arcsec \citep{2018A&A...616A...2L}. The final corrected distance modulus is $(m-M)_0=9.69$ mag. For an initial composition of $Z=0.015$ ([Fe/H]$\sim$0), $Y=0.275$ (our corresponding He value),
the best fit has been obtained adopting an extinction $A_\mathrm{V}=0.1$ mag and an age of $\log (t/\mathrm{yr})=9.57$. Plotted in the figure are both a non-rotating isochrone (dashed red line) and one for a slow rotation (solid red line). For both isochrones we use the same best fit parameters because the adopted low rotation affects only marginally the region above the cluster turn-off. 

In the same figure, we show also the results we obtain using \textsc{parsec V1.2S}, labeled ``PS1".
For the model represented by the green dashed line we adopt the same fit parameters of the PS2 solutions. The inset in the figure zooms the turn-off region to highlight the differences between the isochrones.
We see that this \textsc{parsec V1.2S} isochrone has a more pronounced hook, at the same fitting parameters. This is an evident feature of models computed with a larger core overshooting parameter. Instead, the cyan dashed isochrone was 
drawn to reproduce the fit obtained with PS1 models, keeping fixed only the metallicity and letting the other parameters to vary within reasonable uncertainties.
For this second PS1 model, that runs almost on top of the PS2 isochrones,  we 
adopt a slightly shorter distance modulus, $(m-M)_0=9.62$, a larger extinction, $A_\mathrm{V}=0.15$ mag, and a 5\% older age $\log (t/\mathrm{yr})=9.59$.
We note that all the four isochrones run almost superimposed onto one another in the subgiant branch which, being an almost horizontal feature in the CMD, is a robust indicator of the apparent distance. The new fitting parameters result from the need to diminish the hook extension that, with \textsc{parsec V1.2S}, can be done only by using a slightly older age for a fixed metallicity. 
The variation of the distance modulus and the attenuation almost compensate each other but the latter is also needed to improve the fitting of the colours of the turn-off region. The differences of the parameters between the latter fit and the PS2 ones should be representative of the differences obtained by using the new version of \textsc{parsec} instead of the previous V1.2S version, in this age domain.
  
Recently, it has been shown that M67 harbours
an interesting spectroscopic binary system located near the turn-off region, WOCS 11028, that challenges theoretical models \cite[see][for a thorough discussion]{2021AJ....161...59S}. Briefly, the mass of the primary component is estimated to be M$_{WOCSa}=1.222\pm 0.006~\Msun$ while, current predictions using different stellar evolution codes (including \textsc{parsec V1.2S}), 
give values that are lower by $\delta m=0.05$ \Msun, i.e. about 8$\sigma$ lower. 
We confirm that we get the same result also with the new version of \textsc{parsec} and leave this problem to a more exhaustive investigation using new \textsc{parsec} models at varying initial metallicity and He content \cite[see also][]{2021AJ....161...59S}.

Another object we present in this paper as a preliminary check of the new models is the young open cluster NGC~6633,  also present in the GDR2's catalogue. High-resolution spectroscopy for NGC6633 comes from the analysis by 
\citet{2021A&A...652A..25C} who studied the age metallicity relation of the Milky Way using 47 open clusters observed with \textit{Gaia}. \citet{2019A&A...623A.108B} derived for NGC~6633
$(m-M)_0=7.866^{+0.024}_{-0.025}$ mag, $\log (t/\mathrm{yr}) = 8.888^{+0.006}_{-0.032}$ and $A_\mathrm{V}=0.451^{+0.025}_{-0.02}$ mag.
With the same procedure used for M67, 
we fit the CMD of NGC~6633 with the new isochrones adopting $Z=0.012$, $Y=0.270$, $A_\mathrm{V}=0.451$ mag and distance modulus $(m-M)_0=7.841$ mag including $-30\mu$arcsec offset in \textit{Gaia} parallaxes, and the age of $\log (t/\mathrm{yr})=8.85$ (Fig.~\ref{fit_NGC6633}). Both non-rotating and rotating-isochrones are displayed with values indicated in the corresponding labels.  The lower MS is very well fitted while the extended MS turn-off region is fully reproduced by rotating-isochrones taking also into account the effects of inclination angles that, in this cluster, are clearly seen. 
Furthermore, the different rotational velocities in this cluster can also explain the particular feature visible near the red clump. Indeed, if only rotating models were used, as needed by the fit of the TO region, it would have been difficult to explain the position of the three stars that fall clearly below the corresponding He clump, given the corresponding much shorter evolutionary lifetimes. They are instead fully compatible with the He clump of non-rotating models of similar age. Thus even in NGC~6633 there are hints for the presence of at least one population of non-rotating stars and another of fast rotators, as in the case of the young LMC cluster NGC~1866 \citep{2019A&A...631A.128C}.

\begin{acknowledgements}
GC acknowledges financial support from the European Research Council for the ERC Consolidator grant DEMOBLACK, under contract no. 770017. LG and PM acknowledge financial support from Padova University, Department of Physics and Astronomy Research Project 2021 (PRD 2021). AB acknowledges funding from PRIN MIUR 2017 prot. 20173ML3WW 001 and 002, `Opening the ALMA window on the cosmic evolution of gas, stars and supermassive black holes'. CY and XF acknowledge the science research grants from the China Manned Space Project with NO. CMS-CSST-2021-A08. YC acknowledges the financial support from  the National Natural Science Foundation of China (12003001) and the National Key R\&D Program of China (2021YFC2203100). XF thanks the support of China Postdoctoral Science Foundation No. 2020M670023, the National Key R\&D Program of China No. 2019YFA0405500 and the National Natural Science Foundation of China (NSFC) under grant No.11973001, 12090040, and 12090044. PG acknowledges support provided by NASA through grant HST-AR-15023 from the Space Telescope Science Institute, which is operated by the Association of Universities for Research in Astronomy, Inc., under NASA contract NAS5-26555.
\end{acknowledgements}

%%%%%%%%%%%%%%%%%%%% REFERENCES %%%%%%%%%%%%%%%%%%

\bibliographystyle{aa}
\bibliography{references} 

\begin{thebibliography}{134}
\expandafter\ifx\csname natexlab\endcsname\relax\def\natexlab#1{#1}\fi

\bibitem[{{Aaronson} \& {Mould}(1982)}]{1982ApJS...48..161A}
{Aaronson}, M. \& {Mould}, J. 1982, \apjs, 48, 161

\bibitem[{{Allard} {et~al.}(2012){Allard}, {Homeier}, \&
  {Freytag}}]{2012RSPTA.370.2765A}
{Allard}, F., {Homeier}, D., \& {Freytag}, B. 2012, Philosophical Transactions
  of the Royal Society of London Series A, 370, 2765

\bibitem[{{Alongi} {et~al.}(1991){Alongi}, {Bertelli}, {Bressan}, \&
  {Chiosi}}]{1991A&A...244...95A}
{Alongi}, M., {Bertelli}, G., {Bressan}, A., \& {Chiosi}, C. 1991, \aap, 244,
  95

\bibitem[{{Amard} {et~al.}(2019){Amard}, {Palacios}, {Charbonnel}, {Gallet},
  {Georgy}, {Lagarde}, \& {Siess}}]{2019A&A...631A..77A}
{Amard}, L., {Palacios}, A., {Charbonnel}, C., {et~al.} 2019, \aap, 631, A77

\bibitem[{{Aparicio} {et~al.}(1990){Aparicio}, {Bertelli}, {Chiosi}, \&
  {Garcia-Pelayo}}]{1990A&A...240..262A}
{Aparicio}, A., {Bertelli}, G., {Chiosi}, C., \& {Garcia-Pelayo}, J.~M. 1990,
  \aap, 240, 262

\bibitem[{{Arenou} {et~al.}(2018){Arenou}, {Luri}, {Babusiaux}, {Fabricius},
  {Helmi}, {Muraveva}, {Robin}, {Spoto}, {Vallenari}, {Antoja},
  {Cantat-Gaudin}, {Jordi}, {Leclerc}, {Reyl{\'e}}, {Romero-G{\'o}mez}, {Shih},
  {Soria}, {Barache}, {Bossini}, {Bragaglia}, {Breddels}, {Fabrizio},
  {Lambert}, {Marrese}, {Massari}, {Moitinho}, {Robichon}, {Ruiz-Dern},
  {Sordo}, {Veljanoski}, {Eyer}, {Jasniewicz}, {Pancino}, {Soubiran}, {Spagna},
  {Tanga}, {Turon}, \& {Zurbach}}]{2018A&A...616A..17A}
{Arenou}, F., {Luri}, X., {Babusiaux}, C., {et~al.} 2018, \aap, 616, A17

\bibitem[{{Asplund} {et~al.}(2006){Asplund}, {Grevesse}, \& {Jacques
  Sauval}}]{2006NuPhA.777....1A}
{Asplund}, M., {Grevesse}, N., \& {Jacques Sauval}, A. 2006, \nphysa, 777, 1

\bibitem[{{Asplund} {et~al.}(2009){Asplund}, {Grevesse}, {Sauval}, \&
  {Scott}}]{2009ARA&A..47..481A}
{Asplund}, M., {Grevesse}, N., {Sauval}, A.~J., \& {Scott}, P. 2009, \araa, 47,
  481

\bibitem[{{Bertelli} {et~al.}(1990{\natexlab{a}}){Bertelli}, {Betto},
  {Bressan}, {Chiosi}, {Nasi}, \& {Vallenari}}]{Bertelli1990}
{Bertelli}, G., {Betto}, R., {Bressan}, A., {et~al.} 1990{\natexlab{a}}, \aaps,
  85, 845

\bibitem[{{Bertelli} {et~al.}(1990{\natexlab{b}}){Bertelli}, {Betto},
  {Bressan}, {Chiosi}, {Nasi}, \& {Vallenari}}]{1990A&AS...85..845B}
{Bertelli}, G., {Betto}, R., {Bressan}, A., {et~al.} 1990{\natexlab{b}}, \aaps,
  85, 845

\bibitem[{{Bertelli} {et~al.}(1986){Bertelli}, {Bressan}, {Chiosi}, \&
  {Angerer}}]{1986A&AS...66..191B}
{Bertelli}, G., {Bressan}, A., {Chiosi}, C., \& {Angerer}, K. 1986, \aaps, 66,
  191

\bibitem[{{Bertelli} {et~al.}(1984){Bertelli}, {Bressan}, \&
  {Chiosi}}]{Bertelli1984}
{Bertelli}, G., {Bressan}, A.~G., \& {Chiosi}, C. 1984, \aap, 130, 279

\bibitem[{{Bertelli} {et~al.}(2008){Bertelli}, {Girardi}, {Marigo}, \&
  {Nasi}}]{2008A&A...484..815B}
{Bertelli}, G., {Girardi}, L., {Marigo}, P., \& {Nasi}, E. 2008, \aap, 484, 815

\bibitem[{{Bjorkman} \& {Cassinelli}(1993)}]{1993ApJ...409..429B}
{Bjorkman}, J.~E. \& {Cassinelli}, J.~P. 1993, \apj, 409, 429

\bibitem[{{Bloecker}(1995)}]{1995A&A...297..727B}
{Bloecker}, T. 1995, \aap, 297, 727

\bibitem[{{B{\"o}hm}(1958)}]{1958ZA.....46..245B}
{B{\"o}hm}, K.~H. 1958, \zap, 46, 245

\bibitem[{{B{\"{o}}hm-Vitense}(1958)}]{Bohm-Vitense1958}
{B{\"{o}}hm-Vitense}, E. 1958, \zap, 46, 108

\bibitem[{{Bossini} {et~al.}(2015){Bossini}, {Miglio}, {Salaris},
  {Pietrinferni}, {Montalb{\'a}n}, {Bressan}, {Noels}, {Cassisi}, {Girardi}, \&
  {Marigo}}]{Bossini2015MNRAS}
{Bossini}, D., {Miglio}, A., {Salaris}, M., {et~al.} 2015, \mnras, 453, 2290

\bibitem[{{Bossini} {et~al.}(2019){Bossini}, {Vallenari}, {Bragaglia},
  {Cantat-Gaudin}, {Sordo}, {Balaguer-N{\'u}{\~n}ez}, {Jordi}, {Moitinho},
  {Soubiran}, {Casamiquela}, {Carrera}, \& {Heiter}}]{2019A&A...623A.108B}
{Bossini}, D., {Vallenari}, A., {Bragaglia}, A., {et~al.} 2019, \aap, 623, A108

\bibitem[{{Bressan} {et~al.}(1986){Bressan}, {Bertelli}, \&
  {Chiosi}}]{Bressan86}
{Bressan}, A., {Bertelli}, G., \& {Chiosi}, C. 1986, \memsai, 57, 411

\bibitem[{{Bressan} {et~al.}(1993){Bressan}, {Fagotto}, {Bertelli}, \&
  {Chiosi}}]{Bressan1993}
{Bressan}, A., {Fagotto}, F., {Bertelli}, G., \& {Chiosi}, C. 1993, \aaps, 100,
  647

\bibitem[{{Bressan} {et~al.}(2012){Bressan}, {Marigo}, {Girardi}, {Salasnich},
  {Dal Cero}, {Rubele}, \& {Nanni}}]{2012MNRAS.427..127B}
{Bressan}, A., {Marigo}, P., {Girardi}, L., {et~al.} 2012, \mnras, 427, 127

\bibitem[{{Bressan} {et~al.}(1981){Bressan}, {Chiosi}, \&
  {Bertelli}}]{1981A&A...102...25B}
{Bressan}, A.~G., {Chiosi}, C., \& {Bertelli}, G. 1981, \aap, 102, 25

\bibitem[{{Caffau} {et~al.}(2011){Caffau}, {Ludwig}, {Steffen}, {Freytag}, \&
  {Bonifacio}}]{2011SoPh..268..255C}
{Caffau}, E., {Ludwig}, H.~G., {Steffen}, M., {Freytag}, B., \& {Bonifacio}, P.
  2011, \solphys, 268, 255

\bibitem[{{Cantat-Gaudin} {et~al.}(2018){Cantat-Gaudin}, {Jordi}, {Vallenari},
  {Bragaglia}, {Balaguer-N{\'u}{\~n}ez}, {Soubiran}, {Bossini}, {Moitinho},
  {Castro-Ginard}, {Krone-Martins}, {Casamiquela}, {Sordo}, \&
  {Carrera}}]{2018A&A...618A..93C}
{Cantat-Gaudin}, T., {Jordi}, C., {Vallenari}, A., {et~al.} 2018, \aap, 618,
  A93

\bibitem[{{Casamiquela} {et~al.}(2021){Casamiquela}, {Soubiran}, {Jofr{\'e}},
  {Chiappini}, {Lagarde}, {Tarricq}, {Carrera}, {Jordi},
  {Balaguer-N{\'u}{\~n}ez}, {Carbajo-Hijarrubia}, \&
  {Blanco-Cuaresma}}]{2021A&A...652A..25C}
{Casamiquela}, L., {Soubiran}, C., {Jofr{\'e}}, P., {et~al.} 2021, \aap, 652,
  A25

\bibitem[{{Cassisi} {et~al.}(2002){Cassisi}, {Salaris}, \&
  {Bono}}]{2002ApJ...565.1231C}
{Cassisi}, S., {Salaris}, M., \& {Bono}, G. 2002, \apj, 565, 1231

\bibitem[{{Catal{\'a}n} {et~al.}(2008){Catal{\'a}n}, {Isern},
  {Garc{\'\i}a-Berro}, \& {Ribas}}]{2008MNRAS.387.1693C}
{Catal{\'a}n}, S., {Isern}, J., {Garc{\'\i}a-Berro}, E., \& {Ribas}, I. 2008,
  \mnras, 387, 1693

\bibitem[{{Chaboyer} \& {Zahn}(1992)}]{1992A&A...253..173C}
{Chaboyer}, B. \& {Zahn}, J.~P. 1992, \aap, 253, 173

\bibitem[{{Chen} {et~al.}(2015){Chen}, {Bressan}, {Girardi}, {Marigo}, {Kong},
  \& {Lanza}}]{2015MNRAS.452.1068C}
{Chen}, Y., {Bressan}, A., {Girardi}, L., {et~al.} 2015, \mnras, 452, 1068

\bibitem[{{Chen} {et~al.}(2014){Chen}, {Girardi}, {Bressan}, {Marigo},
  {Barbieri}, \& {Kong}}]{2014MNRAS.444.2525C}
{Chen}, Y., {Girardi}, L., {Bressan}, A., {et~al.} 2014, \mnras, 444, 2525

\bibitem[{{Chen} {et~al.}(2019){Chen}, {Girardi}, {Fu}, {Bressan}, {Aringer},
  {Dal Tio}, {Pastorelli}, {Marigo}, {Costa}, \& {Zhang}}]{2019A&A...632A.105C}
{Chen}, Y., {Girardi}, L., {Fu}, X., {et~al.} 2019, \aap, 632, A105

\bibitem[{{Chieffi} \& {Limongi}(2013{\natexlab{a}})}]{2013ApJ...764...21C}
{Chieffi}, A. \& {Limongi}, M. 2013{\natexlab{a}}, \apj, 764, 21

\bibitem[{{Chieffi} \& {Limongi}(2013{\natexlab{b}})}]{Chieffi2013}
{Chieffi}, A. \& {Limongi}, M. 2013{\natexlab{b}}, \apj, 764, 21

\bibitem[{{Chieffi} \& {Limongi}(2017)}]{Chieffi2017}
{Chieffi}, A. \& {Limongi}, M. 2017, \apj, 836, 79

\bibitem[{{Choi} {et~al.}(2016){Choi}, {Dotter}, {Conroy}, {Cantiello},
  {Paxton}, \& {Johnson}}]{2016ApJ...823..102C}
{Choi}, J., {Dotter}, A., {Conroy}, C., {et~al.} 2016, \apj, 823, 102

\bibitem[{{Christensen-Dalsgaard} {et~al.}(2011){Christensen-Dalsgaard},
  {Monteiro}, {Rempel}, \& {Thompson}}]{2011MNRAS.414.1158C}
{Christensen-Dalsgaard}, J., {Monteiro}, M.~J.~P.~F.~G., {Rempel}, M., \&
  {Thompson}, M.~J. 2011, \mnras, 414, 1158

\bibitem[{{Claret} \& {Torres}(2016)}]{2016A&A...592A..15C}
{Claret}, A. \& {Torres}, G. 2016, \aap, 592, A15

\bibitem[{{Claret} \& {Torres}(2017)}]{2017ApJ...849...18C}
{Claret}, A. \& {Torres}, G. 2017, \apj, 849, 18

\bibitem[{{Claret} \& {Torres}(2018)}]{2018ApJ...859..100C}
{Claret}, A. \& {Torres}, G. 2018, \apj, 859, 100

\bibitem[{{Claret} \& {Torres}(2019)}]{2019ApJ...876..134C}
{Claret}, A. \& {Torres}, G. 2019, \apj, 876, 134

\bibitem[{{Costa} {et~al.}(2022){Costa}, {Ballone}, {Mapelli}, \&
  {Bressan}}]{Costa2022arXiv220403492C}
{Costa}, G., {Ballone}, A., {Mapelli}, M., \& {Bressan}, A. 2022, arXiv
  e-prints, arXiv:2204.03492

\bibitem[{{Costa} {et~al.}(2021){Costa}, {Bressan}, {Mapelli}, {Marigo},
  {Iorio}, \& {Spera}}]{Costa2021MNRAS.501.4514C}
{Costa}, G., {Bressan}, A., {Mapelli}, M., {et~al.} 2021, \mnras, 501, 4514

\bibitem[{{Costa} {et~al.}(2019{\natexlab{a}}){Costa}, {Girardi}, {Bressan},
  {Chen}, {Goudfrooij}, {Marigo}, {Rodrigues}, \&
  {Lanza}}]{2019A&A...631A.128C}
{Costa}, G., {Girardi}, L., {Bressan}, A., {et~al.} 2019{\natexlab{a}}, \aap,
  631, A128

\bibitem[{{Costa} {et~al.}(2019{\natexlab{b}}){Costa}, {Girardi}, {Bressan},
  {Marigo}, {Rodrigues}, {Chen}, {Lanza}, \&
  {Goudfrooij}}]{2019MNRAS.485.4641C}
{Costa}, G., {Girardi}, L., {Bressan}, A., {et~al.} 2019{\natexlab{b}}, \mnras,
  485, 4641

\bibitem[{{Cranmer} \& {Saar}(2011)}]{2011ApJ...741...54C}
{Cranmer}, S.~R. \& {Saar}, S.~H. 2011, \apj, 741, 54

\bibitem[{{Cranmer} {et~al.}(2007){Cranmer}, {van Ballegooijen}, \&
  {Edgar}}]{2007ApJS..171..520C}
{Cranmer}, S.~R., {van Ballegooijen}, A.~A., \& {Edgar}, R.~J. 2007, \apjs,
  171, 520

\bibitem[{{D'Antona} {et~al.}(2017){D'Antona}, {Milone}, {Tailo}, {Ventura},
  {Vesperini}, \& {di Criscienzo}}]{2017NatAs...1E.186D}
{D'Antona}, F., {Milone}, A.~P., {Tailo}, M., {et~al.} 2017, Nature Astronomy,
  1, 0186

\bibitem[{{de Jager} {et~al.}(1988){de Jager}, {Nieuwenhuijzen}, \& {van der
  Hucht}}]{1988A&AS...72..259D}
{de Jager}, C., {Nieuwenhuijzen}, H., \& {van der Hucht}, K.~A. 1988, \aaps,
  72, 259

\bibitem[{{Demarque} {et~al.}(2004){Demarque}, {Woo}, {Kim}, \&
  {Yi}}]{2004ApJS..155..667D}
{Demarque}, P., {Woo}, J.-H., {Kim}, Y.-C., \& {Yi}, S.~K. 2004, \apjs, 155,
  667

\bibitem[{{Dupree} {et~al.}(2017){Dupree}, {Dotter}, {Johnson}, {Marino},
  {Milone}, {Bailey}, {Crane}, {Mateo}, \& {Olszewski}}]{2017ApJ...846L...1D}
{Dupree}, A.~K., {Dotter}, A., {Johnson}, C.~I., {et~al.} 2017, \apjl, 846, L1

\bibitem[{{Eggenberger} {et~al.}(2010){Eggenberger}, {Miglio}, {Montalban},
  {Moreira}, {Noels}, {Meynet}, \& {Maeder}}]{2010A&A...509A..72E}
{Eggenberger}, P., {Miglio}, A., {Montalban}, J., {et~al.} 2010, \aap, 509, A72

\bibitem[{{Ekstr{\"o}m} {et~al.}(2012){Ekstr{\"o}m}, {Georgy}, {Eggenberger},
  {Meynet}, {Mowlavi}, {Wyttenbach}, {Granada}, {Decressin}, {Hirschi},
  {Frischknecht}, {Charbonnel}, \& {Maeder}}]{2012A&A...537A.146E}
{Ekstr{\"o}m}, S., {Georgy}, C., {Eggenberger}, P., {et~al.} 2012, \aap, 537,
  A146

\bibitem[{{Espinosa Lara} \& {Rieutord}(2007)}]{2007A&A...470.1013E}
{Espinosa Lara}, F. \& {Rieutord}, M. 2007, \aap, 470, 1013

\bibitem[{{Evans} {et~al.}(2018){Evans}, {Riello}, {De Angeli}, {Carrasco},
  {Montegriffo}, {Fabricius}, {Jordi}, {Palaversa}, {Diener}, {Busso},
  {Cacciari}, {van Leeuwen}, {Burgess}, {Davidson}, {Harrison}, {Hodgkin},
  {Pancino}, {Richards}, {Altavilla}, {Balaguer-N{\'u}{\~n}ez}, {Barstow},
  {Bellazzini}, {Brown}, {Castellani}, {Cocozza}, {De Luise}, {Delgado},
  {Ducourant}, {Galleti}, {Gilmore}, {Giuffrida}, {Holl}, {Kewley}, {Koposov},
  {Marinoni}, {Marrese}, {Osborne}, {Piersimoni}, {Portell}, {Pulone},
  {Ragaini}, {Sanna}, {Terrett}, {Walton}, {Wevers}, \&
  {Wyrzykowski}}]{2018A&A...616A...4E}
{Evans}, D.~W., {Riello}, M., {De Angeli}, F., {et~al.} 2018, \aap, 616, A4

\bibitem[{{Fagotto} {et~al.}(1994{\natexlab{a}}){Fagotto}, {Bressan},
  {Bertelli}, \& {Chiosi}}]{Fagotto1994a}
{Fagotto}, F., {Bressan}, A., {Bertelli}, G., \& {Chiosi}, C.
  1994{\natexlab{a}}, \aaps, 104, 365

\bibitem[{{Fagotto} {et~al.}(1994{\natexlab{b}}){Fagotto}, {Bressan},
  {Bertelli}, \& {Chiosi}}]{Fagotto1994b}
{Fagotto}, F., {Bressan}, A., {Bertelli}, G., \& {Chiosi}, C.
  1994{\natexlab{b}}, \aaps, 105, 29

\bibitem[{{Freedman} {et~al.}(2019){Freedman}, {Madore}, {Hatt}, {Hoyt},
  {Jang}, {Beaton}, {Burns}, {Lee}, {Monson}, {Neeley}, {Phillips}, {Rich}, \&
  {Seibert}}]{2019ApJ...882...34F}
{Freedman}, W.~L., {Madore}, B.~F., {Hatt}, D., {et~al.} 2019, \apj, 882, 34

\bibitem[{{Freedman} {et~al.}(2020){Freedman}, {Madore}, {Hoyt}, {Jang},
  {Beaton}, {Lee}, {Monson}, {Neeley}, \& {Rich}}]{2020ApJ...891...57F}
{Freedman}, W.~L., {Madore}, B.~F., {Hoyt}, T., {et~al.} 2020, \apj, 891, 57

\bibitem[{{Friend} \& {Abbott}(1986)}]{1986ApJ...311..701F}
{Friend}, D.~B. \& {Abbott}, D.~C. 1986, \apj, 311, 701

\bibitem[{{Fu} {et~al.}(2018){Fu}, {Bressan}, {Marigo}, {Girardi},
  {Montalb{\'a}n}, {Chen}, \& {Nanni}}]{2018MNRAS.476..496F}
{Fu}, X., {Bressan}, A., {Marigo}, P., {et~al.} 2018, \mnras, 476, 496

\bibitem[{{Gaia Collaboration} {et~al.}(2018{\natexlab{a}}){Gaia
  Collaboration}, {Babusiaux}, {van Leeuwen}, {Barstow}, {Jordi}, {Vallenari},
  {Bossini}, {Bressan}, {Cantat-Gaudin}, {van Leeuwen}, {Brown}, {Prusti}, {de
  Bruijne}, {Bailer-Jones}, {Biermann}, {Evans}, {Eyer}, {Jansen}, {Klioner},
  {Lammers}, {Lindegren}, {Luri}, {Mignard}, {Panem}, {Pourbaix}, {Randich},
  {Sartoretti}, {Siddiqui}, {Soubiran}, {Walton}, {Arenou}, {Bastian},
  {Cropper}, {Drimmel}, {Katz}, {Lattanzi}, {Bakker}, {Cacciari},
  {Casta{\~n}eda}, {Chaoul}, {Cheek}, {De Angeli}, {Fabricius}, {Guerra},
  {Holl}, {Masana}, {Messineo}, {Mowlavi}, {Nienartowicz}, {Panuzzo},
  {Portell}, {Riello}, {Seabroke}, {Tanga}, {Th{\'e}venin}, {Gracia-Abril},
  {Comoretto}, {Garcia-Reinaldos}, {Teyssier}, {Altmann}, {Andrae}, {Audard},
  {Bellas-Velidis}, {Benson}, {Berthier}, {Blomme}, {Burgess}, {Busso},
  {Carry}, {Cellino}, {Clementini}, {Clotet}, {Creevey}, {Davidson}, {De
  Ridder}, {Delchambre}, {Dell'Oro}, {Ducourant},
  {Fern{\'a}ndez-Hern{\'a}ndez}, {Fouesneau}, {Fr{\'e}mat}, {Galluccio},
  {Garc{\'\i}a-Torres}, {Gonz{\'a}lez-N{\'u}{\~n}ez}, {Gonz{\'a}lez-Vidal},
  {Gosset}, {Guy}, {Halbwachs}, {Hambly}, {Harrison}, {Hern{\'a}ndez},
  {Hestroffer}, {Hodgkin}, {Hutton}, {Jasniewicz}, {Jean-Antoine-Piccolo},
  {Jordan}, {Korn}, {Krone-Martins}, {Lanzafame}, {Lebzelter}, {L{\"o}ffler},
  {Manteiga}, {Marrese}, {Mart{\'\i}n-Fleitas}, {Moitinho}, {Mora}, {Muinonen},
  {Osinde}, {Pancino}, {Pauwels}, {Petit}, {Recio-Blanco}, {Richards},
  {Rimoldini}, {Robin}, {Sarro}, {Siopis}, {Smith}, {Sozzetti}, {S{\"u}veges},
  {Torra}, {van Reeven}, {Abbas}, {Abreu Aramburu}, {Accart}, {Aerts},
  {Altavilla}, {{\'A}lvarez}, {Alvarez}, {Alves}, {Anderson}, {Andrei},
  {Anglada Varela}, {Antiche}, {Antoja}, {Arcay}, {Astraatmadja}, {Bach},
  {Baker}, {Balaguer-N{\'u}{\~n}ez}, {Balm}, {Barache}, {Barata}, {Barbato},
  {Barblan}, {Barklem}, {Barrado}, {Barros}, {Bartholom{\'e} Mu{\~n}oz},
  {Bassilana}, {Becciani}, {Bellazzini}, {Berihuete}, {Bertone}, {Bianchi},
  {Bienaym{\'e}}, {Blanco-Cuaresma}, {Boch}, {Boeche}, {Bombrun}, {Borrachero},
  {Bouquillon}, {Bourda}, {Bragaglia}, {Bramante}, {Breddels}, {Brouillet},
  {Br{\"u}semeister}, {Brugaletta}, {Bucciarelli}, {Burlacu}, {Busonero},
  {Butkevich}, {Buzzi}, {Caffau}, {Cancelliere}, {Cannizzaro}, {Carballo},
  {Carlucci}, {Carrasco}, {Casamiquela}, {Castellani}, {Castro-Ginard},
  {Charlot}, {Chemin}, {Chiavassa}, {Cocozza}, {Costigan}, {Cowell}, {Crifo},
  {Crosta}, {Crowley}, {Cuypers}, {Dafonte}, {Damerdji}, {Dapergolas}, {David},
  {David}, {de Laverny}, {De Luise}, {De March}, {de Martino}, {de Souza}, {de
  Torres}, {Debosscher}, {del Pozo}, {Delbo}, {Delgado}, {Delgado}, {Diakite},
  {Diener}, {Distefano}, {Dolding}, {Drazinos}, {Dur{\'a}n}, {Edvardsson},
  {Enke}, {Eriksson}, {Esquej}, {Eynard Bontemps}, {Fabre}, {Fabrizio},
  {Faigler}, {Falc{\~a}o}, {Farr{\`a}s Casas}, {Federici}, {Fedorets},
  {Fernique}, {Figueras}, {Filippi}, {Findeisen}, {Fonti}, {Fraile}, {Fraser},
  {Fr{\'e}zouls}, {Gai}, {Galleti}, {Garabato}, {Garc{\'\i}a-Sedano},
  {Garofalo}, {Garralda}, {Gavel}, {Gavras}, {Gerssen}, {Geyer}, {Giacobbe},
  {Gilmore}, {Girona}, {Giuffrida}, {Glass}, {Gomes}, {Granvik}, {Gueguen},
  {Guerrier}, {Guiraud}, {Guti{\'e}}, {Haigron}, {Hatzidimitriou}, {Hauser},
  {Haywood}, {Heiter}, {Helmi}, {Heu}, {Hilger}, {Hobbs}, {Hofmann}, {Holland},
  {Huckle}, {Hypki}, {Icardi}, {Jan{\ss}en}, {Jevardat de Fombelle}, {Jonker},
  {Juh{\'a}sz}, {Julbe}, {Karampelas}, {Kewley}, {Klar}, {Kochoska}, {Kohley},
  {Kolenberg}, {Kontizas}, {Kontizas}, {Koposov}, {Kordopatis},
  {Kostrzewa-Rutkowska}, {Koubsky}, {Lambert}, {Lanza}, {Lasne}, {Lavigne}, {Le
  Fustec}, {Le Poncin-Lafitte}, {Lebreton}, {Leccia}, {Leclerc},
  {Lecoeur-Taibi}, {Lenhardt}, {Leroux}, {Liao}, {Licata}, {Lindstr{\o}m},
  {Lister}, {Livanou}, {Lobel}, {L{\'o}pez}, {Managau}, {Mann}, {Mantelet},
  {Marchal}, {Marchant}, {Marconi}, {Marinoni}, {Marschalk{\'o}}, {Marshall},
  {Martino}, {Marton}, {Mary}, {Massari}, {Matijevi{\v{c}}}, {Mazeh},
  {McMillan}, {Messina}, {Michalik}, {Millar}, {Molina}, {Molinaro},
  {Moln{\'a}r}, {Montegriffo}, {Mor}, {Morbidelli}, {Morel}, {Morris},
  {Mulone}, {Muraveva}, {Musella}, {Nelemans}, {Nicastro}, {Noval},
  {O'Mullane}, {Ord{\'e}novic}, {Ord{\'o}{\~n}ez-Blanco}, {Osborne}, {Pagani},
  {Pagano}, {Pailler}, {Palacin}, {Palaversa}, {Panahi}, {Pawlak},
  {Piersimoni}, {Pineau}, {Plachy}, {Plum}, {Poggio}, {Poujoulet},
  {Pr{\v{s}}a}, {Pulone}, {Racero}, {Ragaini}, {Rambaux}, {Ramos-Lerate},
  {Regibo}, {Reyl{\'e}}, {Riclet}, {Ripepi}, {Riva}, {Rivard}, {Rixon},
  {Roegiers}, {Roelens}, {Romero-G{\'o}mez}, {Rowell}, {Royer}, {Ruiz-Dern},
  {Sadowski}, {Sagrist{\`a} Sell{\'e}s}, {Sahlmann}, {Salgado}, {Salguero},
  {Sanna}, {Santana-Ros}, {Sarasso}, {Savietto}, {Schultheis}, {Sciacca},
  {Segol}, {Segovia}, {S{\'e}gransan}, {Shih}, {Siltala}, {Silva}, {Smart},
  {Smith}, {Solano}, {Solitro}, {Sordo}, {Soria Nieto}, {Souchay}, {Spagna},
  {Spoto}, {Stampa}, {Steele}, {Steidelm{\"u}ller}, {Stephenson}, {Stoev},
  {Suess}, {Surdej}, {Szabados}, {Szegedi-Elek}, {Tapiador}, {Taris}, {Tauran},
  {Taylor}, {Teixeira}, {Terrett}, {Teyssandier}, {Thuillot}, {Titarenko},
  {Torra Clotet}, {Turon}, {Ulla}, {Utrilla}, {Uzzi}, {Vaillant}, {Valentini},
  {Valette}, {van Elteren}, {Van Hemelryck}, {Vaschetto}, {Vecchiato},
  {Veljanoski}, {Viala}, {Vicente}, {Vogt}, {von Essen}, {Voss}, {Votruba},
  {Voutsinas}, {Walmsley}, {Weiler}, {Wertz}, {Wevers}, {Wyrzykowski},
  {Yoldas}, {{\v{Z}}erjal}, {Ziaeepour}, {Zorec}, {Zschocke}, {Zucker},
  {Zurbach}, \& {Zwitter}}]{2018A&A...616A..10G}
{Gaia Collaboration}, {Babusiaux}, C., {van Leeuwen}, F., {et~al.}
  2018{\natexlab{a}}, \aap, 616, A10

\bibitem[{{Gaia Collaboration} {et~al.}(2018{\natexlab{b}}){Gaia
  Collaboration}, {Brown}, {Vallenari}, {Prusti}, {de Bruijne}, {Babusiaux},
  {Bailer-Jones}, {Biermann}, {Evans}, {Eyer}, {Jansen}, {Jordi}, {Klioner},
  {Lammers}, {Lindegren}, {Luri}, {Mignard}, {Panem}, {Pourbaix}, {Randich},
  {Sartoretti}, {Siddiqui}, {Soubiran}, {van Leeuwen}, {Walton}, {Arenou},
  {Bastian}, {Cropper}, {Drimmel}, {Katz}, {Lattanzi}, {Bakker}, {Cacciari},
  {Casta{\~n}eda}, {Chaoul}, {Cheek}, {De Angeli}, {Fabricius}, {Guerra},
  {Holl}, {Masana}, {Messineo}, {Mowlavi}, {Nienartowicz}, {Panuzzo},
  {Portell}, {Riello}, {Seabroke}, {Tanga}, {Th{\'e}venin}, {Gracia-Abril},
  {Comoretto}, {Garcia-Reinaldos}, {Teyssier}, {Altmann}, {Andrae}, {Audard},
  {Bellas-Velidis}, {Benson}, {Berthier}, {Blomme}, {Burgess}, {Busso},
  {Carry}, {Cellino}, {Clementini}, {Clotet}, {Creevey}, {Davidson}, {De
  Ridder}, {Delchambre}, {Dell'Oro}, {Ducourant},
  {Fern{\'a}ndez-Hern{\'a}ndez}, {Fouesneau}, {Fr{\'e}mat}, {Galluccio},
  {Garc{\'\i}a-Torres}, {Gonz{\'a}lez-N{\'u}{\~n}ez}, {Gonz{\'a}lez-Vidal},
  {Gosset}, {Guy}, {Halbwachs}, {Hambly}, {Harrison}, {Hern{\'a}ndez},
  {Hestroffer}, {Hodgkin}, {Hutton}, {Jasniewicz}, {Jean-Antoine-Piccolo},
  {Jordan}, {Korn}, {Krone-Martins}, {Lanzafame}, {Lebzelter}, {L{\"o}ffler},
  {Manteiga}, {Marrese}, {Mart{\'\i}n-Fleitas}, {Moitinho}, {Mora}, {Muinonen},
  {Osinde}, {Pancino}, {Pauwels}, {Petit}, {Recio-Blanco}, {Richards},
  {Rimoldini}, {Robin}, {Sarro}, {Siopis}, {Smith}, {Sozzetti}, {S{\"u}veges},
  {Torra}, {van Reeven}, {Abbas}, {Abreu Aramburu}, {Accart}, {Aerts},
  {Altavilla}, {{\'A}lvarez}, {Alvarez}, {Alves}, {Anderson}, {Andrei},
  {Anglada Varela}, {Antiche}, {Antoja}, {Arcay}, {Astraatmadja}, {Bach},
  {Baker}, {Balaguer-N{\'u}{\~n}ez}, {Balm}, {Barache}, {Barata}, {Barbato},
  {Barblan}, {Barklem}, {Barrado}, {Barros}, {Barstow}, {Bartholom{\'e}
  Mu{\~n}oz}, {Bassilana}, {Becciani}, {Bellazzini}, {Berihuete}, {Bertone},
  {Bianchi}, {Bienaym{\'e}}, {Blanco-Cuaresma}, {Boch}, {Boeche}, {Bombrun},
  {Borrachero}, {Bossini}, {Bouquillon}, {Bourda}, {Bragaglia}, {Bramante},
  {Breddels}, {Bressan}, {Brouillet}, {Br{\"u}semeister}, {Brugaletta},
  {Bucciarelli}, {Burlacu}, {Busonero}, {Butkevich}, {Buzzi}, {Caffau},
  {Cancelliere}, {Cannizzaro}, {Cantat-Gaudin}, {Carballo}, {Carlucci},
  {Carrasco}, {Casamiquela}, {Castellani}, {Castro-Ginard}, {Charlot},
  {Chemin}, {Chiavassa}, {Cocozza}, {Costigan}, {Cowell}, {Crifo}, {Crosta},
  {Crowley}, {Cuypers}, {Dafonte}, {Damerdji}, {Dapergolas}, {David}, {David},
  {de Laverny}, {De Luise}, {De March}, {de Martino}, {de Souza}, {de Torres},
  {Debosscher}, {del Pozo}, {Delbo}, {Delgado}, {Delgado}, {Di Matteo},
  {Diakite}, {Diener}, {Distefano}, {Dolding}, {Drazinos}, {Dur{\'a}n},
  {Edvardsson}, {Enke}, {Eriksson}, {Esquej}, {Eynard Bontemps}, {Fabre},
  {Fabrizio}, {Faigler}, {Falc{\~a}o}, {Farr{\`a}s Casas}, {Federici},
  {Fedorets}, {Fernique}, {Figueras}, {Filippi}, {Findeisen}, {Fonti},
  {Fraile}, {Fraser}, {Fr{\'e}zouls}, {Gai}, {Galleti}, {Garabato},
  {Garc{\'\i}a-Sedano}, {Garofalo}, {Garralda}, {Gavel}, {Gavras}, {Gerssen},
  {Geyer}, {Giacobbe}, {Gilmore}, {Girona}, {Giuffrida}, {Glass}, {Gomes},
  {Granvik}, {Gueguen}, {Guerrier}, {Guiraud}, {Guti{\'e}rrez-S{\'a}nchez},
  {Haigron}, {Hatzidimitriou}, {Hauser}, {Haywood}, {Heiter}, {Helmi}, {Heu},
  {Hilger}, {Hobbs}, {Hofmann}, {Holland}, {Huckle}, {Hypki}, {Icardi},
  {Jan{\ss}en}, {Jevardat de Fombelle}, {Jonker}, {Juh{\'a}sz}, {Julbe},
  {Karampelas}, {Kewley}, {Klar}, {Kochoska}, {Kohley}, {Kolenberg},
  {Kontizas}, {Kontizas}, {Koposov}, {Kordopatis}, {Kostrzewa-Rutkowska},
  {Koubsky}, {Lambert}, {Lanza}, {Lasne}, {Lavigne}, {Le Fustec}, {Le
  Poncin-Lafitte}, {Lebreton}, {Leccia}, {Leclerc}, {Lecoeur-Taibi},
  {Lenhardt}, {Leroux}, {Liao}, {Licata}, {Lindstr{\o}m}, {Lister}, {Livanou},
  {Lobel}, {L{\'o}pez}, {Managau}, {Mann}, {Mantelet}, {Marchal}, {Marchant},
  {Marconi}, {Marinoni}, {Marschalk{\'o}}, {Marshall}, {Martino}, {Marton},
  {Mary}, {Massari}, {Matijevi{\v{c}}}, {Mazeh}, {McMillan}, {Messina},
  {Michalik}, {Millar}, {Molina}, {Molinaro}, {Moln{\'a}r}, {Montegriffo},
  {Mor}, {Morbidelli}, {Morel}, {Morris}, {Mulone}, {Muraveva}, {Musella},
  {Nelemans}, {Nicastro}, {Noval}, {O'Mullane}, {Ord{\'e}novic},
  {Ord{\'o}{\~n}ez-Blanco}, {Osborne}, {Pagani}, {Pagano}, {Pailler},
  {Palacin}, {Palaversa}, {Panahi}, {Pawlak}, {Piersimoni}, {Pineau}, {Plachy},
  {Plum}, {Poggio}, {Poujoulet}, {Pr{\v{s}}a}, {Pulone}, {Racero}, {Ragaini},
  {Rambaux}, {Ramos-Lerate}, {Regibo}, {Reyl{\'e}}, {Riclet}, {Ripepi}, {Riva},
  {Rivard}, {Rixon}, {Roegiers}, {Roelens}, {Romero-G{\'o}mez}, {Rowell},
  {Royer}, {Ruiz-Dern}, {Sadowski}, {Sagrist{\`a} Sell{\'e}s}, {Sahlmann},
  {Salgado}, {Salguero}, {Sanna}, {Santana-Ros}, {Sarasso}, {Savietto},
  {Schultheis}, {Sciacca}, {Segol}, {Segovia}, {S{\'e}gransan}, {Shih},
  {Siltala}, {Silva}, {Smart}, {Smith}, {Solano}, {Solitro}, {Sordo}, {Soria
  Nieto}, {Souchay}, {Spagna}, {Spoto}, {Stampa}, {Steele},
  {Steidelm{\"u}ller}, {Stephenson}, {Stoev}, {Suess}, {Surdej}, {Szabados},
  {Szegedi-Elek}, {Tapiador}, {Taris}, {Tauran}, {Taylor}, {Teixeira},
  {Terrett}, {Teyssandier}, {Thuillot}, {Titarenko}, {Torra Clotet}, {Turon},
  {Ulla}, {Utrilla}, {Uzzi}, {Vaillant}, {Valentini}, {Valette}, {van Elteren},
  {Van Hemelryck}, {van Leeuwen}, {Vaschetto}, {Vecchiato}, {Veljanoski},
  {Viala}, {Vicente}, {Vogt}, {von Essen}, {Voss}, {Votruba}, {Voutsinas},
  {Walmsley}, {Weiler}, {Wertz}, {Wevers}, {Wyrzykowski}, {Yoldas},
  {{\v{Z}}erjal}, {Ziaeepour}, {Zorec}, {Zschocke}, {Zucker}, {Zurbach}, \&
  {Zwitter}}]{2018A&A...616A...1G}
{Gaia Collaboration}, {Brown}, A.~G.~A., {Vallenari}, A., {et~al.}
  2018{\natexlab{b}}, \aap, 616, A1

\bibitem[{{Georgy} {et~al.}(2013){Georgy}, {Ekstr{\"o}m}, {Eggenberger},
  {Meynet}, {Haemmerl{\'e}}, {Maeder}, {Granada}, {Groh}, {Hirschi}, {Mowlavi},
  {Yusof}, {Charbonnel}, {Decressin}, \& {Barblan}}]{2013A&A...558A.103G}
{Georgy}, C., {Ekstr{\"o}m}, S., {Eggenberger}, P., {et~al.} 2013, \aap, 558,
  A103

\bibitem[{{Girardi} {et~al.}(2002){Girardi}, {Bertelli}, {Bressan}, {Chiosi},
  {Groenewegen}, {Marigo}, {Salasnich}, \& {Weiss}}]{2002A&A...391..195G}
{Girardi}, L., {Bertelli}, G., {Bressan}, A., {et~al.} 2002, \aap, 391, 195

\bibitem[{{Girardi} {et~al.}(2000{\natexlab{a}}){Girardi}, {Bressan},
  {Bertelli}, \& {Chiosi}}]{Girardi2000}
{Girardi}, L., {Bressan}, A., {Bertelli}, G., \& {Chiosi}, C.
  2000{\natexlab{a}}, \aaps, 141, 371

\bibitem[{{Girardi} {et~al.}(2000{\natexlab{b}}){Girardi}, {Bressan},
  {Bertelli}, \& {Chiosi}}]{2000A&AS..141..371G}
{Girardi}, L., {Bressan}, A., {Bertelli}, G., \& {Chiosi}, C.
  2000{\natexlab{b}}, \aaps, 141, 371

\bibitem[{{Girardi} {et~al.}(2019){Girardi}, {Costa}, {Chen}, {Goudfrooij},
  {Bressan}, {Marigo}, \& {Bellini}}]{2019MNRAS.488..696G}
{Girardi}, L., {Costa}, G., {Chen}, Y., {et~al.} 2019, \mnras, 488, 696

\bibitem[{{Girardi} {et~al.}(2008){Girardi}, {Dalcanton}, {Williams}, {de
  Jong}, {Gallart}, {Monelli}, {Groenewegen}, {Holtzman}, {Olsen}, {Seth},
  {Weisz}, \& {ANGST/ANGRRR Collaboration}}]{2008PASP..120..583G}
{Girardi}, L., {Dalcanton}, J., {Williams}, B., {et~al.} 2008, \pasp, 120, 583

\bibitem[{{Girardi} {et~al.}(2005){Girardi}, {Groenewegen}, {Hatziminaoglou},
  \& {da Costa}}]{2005A&A...436..895G}
{Girardi}, L., {Groenewegen}, M.~A.~T., {Hatziminaoglou}, E., \& {da Costa}, L.
  2005, \aap, 436, 895

\bibitem[{{Godoy-Rivera} {et~al.}(2021){Godoy-Rivera}, {Pinsonneault}, \&
  {Rebull}}]{Godoy-Rivera2021ApJS}
{Godoy-Rivera}, D., {Pinsonneault}, M.~H., \& {Rebull}, L.~M. 2021, \apjs, 257,
  46

\bibitem[{{Grevesse} \& {Sauval}(1998)}]{1998SSRv...85..161G}
{Grevesse}, N. \& {Sauval}, A.~J. 1998, \ssr, 85, 161

\bibitem[{{Heger} {et~al.}(2000){Heger}, {Langer}, \&
  {Woosley}}]{2000ApJ...528..368H}
{Heger}, A., {Langer}, N., \& {Woosley}, S.~E. 2000, \apj, 528, 368

\bibitem[{{Hidalgo} {et~al.}(2018){Hidalgo}, {Pietrinferni}, {Cassisi},
  {Salaris}, {Mucciarelli}, {Savino}, {Aparicio}, {Silva Aguirre}, \&
  {Verma}}]{2018ApJ...856..125H}
{Hidalgo}, S.~L., {Pietrinferni}, A., {Cassisi}, S., {et~al.} 2018, \apj, 856,
  125

\bibitem[{{Higl} {et~al.}(2021){Higl}, {M{\"u}ller}, \&
  {Weiss}}]{2021A&A...646A.133H}
{Higl}, J., {M{\"u}ller}, E., \& {Weiss}, A. 2021, \aap, 646, A133

\bibitem[{{Iglesias} \& {Rogers}(1996)}]{1996ApJ...464..943I}
{Iglesias}, C.~A. \& {Rogers}, F.~J. 1996, \apj, 464, 943

\bibitem[{{Itoh} {et~al.}(2008){Itoh}, {Uchida}, {Sakamoto}, {Kohyama}, \&
  {Nozawa}}]{2008ApJ...677..495I}
{Itoh}, N., {Uchida}, S., {Sakamoto}, Y., {Kohyama}, Y., \& {Nozawa}, S. 2008,
  \apj, 677, 495

\bibitem[{{Jermyn} {et~al.}(2018){Jermyn}, {Tout}, \& {Chitre}}]{Jermyn2018}
{Jermyn}, A.~S., {Tout}, C.~A., \& {Chitre}, S.~M. 2018, \mnras, 480, 5427

\bibitem[{{Kalirai} {et~al.}(2008){Kalirai}, {Hansen}, {Kelson}, {Reitzel},
  {Rich}, \& {Richer}}]{2008ApJ...676..594K}
{Kalirai}, J.~S., {Hansen}, B. M.~S., {Kelson}, D.~D., {et~al.} 2008, \apj,
  676, 594

\bibitem[{{Kippenhahn} {et~al.}(1970){Kippenhahn}, {Meyer-Hofmeister}, \&
  {Thomas}}]{1970A&A.....5..155K}
{Kippenhahn}, R., {Meyer-Hofmeister}, E., \& {Thomas}, H.~C. 1970, \aap, 5, 155

\bibitem[{{Kippenhahn} \& {Thomas}(1970)}]{1970stro.coll...20K}
{Kippenhahn}, R. \& {Thomas}, H.~C. 1970, in IAU Colloq. 4: Stellar Rotation,
  ed. A.~{Slettebak}, 20

\bibitem[{{Kippenhahn} {et~al.}(2012){Kippenhahn}, {Weigert}, \&
  {Weiss}}]{2012sse..book.....K}
{Kippenhahn}, R., {Weigert}, A., \& {Weiss}, A. 2012, {Stellar Structure and
  Evolution}

\bibitem[{{Komatsu} {et~al.}(2011){Komatsu}, {Smith}, {Dunkley}, {Bennett},
  {Gold}, {Hinshaw}, {Jarosik}, {Larson}, {Nolta}, {Page}, {Spergel},
  {Halpern}, {Hill}, {Kogut}, {Limon}, {Meyer}, {Odegard}, {Tucker}, {Weiland},
  {Wollack}, \& {Wright}}]{2011ApJS..192...18K}
{Komatsu}, E., {Smith}, K.~M., {Dunkley}, J., {et~al.} 2011, \apjs, 192, 18

\bibitem[{{Lindegren} {et~al.}(2018){Lindegren}, {Hern{\'a}ndez}, {Bombrun},
  {Klioner}, {Bastian}, {Ramos-Lerate}, {de Torres}, {Steidelm{\"u}ller},
  {Stephenson}, {Hobbs}, {Lammers}, {Biermann}, {Geyer}, {Hilger}, {Michalik},
  {Stampa}, {McMillan}, {Casta{\~n}eda}, {Clotet}, {Comoretto}, {Davidson},
  {Fabricius}, {Gracia}, {Hambly}, {Hutton}, {Mora}, {Portell}, {van Leeuwen},
  {Abbas}, {Abreu}, {Altmann}, {Andrei}, {Anglada}, {Balaguer-N{\'u}{\~n}ez},
  {Barache}, {Becciani}, {Bertone}, {Bianchi}, {Bouquillon}, {Bourda},
  {Br{\"u}semeister}, {Bucciarelli}, {Busonero}, {Buzzi}, {Cancelliere},
  {Carlucci}, {Charlot}, {Cheek}, {Crosta}, {Crowley}, {de Bruijne}, {de
  Felice}, {Drimmel}, {Esquej}, {Fienga}, {Fraile}, {Gai}, {Garralda},
  {Gonz{\'a}lez-Vidal}, {Guerra}, {Hauser}, {Hofmann}, {Holl}, {Jordan},
  {Lattanzi}, {Lenhardt}, {Liao}, {Licata}, {Lister}, {L{\"o}ffler},
  {Marchant}, {Martin-Fleitas}, {Messineo}, {Mignard}, {Morbidelli}, {Poggio},
  {Riva}, {Rowell}, {Salguero}, {Sarasso}, {Sciacca}, {Siddiqui}, {Smart},
  {Spagna}, {Steele}, {Taris}, {Torra}, {van Elteren}, {van Reeven}, \&
  {Vecchiato}}]{2018A&A...616A...2L}
{Lindegren}, L., {Hern{\'a}ndez}, J., {Bombrun}, A., {et~al.} 2018, \aap, 616,
  A2

\bibitem[{{Lodders} {et~al.}(2009){Lodders}, {Palme}, \&
  {Gail}}]{2009LanB...4B..712L}
{Lodders}, K., {Palme}, H., \& {Gail}, H.~P. 2009, Landolt B\&ouml;rnstein, 4B,
  712

\bibitem[{{Maeder}(1975)}]{1975A&A....40..303M}
{Maeder}, A. 1975, \aap, 40, 303

\bibitem[{{Maeder}(2009)}]{2009pfer.book.....M}
{Maeder}, A. 2009, {Physics, Formation and Evolution of Rotating Stars}

\bibitem[{{Maeder} \& {Meynet}(2000)}]{2000A&A...361..159M}
{Maeder}, A. \& {Meynet}, G. 2000, \aap, 361, 159

\bibitem[{{Marigo} \& {Aringer}(2009)}]{2009A&A...508.1539M}
{Marigo}, P. \& {Aringer}, B. 2009, \aap, 508, 1539

\bibitem[{{Marigo} {et~al.}(2013){Marigo}, {Bressan}, {Nanni}, {Girardi}, \&
  {Pumo}}]{2013MNRAS.434..488M}
{Marigo}, P., {Bressan}, A., {Nanni}, A., {Girardi}, L., \& {Pumo}, M.~L. 2013,
  \mnras, 434, 488

\bibitem[{{Marigo} {et~al.}(2017){Marigo}, {Girardi}, {Bressan}, {Rosenfield},
  {Aringer}, {Chen}, {Dussin}, {Nanni}, {Pastorelli}, {Rodrigues}, {Trabucchi},
  {Bladh}, {Dalcanton}, {Groenewegen}, {Montalb{\'a}n}, \&
  {Wood}}]{2017ApJ...835...77M}
{Marigo}, P., {Girardi}, L., {Bressan}, A., {et~al.} 2017, \apj, 835, 77

\bibitem[{{McQuillan} {et~al.}(2014){McQuillan}, {Mazeh}, \&
  {Aigrain}}]{2014ApJS..211...24M}
{McQuillan}, A., {Mazeh}, T., \& {Aigrain}, S. 2014, \apjs, 211, 24

\bibitem[{{Meynet} \& {Maeder}(1997)}]{1997A&A...321..465M}
{Meynet}, G. \& {Maeder}, A. 1997, \aap, 321, 465

\bibitem[{{Meynet} {et~al.}(1994){Meynet}, {Maeder}, {Schaller}, {Schaerer}, \&
  {Charbonnel}}]{Meynet_etal94}
{Meynet}, G., {Maeder}, A., {Schaller}, G., {Schaerer}, D., \& {Charbonnel}, C.
  1994, \aaps, 103, 97

\bibitem[{{Miglio} {et~al.}(2012){Miglio}, {Brogaard}, {Stello}, {Chaplin},
  {D'Antona}, {Montalb{\'a}n}, {Basu}, {Bressan}, {Grundahl}, {Pinsonneault},
  {Serenelli}, {Elsworth}, {Hekker}, {Kallinger}, {Mosser}, {Ventura},
  {Bonanno}, {Noels}, {Silva Aguirre}, {Szabo}, {Li}, {McCauliff}, {Middour},
  \& {Kjeldsen}}]{2012MNRAS.419.2077M}
{Miglio}, A., {Brogaard}, K., {Stello}, D., {et~al.} 2012, \mnras, 419, 2077

\bibitem[{{Moc{\'a}k} {et~al.}(2008){Moc{\'a}k}, {M{\"u}ller}, {Weiss}, \&
  {Kifonidis}}]{2008A&A...490..265M}
{Moc{\'a}k}, M., {M{\"u}ller}, E., {Weiss}, A., \& {Kifonidis}, K. 2008, \aap,
  490, 265

\bibitem[{{Mowlavi} {et~al.}(2012){Mowlavi}, {Eggenberger}, {Meynet},
  {Ekstr{\"o}m}, {Georgy}, {Maeder}, {Charbonnel}, \&
  {Eyer}}]{2012A&A...541A..41M}
{Mowlavi}, N., {Eggenberger}, P., {Meynet}, G., {et~al.} 2012, \aap, 541, A41

\bibitem[{{Noll} {et~al.}(2021){Noll}, {Deheuvels}, \&
  {Ballot}}]{2021A&A...647A.187N}
{Noll}, A., {Deheuvels}, S., \& {Ballot}, J. 2021, \aap, 647, A187

\bibitem[{{Paxton} {et~al.}(2011){Paxton}, {Bildsten}, {Dotter}, {Herwig},
  {Lesaffre}, \& {Timmes}}]{Paxton2011ApJS}
{Paxton}, B., {Bildsten}, L., {Dotter}, A., {et~al.} 2011, \apjs, 192, 3

\bibitem[{{Paxton} {et~al.}(2018){Paxton}, {Schwab}, {Bauer}, {Bildsten},
  {Blinnikov}, {Duffell}, {Farmer}, {Goldberg}, {Marchant}, {Sorokina},
  {Thoul}, {Townsend}, \& {Timmes}}]{Paxton2018ApJS}
{Paxton}, B., {Schwab}, J., {Bauer}, E.~B., {et~al.} 2018, \apjs, 234, 34

\bibitem[{{Pietrinferni} {et~al.}(2004){Pietrinferni}, {Cassisi}, {Salaris}, \&
  {Castelli}}]{2004ApJ...612..168P}
{Pietrinferni}, A., {Cassisi}, S., {Salaris}, M., \& {Castelli}, F. 2004, \apj,
  612, 168

\bibitem[{{Pols} {et~al.}(1998){Pols}, {Schr{\"o}der}, {Hurley}, {Tout}, \&
  {Eggleton}}]{Pols1998}
{Pols}, O.~R., {Schr{\"o}der}, K.-P., {Hurley}, J.~R., {Tout}, C.~A., \&
  {Eggleton}, P.~P. 1998, \mnras, 298, 525

\bibitem[{{Potter} {et~al.}(2012){Potter}, {Tout}, \&
  {Eldridge}}]{2012MNRAS.419..748P}
{Potter}, A.~T., {Tout}, C.~A., \& {Eldridge}, J.~J. 2012, \mnras, 419, 748

\bibitem[{{Reimers}(1975)}]{1975MSRSL...8..369R}
{Reimers}, D. 1975, Memoires of the Societe Royale des Sciences de Liege, 8,
  369

\bibitem[{{Reimers}(1977)}]{1977A&A....61..217R}
{Reimers}, D. 1977, \aap, 61, 217

\bibitem[{{Renzini} \& {Fusi Pecci}(1988)}]{1988ARA&A..26..199R}
{Renzini}, A. \& {Fusi Pecci}, F. 1988, \araa, 26, 199

\bibitem[{{Riello} {et~al.}(2021){Riello}, {de Angeli}, {Evans}, {Montegriffo},
  {Carrasco}, {Busso}, {Palaversa}, {Burgess}, {Diener}, {Davidson}, {Rowell},
  {Fabricius}, {Jordi}, {Bellazzini}, {Pancino}, {Harrison}, {Cacciari}, {van
  Leeuwen}, {Hambly}, {Hodgkin}, {Osborne}, {Altavilla}, {Barstow}, {Brown},
  {Castellani}, {Cowell}, {de Luise}, {Gilmore}, {Giuffrida}, {Hidalgo},
  {Holland}, {Marinoni}, {Pagani}, {Piersimoni}, {Pulone}, {Ragaini}, {Rainer},
  {Richards}, {Sanna}, {Walton}, {Weiler}, \& {Yoldas}}]{2021yCat..36490003R}
{Riello}, M., {de Angeli}, F., {Evans}, D.~W., {et~al.} 2021, VizieR Online
  Data Catalog, J/A+A/649/A3

\bibitem[{{Rosenfield} {et~al.}(2014){Rosenfield}, {Marigo}, {Girardi},
  {Dalcanton}, {Bressan}, {Gullieuszik}, {Weisz}, {Williams}, {Dolphin}, \&
  {Aringer}}]{2014ApJ...790...22R}
{Rosenfield}, P., {Marigo}, P., {Girardi}, L., {et~al.} 2014, \apj, 790, 22

\bibitem[{{Roxburgh}(1978)}]{Roxburgh1978}
{Roxburgh}, I.~W. 1978, \aap, 65, 281

\bibitem[{{Royer} {et~al.}(2007){Royer}, {Zorec}, \&
  {G{\'o}mez}}]{2007A&A...463..671R}
{Royer}, F., {Zorec}, J., \& {G{\'o}mez}, A.~E. 2007, \aap, 463, 671

\bibitem[{{Salaris} {et~al.}(2009){Salaris}, {Serenelli}, {Weiss}, \& {Miller
  Bertolami}}]{2009ApJ...692.1013S}
{Salaris}, M., {Serenelli}, A., {Weiss}, A., \& {Miller Bertolami}, M. 2009,
  \apj, 692, 1013

\bibitem[{{Sandquist} {et~al.}(2021){Sandquist}, {Latham}, {Mathieu}, {Leiner},
  {Vanderburg}, {Stello}, {Orosz}, {Bedin}, {Libralato}, {Malavolta}, \&
  {Nardiello}}]{2021AJ....161...59S}
{Sandquist}, E.~L., {Latham}, D.~W., {Mathieu}, R.~D., {et~al.} 2021, \aj, 161,
  59

\bibitem[{{Sarajedini} {et~al.}(2009){Sarajedini}, {Dotter}, \&
  {Kirkpatrick}}]{2009ApJ...698.1872S}
{Sarajedini}, A., {Dotter}, A., \& {Kirkpatrick}, A. 2009, \apj, 698, 1872

\bibitem[{{Saslaw} \& {Schwarzschild}(1965)}]{1965ApJ...142.1468S}
{Saslaw}, W.~C. \& {Schwarzschild}, M. 1965, \apj, 142, 1468

\bibitem[{{Schr{\"o}der} \& {Cuntz}(2005)}]{2005ApJ...630L..73S}
{Schr{\"o}der}, K.~P. \& {Cuntz}, M. 2005, \apjl, 630, L73

\bibitem[{{Schwarzschild}(1958)}]{Schwarzschild1958}
{Schwarzschild}, M. 1958, {Structure and evolution of the stars.} (Princeton,
  Princeton University Press, 1958.)

\bibitem[{{Song} {et~al.}(2020){Song}, {Alexeeva}, {Sitnova}, {Wang}, {Grupp},
  \& {Zhao}}]{2020A&A...635A.176S}
{Song}, N., {Alexeeva}, S., {Sitnova}, T., {et~al.} 2020, \aap, 635, A176

\bibitem[{{Sonoi} {et~al.}(2019){Sonoi}, {Ludwig}, {Dupret}, {Montalb{\'a}n},
  {Samadi}, {Belkacem}, {Caffau}, \& {Goupil}}]{2019A&A...621A..84S}
{Sonoi}, T., {Ludwig}, H.~G., {Dupret}, M.~A., {et~al.} 2019, \aap, 621, A84

\bibitem[{{Spada} {et~al.}(2017){Spada}, {Demarque}, {Kim}, {Boyajian}, \&
  {Brewer}}]{2017ApJ...838..161S}
{Spada}, F., {Demarque}, P., {Kim}, Y.~C., {Boyajian}, T.~S., \& {Brewer},
  J.~M. 2017, \apj, 838, 161

\bibitem[{{Spera} {et~al.}(2019){Spera}, {Mapelli}, {Giacobbo}, {Trani},
  {Bressan}, \& {Costa}}]{Spera2019MNRAS.485..889S}
{Spera}, M., {Mapelli}, M., {Giacobbo}, N., {et~al.} 2019, \mnras, 485, 889

\bibitem[{{Stello} {et~al.}(2016){Stello}, {Vanderburg}, {Casagrande},
  {Gilliland}, {Silva Aguirre}, {Sandquist}, {Leiner}, {Mathieu}, \&
  {Soderblom}}]{2016ApJ...832..133S}
{Stello}, D., {Vanderburg}, A., {Casagrande}, L., {et~al.} 2016, \apj, 832, 133

\bibitem[{{Talon} \& {Zahn}(1997)}]{1997A&A...317..749T}
{Talon}, S. \& {Zahn}, J.~P. 1997, \aap, 317, 749

\bibitem[{{Tang} {et~al.}(2014){Tang}, {Bressan}, {Rosenfield}, {Slemer},
  {Marigo}, {Girardi}, \& {Bianchi}}]{2014MNRAS.445.4287T}
{Tang}, J., {Bressan}, A., {Rosenfield}, P., {et~al.} 2014, \mnras, 445, 4287

\bibitem[{{Torres} {et~al.}(2014){Torres}, {Vaz}, {Sandberg Lacy}, \&
  {Claret}}]{2014AJ....147...36T}
{Torres}, G., {Vaz}, L. P.~R., {Sandberg Lacy}, C.~H., \& {Claret}, A. 2014,
  \aj, 147, 36

\bibitem[{{Ventura} {et~al.}(1998){Ventura}, {Zeppieri}, {Mazzitelli}, \&
  {D'Antona}}]{1998A&A...334..953V}
{Ventura}, P., {Zeppieri}, A., {Mazzitelli}, I., \& {D'Antona}, F. 1998, \aap,
  334, 953

\bibitem[{{Viani} {et~al.}(2018){Viani}, {Basu}, {Ong J.}, {Bonaca}, \&
  {Chaplin}}]{2018ApJ...858...28V}
{Viani}, L.~S., {Basu}, S., {Ong J.}, M.~J., {Bonaca}, A., \& {Chaplin}, W.~J.
  2018, \apj, 858, 28

\bibitem[{{Vink} {et~al.}(2001){Vink}, {de Koter}, \&
  {Lamers}}]{2001A&A...369..574V}
{Vink}, J.~S., {de Koter}, A., \& {Lamers}, H.~J.~G.~L.~M. 2001, \aap, 369, 574

\bibitem[{{Viscasillas V{\'a}zquez} {et~al.}(2022){Viscasillas V{\'a}zquez},
  {Magrini}, {Casali}, {Tautvai{\v{s}}ien{\.{e}}}, {Spina}, {Van der Swaelmen},
  {Randich}, {Bensby}, {Bragaglia}, {Friel}, {Feltzing}, {Sacco}, {Turchi},
  {Jim{\'e}nez-Esteban}, {D'Orazi}, {Delgado-Mena}, {Mikolaitis},
  {Drazdauskas}, {Minkevi{\v{c}}i{\={u}}t{\.{e}}}, {Stonkut{\.{e}}},
  {Bagdonas}, {Montes}, {Guiglion}, {Baratella}, {Tabernero}, {Gilmore},
  {Alfaro}, {Francois}, {Korn}, {Smiljanic}, {Bergemann}, {Franciosini},
  {Gonneau}, {Hourihane}, {Worley}, \& {Zaggia}}]{2022A&A...660A.135V}
{Viscasillas V{\'a}zquez}, C., {Magrini}, L., {Casali}, G., {et~al.} 2022,
  \aap, 660, A135

\bibitem[{{von Steiger} \& {Zurbuchen}(2016)}]{2016ApJ...816...13V}
{von Steiger}, R. \& {Zurbuchen}, T.~H. 2016, \apj, 816, 13

\bibitem[{{von Zeipel}(1924{\natexlab{a}})}]{1924MNRAS..84..665V}
{von Zeipel}, H. 1924{\natexlab{a}}, \mnras, 84, 665

\bibitem[{{von Zeipel}(1924{\natexlab{b}})}]{1924MNRAS..84..684V}
{von Zeipel}, H. 1924{\natexlab{b}}, \mnras, 84, 684

\bibitem[{{Weiss} \& {Schlattl}(2008)}]{Weiss2008}
{Weiss}, A. \& {Schlattl}, H. 2008, \apss, 316, 99

\bibitem[{{Xu} {et~al.}(2013){Xu}, {Goriely}, {Jorissen}, {Chen}, \&
  {Arnould}}]{2013A&A...549A.106X}
{Xu}, Y., {Goriely}, S., {Jorissen}, A., {Chen}, G.~L., \& {Arnould}, M. 2013,
  \aap, 549, A106

\bibitem[{{Zahn}(1992)}]{1992A&A...265..115Z}
{Zahn}, J.~P. 1992, \aap, 265, 115

\end{thebibliography}

%%%%%%%%%%%%%%%%%%%%%%%%%%%%%%%%%%%%
\makeatletter
% define \thebiblio (same as thebibliography, but
% without the section heading)
\def\thebiblio#1{%
 \list{}{\usecounter{dummy}%
         \labelwidth\z@
         \leftmargin 1.5em
         \itemsep \z@
         \itemindent-\leftmargin}
 \reset@font\small
 \parindent\z@
 \parskip\z@ plus .1pt\relax
 \def\newblock{\hskip .11em plus .33em minus .07em}
 \sloppy\clubpenalty4000\widowpenalty4000
 \sfcode`\.=1000\relax
}
\let\endthebiblio=\endlist
\makeatother

\label{lastpage}

\end{document}